\makeatletter
\def\input@path{{latex/}}
\makeatother
\documentclass[cernpreprint,texlive=2016,UKenglish,texmf]{atlasdoc}
 
\usepackage[subcaption=true,block=none]{atlaspackage}
\usepackage[backref=false]{atlasbiblatex}
 
\usepackage{hyperref}
\usepackage[export]{adjustbox}
 
\usepackage[BSM=true] {atlasphysics}
 
\DeclareFieldFormat[incollection]{title}{\emph{#1}}
\DeclareFieldFormat[inproceedings]{title}{\emph{#1}}
 
\graphicspath{{logos/}{figures/}}
 
\usepackage{ANA-SUSY-2019-04-PAPER-defs}
\usepackage{multirow}
\usepackage{cancel}

\usepackage[color=orange,]{todonotes}

\AtlasTitle{Search for R-parity-violating supersymmetry in a final state containing leptons and many jets with the ATLAS experiment using $\sqrt{s} = \SI{13}{TeV}$ proton--proton collision data}
 
\AtlasAbstract{
A search for R-parity-violating supersymmetry in final states characterized by high jet multiplicity, at least one isolated light lepton and either zero or at least three $b$-tagged jets is presented.
The search uses $\SI{139}{\text{fb}^{-1}}$ of $\sqrt{s} = \SI{13}{TeV}$ proton--proton collision data collected by the ATLAS experiment during Run 2 of the Large Hadron Collider.
The results are interpreted in the context of R-parity-violating supersymmetry models that feature gluino production, top-squark production, or electroweakino production.
The dominant sources of background are estimated using a data-driven model, based on observables at medium jet multiplicity,
to predict the $b$-tagged jet multiplicity distribution at the higher jet multiplicities used in the search.
Machine-learning techniques are used to reach sensitivity to electroweakino production, extending the data-driven background estimation to the shape of the machine-learning discriminant.
No significant excess over the Standard Model expectation is observed and exclusion limits at the 95\% confidence level are extracted,
reaching as high as \SI{2.4}{TeV} in gluino mass, \SI{1.35}{TeV} in top-squark mass, and 320 (365)~\gev\ in higgsino (wino) mass.
}
 
\PreprintIdNumber{CERN-EP-2021-066}
\AtlasJournalRef{Eur. Phys. J. C 81 (2021) 1023}
\AtlasDOI{10.1140/epjc/s10052-021-09761-x}
\hypersetup{pdftitle={ATLAS document},pdfauthor={The ATLAS Collaboration}}

\begin{document}
 
\maketitle

\section{Introduction}
Supersymmetry (SUSY)~\cite{Golfand:1971iw,Volkov:1973ix,Wess:1974tw,Wess:1974jb,Ferrara:1974pu,Salam:1974ig}
is a theoretical extension of the Standard Model (SM) which fundamentally relates fermions and bosons.
It is an alluring theoretical possibility given its potential to solve the hierarchy problem~\cite{Sakai:1981gr,Dimopoulos:1981yj,Ibanez:1981yh,Dimopoulos:1981zb}.
An ad hoc conserved quantity, R-parity~\cite{Farrar:1978xj}, is often introduced in SUSY models to avoid rapid proton decay,
rendering the lightest supersymmetric particle (LSP) stable and therefore a potential dark-matter candidate~\cite{Goldberg:1983nd,Ellis:1983ew}.
There is no fundamental theoretical reason to impose strict R-parity conservation,
and R-parity-violating (RPV) SUSY models are well motivated, with fewer experimental constraints than many R-parity-conserving (RPC) models~\cite{Evans2014,Batell:2013zwa},
and allow for more natural supersymmetric mass spectra.

This article presents a search for pair production of supersymmetric particles with subsequent RPV decays in
a final state with at least one isolated lepton (electron or muon), at least eight to fifteen jets (depending on the jet transverse
momentum threshold), several of which may contain $b$-flavoured hadrons ($b$-jets), and with no requirement on the missing transverse momentum.
Such a final state is commonly predicted in RPV models with either baryon-number-violating~\cite{DAmbrosio:2002ex,Csaki:2011ge} or lepton-number-violating couplings~\cite{Bratchikov:2005vp}.
Events are assigned to one of two categories according to their lepton content. The first category contains events with two leptons with the same electric charge (\ll),
while the second category contains all other events and is dominated by single-lepton events (\l).
Electrons and muons from $\tau$-lepton decays are also considered.
A multi-bin fit in each lepton category to the two-dimensional space of jet multiplicity
and $b$-jet multiplicity is used to constrain parameters of benchmark RPV simplified signal models~\cite{Alwall:2008ve,Alwall:2008ag,Alves:2011wf}.
A third variable, based on a machine-learning 
discriminant, is introduced to improve the sensitivity of the search when only testing for
the electroweak production of supersymmetric particles.
This search has potential sensitivity to a large number of beyond the Standard Model (BSM) physics models,
and model-independent limits on the possible contribution of BSM physics to several single-bin signal regions are shown.

The dominant Standard Model background in the \l category arises
from top-quark pair production and $W/Z$~+~jets production, with at least one lepton produced in the vector-boson decay.
In the \ll category, the production of a top-quark pair in association with a $W$ boson (\ttW), or
with a misidentified lepton, constitutes the main background.
The theoretical modelling of these backgrounds at high jet multiplicity suffers
from large uncertainties, so they are estimated from the data by extrapolating the
$b$-jet multiplicity distribution extracted at moderate jet multiplicities to the high jet multiplicities of the search region.

This analysis is an update to a previous ATLAS search~\cite{Aaboud:2017faq} for new phenomena in a final state with a lepton and high jet multiplicity,
which was performed with \SI{36}{fb^{-1}} of $\sqrt s = \SI{13}{\TeV}$ proton--proton collision data.
It improves upon the previous result owing to the larger luminosity, the dedicated categorization and analysis of events with two leptons with the same electric charge,
and the introduction of multivariate discriminants.
This search represents the first LHC result to obtain sensitivity to electroweak production of SUSY particles promptly decaying to quarks, as predicted in baryon-number-violating RPV models.
Previous searches targeting similar RPV SUSY models have been carried out by the ATLAS and CMS collaborations~\cite{Aad:2020uwr,Aad:2019ftg,Aaboud:2018zeb,Aaboud:2018lpl,SUSY-2016-09,Aaboud:2017opj,SUSY-2016-14,Sirunyan:2017dhe,Khachatryan:2016iqn,Chatrchyan:2013xsw,CMS:2021knz}.
The result is also sensitive to SM four-top-quark production, and a validation of the background estimation methods is performed by fitting the normalization of the four-top-quark process relative to the Standard Model value.
Previous searches for four-top-quark
production were carried out by the ATLAS~\cite{Aad:2020klt} and CMS~\cite{Sirunyan:2017tep} collaborations.


\section{Signal models}
\label{sec:signalModels}
\begin{figure}[t!]
\centering
\begin{subfigure}[t]{0.32\linewidth}
\includegraphics[width=1\textwidth]{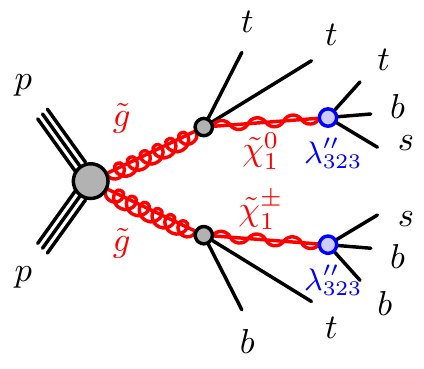}
\caption{}
\label{fig:Gtt}
\end{subfigure}
\begin{subfigure}[t]{0.32\linewidth}
\includegraphics[width=1\textwidth]{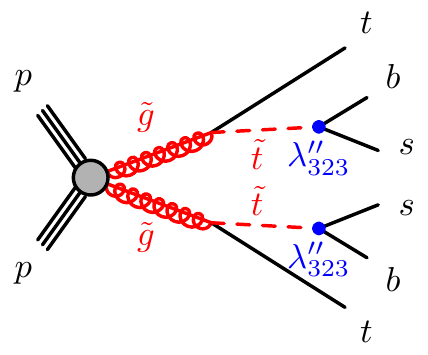}
\caption{}\label{fig:Gtbs}
\end{subfigure}
\begin{subfigure}[t]{0.32\linewidth}
\includegraphics[width=1\textwidth]{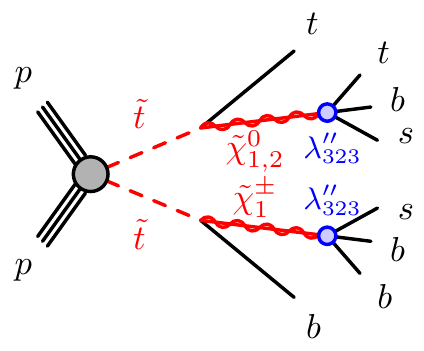}
\caption{}\label{fig:stop}
\end{subfigure}
\begin{subfigure}[t]{0.32\linewidth}
\includegraphics[width=1\textwidth]{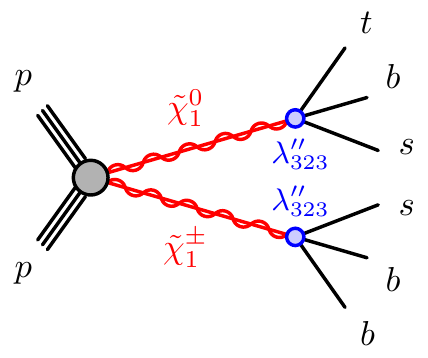}
\caption{}\label{fig:C1N1}
\end{subfigure}
\begin{subfigure}[t]{0.32\linewidth}
\includegraphics[width=1\textwidth]{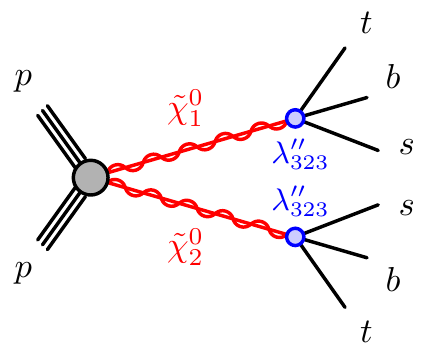}
\caption{}\label{fig:N1N2}
\end{subfigure}
\begin{subfigure}[t]{0.32\linewidth}
\includegraphics[width=1\textwidth]{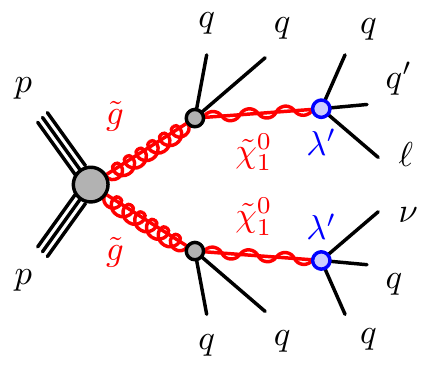}
\caption{}\label{fig:LQD}
\end{subfigure}
\caption{Examples of signal diagrams for the simplified RPV models considered in this article. Cases where both of the gluinos (or the stops) decay in the same way are also considered,
and $\chinoonepm\ninotwo$ pair production is also considered for the higgsino LSP type. For simplicity particles and anti-particles are shown using the same symbols, omitting the anti-particle notation.
}
\label{fig:diagram}
\end{figure}

Simulated signal events from five SUSY benchmark simplified models (representative production diagrams shown in Figure~\ref{fig:diagram}) are used to guide the analysis selections and
to estimate the expected signal yields for different signal-mass hypotheses used to interpret
the analysis results. In all models considered, the RPV couplings and the SUSY particle masses are chosen to ensure prompt decays of the SUSY particles.
The supersymmetric particle content of the models is the partner of the SM gluon (gluino), the partner of the right-handed top quark (stop), and one or more electroweakinos.
The electroweakinos are massive fermions resulting from the
mixing between the partners of SM electroweak and Higgs bosons.\footnote{In SUSY, the Higgs sector is enriched by the presence of an additional complex doublet.}
Three different possibilities for the electroweakino composition are tested: pure bino, pure wino or pure higgsino.
In all cases the lightest neutralino (\ninoone) is the LSP.
When considering a wino (higgsino) LSP, the corresponding chargino \chinoonepm\ (and second neutralino \ninotwo)
is assumed to be effectively mass degenerate with the LSP, as predicted by theory~\cite{Ibe:2012sx,Nagata:2014wma},
and share the same composition as the LSP.
All the electroweakinos that are present under the hypothesis of a given composition are considered in both the production and decay processes.
All other electroweakinos are assumed to be decoupled and not considered in the model.
The gluino and stop branching ratios, as well as the electroweakino production cross-section, are determined by the nature of the electroweakino.
Table~\ref{tab:BRs_xsec} summarizes the gluino and stop branching ratios, and shows example cross-sections for direct electroweakino production~\cite{Beenakker_corr:1999xh,Bozzi:2007qr,Fuks:2013vua,Fuks:2013lya,Fiaschi:2018xdm}, for each electroweakino type.
In each case, the electroweakinos decay through a non-zero RPV coupling large enough to ensure prompt decays for the particle masses considered,
and small enough to avoid more complex decay patterns involving mixtures of both RPC and RPV decays that are not considered here.
Within this scenario the analysis results are independent of the value of the coupling.

\begin{table}[t!]
\centering
\caption{
Stop and gluino branching ratios, as well as cross-sections for direct electroweakino production, as a function of the LSP type.
For a pure bino/wino/higgsino LSP, the electroweakino states considered are \ninoone/\raise.4ex\hbox{$\tilde\chi^0_{1}\tilde\chi^\pm_{1}$}/\raise.4ex\hbox{$\tilde\chi^0_{1}\tilde\chi^\pm_{1}\tilde\chi^0_{2}$}, respectively.
When relevant, decays to \ninoone and \ninotwo are merged as they are assumed to be mass degenerate and both decay in the same way.
The production cross-sections are given for an electroweakino mass of $m(\ninoone,\chinoonepm,\ninotwo)=\SI{300}{\GeV}$.
}
\vspace{0.2cm}
\def\arraystretch{1.4}
\begin{tabular}{| l | c c | c c c | c c c |}
\hline
LSP      & \multicolumn{4}{c}{ Branching ratios: } & & \multicolumn{3}{c|}{Cross-section [fb]} \\
type     & \multicolumn{2}{c|}{ stop } & \multicolumn{3}{c|}{ gluino } &\multicolumn{3}{c|}{for direct production} \\
& $t \ninoonetwo$ & $b \chinoonepm$ & $tt \ninoonetwo$ & $bb \ninoonetwo$ &  $tb \chinoonepm$ & $\chinoonepm\ninoone$  &  $\chinoonepm\ninotwo$  & $\ninotwo\ninoone$  \\
\hline
Bino     & 100\% &  ~~0\%  & 100\% &   ~~0\%  &   ~~0\%  & ~~~~0   & ~~0   & ~~0 \\
Wino     & ~~33\%  &  67\% & ~~17\%  &   17\% &   66\% & 387 & ~~0   & ~~0 \\
Higgsino & ~~50\%  &  50\% & ~~50\%  &   ~~0\%  &   50\% & ~~91  & 91 & 52 \\
\hline
\end{tabular}
\label{tab:BRs_xsec}
\end{table}

Four of the simplified models are inspired by a common natural RPV SUSY model assuming the minimal flavour violation hypothesis~\cite{DAmbrosio:2002ex,Csaki:2011ge}.
The coupling $\lambda_{323}''$ is chosen, as it is predicted to be dominant.\footnote{The results apply equally to the coupling $\lambda_{313}''$ since it leads to the same experimental final state.}
With the chosen model parameters, the electroweakinos decay as $\tilde\chi^0_{1/2} \to t b s$ and $\tilde\chi^\pm_1 \to b b s$, with a branching ratio of 100\%.
At the lowest order in perturbation theory, signal events in these models contain four, six, or eight $b$-jets in the final state, depending on the production mode.
In the first model, gluino production is considered, with decays to heavy-flavour quarks and the electroweakino, which in turn decays via the RPV coupling.
The stop, with a mass assumed to be above the gluino mass, is not considered in the model.
A signal diagram for this model is shown in Figure~\ref{fig:Gtt}.
The second model considers gluino pair production, with each gluino decaying into a top quark and a stop, as shown in Figure~\ref{fig:Gtbs}.
In this model the RPV coupling is assumed to be large, so that the stop decays via an RPV mode into an $s$-quark and a $b$-quark.
The absence of RPC decays of the stop render the electroweakino mass irrelevant in this model.
The third scenario considered involves stop pair production with the stop decaying into an electroweakino and a top or bottom quark,
while the gluino is set to a very high mass and not considered in the model.
An example signal diagram is shown in Figure~\ref{fig:stop}.
In the fourth model, only electroweakino production is considered, with the stop and gluino assigned a very high mass and not considered in the model.
Figure~\ref{fig:C1N1} considers the production of $\chinoonepm\ninoone$. In the case of a higgsino LSP, a similar diagram produces $\chinoonepm\ninotwo$, and a further diagram producing $\ninoone\ninotwo$ is enabled, as shown in Figure~\ref{fig:N1N2}.
Production of $\chinoonepm\ninoonetwo$ contributes only to the \l category, while the production of $\ninoone\ninotwo$ contributes to both the \l and \ll categories.
Production of $\chinoonepm\chinoonepm$ is not considered because the decays produce a final state with no leptons. Production of $\ninoone\ninoone$ vanishes and is not considered. Therefore electroweak production in the pure bino scenario is not considered

The fifth and last simplified RPV model considers gluino pair production, where each gluino decays into
two first- or second-generation quarks ($q \equiv (u,d,s,c)$) and a \ninoone, which is the LSP.
The \ninoone\ decays into two additional first- or second-generation quarks and a charged lepton or a neutrino
($\ninoone \to q \bar{q}' \ell$ or  $\ninoone \to q \bar q  \nu$, labelled as $\ninoone \to q \bar q \ell / \nu$).
The decay proceeds via a $\lambda'$  RPV coupling, where each RPV decay can produce
any of the four first- or second-generation leptons ($e^{\pm},\mu^{\pm},\nu_e,\nu_\mu$) with equal probability.
An example signal diagram is shown in Figure~\ref{fig:LQD}.
Signal decays from this model yield a final state with high jet multiplicity and zero $b$-jets.
 

\section{ATLAS detector}  
\newcommand{\AtlasCoordFootnote}{
ATLAS uses a right-handed coordinate system with its origin at the nominal interaction point (IP)
in the centre of the detector and the \(z\)-axis along the beam pipe.
The \(x\)-axis points from the IP to the centre of the LHC ring,
and the \(y\)-axis points upwards.
Cylindrical coordinates \((r,\phi)\) are used in the transverse plane,
\(\phi\) being the azimuthal angle around the \(z\)-axis.
The pseudorapidity is defined in terms of the polar angle \(\theta\) as \(\eta = -\ln \tan(\theta/2)\).
Angular distance is measured in units of \(\Delta R \equiv \sqrt{(\Delta y)^{2} + (\Delta\phi)^{2}}\).}

The ATLAS experiment~\cite{PERF-2007-01} at the LHC is a multipurpose particle detector
with a forward--backward symmetric cylindrical geometry and nearly \(4\pi\) coverage in
solid angle.\footnote{\AtlasCoordFootnote}
It consists of an inner tracking detector (ID) surrounded by a thin superconducting solenoid
providing a \SI{2}{\tesla} axial magnetic field, electromagnetic (EM) and hadron calorimeters, and a muon spectrometer (MS).
The inner tracking detector covers the pseudorapidity range \(|\eta| < 2.5\).
It consists of silicon pixel, silicon microstrip and transition radiation tracking detectors;
the innermost layer is \SI{33}{mm} from the beamline~\cite{B-layerRef,PIX-2018-001}.
Lead/liquid-argon (LAr) sampling calorimeters provide electromagnetic energy measurements
with high granularity.
A steel/scintillator-tile hadron calorimeter covers the central pseudorapidity range (\(|\eta| < 1.7\)).
The endcap and forward regions are instrumented with LAr calorimeters
for both the EM and hadronic energy measurements up to \(|\eta| = 4.9\).
The muon spectrometer surrounds the calorimeters and is based on
three large superconducting air-core toroidal magnets with eight coils each.
The field integral of the toroids ranges between \num{2.0} and \SI{6.0}{\tesla$\cdot$\metre}
across most of the detector.
The muon spectrometer includes a system of precision tracking chambers and fast detectors for triggering.
A two-level trigger system is used to select events~\cite{TRIG-2012-01}.
The first-level trigger is implemented in hardware and uses a subset of the detector information
to keep the accepted rate below approximately \SI{100}{\kHz}.
This is followed by a software-based trigger that
reduces the accepted event rate to approximately \SI{1}{\kHz} on average
depending on the data-taking conditions.
An extensive software suite~\cite{ATL-SOFT-PUB-2021-001} is used for real and simulated data reconstruction
and analysis, for operation and in the trigger and data acquisition systems of the experiment.

\section{Monte Carlo event simulation} 
\label{sec:mcSimulation}
Signal and background events produced in proton--proton collisions were simulated with various Monte Carlo (MC) generators.
The simulated events are used in the optimization of event selection criteria, in the neural network training,
to estimate systematic uncertainties,
to validate the background estimation procedure employed for the dominant background sources,
and to predict yields for the subdominant background contributions and for possible signals.
The signal and background events were passed through the $\GEANT$~\cite{Agostinelli2003250} simulation
of the ATLAS detector~\cite{SOFT-2010-01} and reconstructed using the same algorithms as are used for the data.

The generation of the simulated event samples includes the effect of multiple proton--proton interactions per bunch crossing,
as well as the impact on the detector response due to interactions from bunch crossings before or after the one containing the hard interaction.
The effect of multiple interactions in the same and neighbouring bunch
crossings (pile-up) is modelled by overlaying the hard-scattering event with simulated inelastic
proton--proton events generated by \textsc{Pythia~8.186}~\cite{Sjostrand:2014zea}
using the \texttt{NNPDF2.3lo} set of parton distribution functions (PDF)~\cite{Ball:2012cx} and the
A3 set of tuned parameters~\cite{ATL-PHYS-PUB-2016-017}.
The MC events are weighted to reproduce the
distribution of the average number of interactions per bunch crossing
($\left<\mu \right>$) observed in the data.
The \evtgen~\cite{EvtGen} program was used to simulate properties of the $b$- and $c$- flavoured hadron decays.

The signal event samples were generated using the \mgamc~\cite{Alwall:2014hca} generator interfaced to \textsc{Pythia~8}
for the modelling of the parton showering, hadronization, and underlying event.
The matrix element (ME) calculation was performed at tree level and includes the emission of up to two additional partons.
The signal samples were processed through a fast simulation of the ATLAS detector~\cite{SOFT-2010-01, ATL-PHYS-PUB-2010-013}.
Gluino and stop signal cross-sections are calculated to approximate next-to-next-to-leading order in the strong coupling constant, adding the resummation of soft gluon emission at next-to-next-to-leading-logarithm accuracy (approximate NNLO+NNLL)~\cite{Beenakker:2016lwe,Beenakker_corr:2014sma,Beenakker:2013mva,Beenakker:2011sf,Beenakker:2009ha,Kulesza:2009kq,Kulesza_corr:2008jb,Beenakker:1996ch,Beenakker:1997ut,Beenakker:2010nq,Beenakker:2016gmf}.
The cross-sections for electroweakino production are calculated to next-to-leading order in the strong coupling constant, adding the resummation of soft gluon emission at next-to-leading-logarithm accuracy (NLO+NLL)~\cite{Beenakker_corr:1999xh,Bozzi:2007qr,Fuks:2013vua,Fuks:2013lya,Fiaschi:2018xdm}.

\begin{table*}[t!]
\begingroup
\begin{center}
\caption{Simulated background event samples:
the corresponding event generator, parton-shower modelling, cross-section normalization,
PDF set and underlying-event parameter tune are shown. The samples marked with (*) are alternative samples used to validate the background estimation method or to assess systematic uncertainties in the modelling. The abbreviation \textsc{MG5}\_a\textsc{MC}\ is used to label the \mgamc generator.
Samples produced with \textsc{Sherpa}\ use the default set of tuned parameters of the generator.
}
\vspace{0.2cm}
\def\arraystretch{1.2}
\setlength{\tabcolsep}{5pt} 
\scriptsize
\begin{tabular}{l l l l l l}
\hline\hline
Physics process & Event & Parton-shower & Cross-section & PDF set & Tune\\
&generator &modelling & normalization & & \\
\hline\hline
$W(\rightarrow \ell\nu)$ + jets              & \textsc{Sherpa}~2.2.1~\cite{Bothmann:2019yzt}    & \textsc{Sherpa}~2.2.1     & NNLO~\cite{Anastasiou:2003ds}  & \texttt{NNPDF3.0nnlo}~\cite{Ball:2014uwa}    & \textsc{Sherpa} \\
$W(\rightarrow \ell\nu)$ + jets (*)              & \textsc{MG5}\_a\textsc{MC}~2.2.2~\cite{Alwall:2014hca}  & \textsc{Pythia 8}{}.186~\cite{Sjostrand:2007gs}   & NNLO  &\texttt{NNPDF3.0nlo}~\cite{Ball:2014uwa}  & A14~\cite{ATL-PHYS-PUB-2014-021}\\
$Z/\gamma^{*}(\rightarrow \ell \ell)$ + jets & \textsc{Sherpa}~2.2.1   & \textsc{Sherpa}~2.2.1       & NNLO~\cite{Anastasiou:2003ds}  & \texttt{NNPDF3.0nnlo}    & \textsc{Sherpa} \\
$t\bar{t}$ + jets             & \textsc{PowhegBox}~v2~\cite{Frixione_corr:2007nw,Nason_corr:2004rx,Frixione_corr:2007vw,Alioli:2010xd} & \textsc{Pythia 8}.230~\cite{Sjostrand:2014zea}  & NNLO+NNLL~\cite{Beneke:2011mq,Cacciari:2011hy,Baernreuther:2012ws,Czakon:2012zr,Czakon:2012pz,Czakon:2013goa,Czakon:2011xx}  & \texttt{NNPDF3.0nlo}  &A14\\
$t\bar{t}$ + jets (*)              &\textsc{PowhegBox}~v2 & \textsc{Herwig}~7.04~\cite{Bahr:2008pv,Bellm:2015jjp}  & NNLO+NNLL  & \texttt{MMHT2014lo}~\cite{Harland-Lang:2014zoa} & H7UE~\cite{Bellm:2015jjp}    \\
$t\bar{t}$ + jets (*)              &\textsc{MG5}\_a\textsc{MC}~2.6.0 & \textsc{Pythia 8}.230  & NNLO+NNLL  & \texttt{NNPDF3.0nlo} & A14    \\
 
Single-top              & \textsc{PowhegBox}~v2~\cite{Alioli:2009je,Re:2010bp,Frederix:2012dh} & \textsc{Pythia 8}.230 & NNLO+NNLL~\cite{Kidonakis:2011wy,Kidonakis:2010ux,Kidonakis:2013zqa_corr}  & \texttt{NNPDF3.0nlo}   & A14\\
 
$t\bar{t}+W/Z$       & \textsc{Sherpa}~2.2.1& \textsc{Sherpa}~2.2.1 & NLO~\cite{Frederix:2017wme}  & \texttt{NNPDF3.0nnlo}& \textsc{Sherpa}\\
$t\bar{t}+\ttbar/t/WW/WZ$       & \textsc{MG5}\_a\textsc{MC}~2.3.3~\cite{Alwall:2014hca}& \textsc{Pythia 8}.230 & NLO~\cite{Frederix:2017wme} (\ttt\ LO)  & \texttt{NNPDF2.3lo}~\cite{Ball:2012cx}& A14 \\
$t\bar{t}t\bar{t}$ (*)  &\textsc{MG5}\_a\textsc{MC}~2.3.3  & \textsc{Herwig}~7.04   & NLO~\cite{Alwall:2014hca}  & \texttt{MMHT2014lo} & H7UE  \\
 
$t\bar{t}H$     &\textsc{PowhegBox}~v2&  \textsc{Pythia 8}.230     & NLO  &  \texttt{NNPDF3.0nlo}& A14 \\
$\ensuremath{tWZ}, \tZ$     & \textsc{MG5}\_a\textsc{MC}~2.3.3&  \textsc{Pythia 8}.212, 8.210     & NLO  &  \texttt{NNPDF3.0nlo}& A14\\
$VV$ and $VVV$    &  \textsc{Sherpa}~2.2.1, 2.2.2, 2.2.4 & \textsc{Sherpa}  & NLO  & \texttt{NNPDF3.0nnlo} & \textsc{Sherpa} \\
 
\hline\hline
\end{tabular}
\label{tab:MC}
\end{center}
\endgroup
\end{table*}

The production of \ttbar, \ttH, and single-top events
was modelled at NLO using the \textsc{PowhegBox}~\cite{Frixione_corr:2007nw,Nason_corr:2004rx,Frixione_corr:2007vw,Alioli:2010xd,Hartanto:2015uka}
generator.
Additional \ttbar\ samples were generated with  \mgamc interfaced with \textsc{Pythia~8}, and with \textsc{PowhegBox} interfaced with \textsc{Herwig~7}~\cite{Bahr:2008pv,Bellm:2015jjp},
for modelling comparisons and evaluation of systematic uncertainties.

The production of \ttV\ ($V$ = $W$, $Z$) events was modelled using the \textsc{Sherpa} generator.
The ME was calculated for up to one additional parton at
NLO and up to two partons at LO using the \textsc{Comix}~\cite{Gleisberg:2008fv} and
\openloops\ libraries~\cite{Cascioli:2011va,Denner:2016kdg}, and merged with the
\textsc{Sherpa} parton shower using the \textsc{MEPS@NLO} prescription~\cite{Schumann:2007mg,Hoeche:2011fd,Hoeche:2012yf,Catani:2001cc,Hoeche:2009rj}.
Alternative $\ttV$ samples produced with the \mgamc generator at NLO were used to evaluate systematic uncertainties associated with the modelling of additional QCD radiation.

The production of \tttt, \ttt, \ensuremath{tWZ}, \tZ, \ttWW, and \ttWZ events was modelled using the \mgamc
generator at NLO, and interfaced with \textsc{Pythia~8}.
An alternative \tttt\ sample showered with \textsc{Herwig~7}, was used to evaluate systematic uncertainties related to the choice of parton-shower model.

The production of an electroweak gauge boson or virtual photon in association with jets ($V$+jets) was simulated with the \textsc{Sherpa}
generator using NLO matrix elements for up to two partons, and LO matrix elements for up to four partons.
Alternative $V$+jets samples used to validate the analysis methods were simulated with \mgamc using LO-accurate MEs with up to four final-state partons.

Event samples with diboson ($VV$) and triboson ($VVV$) final states were simulated with the
\textsc{Sherpa} generator, including off-shell effects and Higgs boson contributions, where appropriate.
The $VV$ processes were simulated using matrix elements at NLO accuracy in QCD for up to one additional parton and at LO accuracy for up to three additional parton emissions.
The production of triboson ($VVV$) events was simulated with the \textsc{Sherpa} generator using factorized gauge-boson decays.
 
A summary of the background samples used, together with the event generator configurations, can be found in Table~\ref{tab:MC}.
 
\section{Event reconstruction and object identification}  
\label{sec:evtReco}
Proton--proton collision data recorded by the ATLAS detector between 2015 and 2018 are used to perform the analysis.
In this period, the LHC delivered colliding beams with a peak instantaneous luminosity up to $L=\SI{2.1e34}{cm^{-2}s^{-1}}$, achieved in 2018, and an average number of $pp$ interactions per bunch crossing of \num{33.7}.
After applying beam, detector, and data-quality criteria
the total integrated luminosity of the dataset is \SI{139}{fb^{-1}}~\cite{DAPR-2018-01}.
The uncertainty in the combined 2015--2018 integrated luminosity is 1.7\% \cite{ATLAS-CONF-2019-021},
obtained using the LUCID-2 detector \cite{LUCID2} for the primary luminosity measurements.
 
Proton--proton interaction vertices are reconstructed from charged-particle tracks with ${\pt>\SI{500}{\MeV}}$~\cite{ATL-PHYS-PUB-2015-026,PERF-2015-01} in the ID.
The presence of at least one such vertex with a minimum of two associated tracks is required,
and the vertex with the largest sum of \pTsq\ of associated tracks is chosen as the primary vertex.

Jet candidates are reconstructed up to $\abseta=4.9$ using the anti-$k_t$ algorithm~\cite{Cacciari:2008gp,Fastjet} with radius parameter $R=0.4$.
It uses particle-flow objects as inputs, combining tracking and calorimetric information as detailed in Ref.~\cite{PERF-2015-09}.
The jets are calibrated using the methodology described in Ref.~\cite{JETM-2018-05}.
Any event that contains jets induced by calorimeter noise or non-collision background,
according to criteria similar to those described in Ref.~\cite{ATLAS-CONF-2015-029}, is removed.
Jets up to $\pt=\SI{60}{\GeV}$ containing a large energy contribution from pile-up interactions are suppressed with the
jet-vertex tagging (JVT) algorithm that uses tracking and primary vertex information to determine if a given jet originates from the primary vertex ~\cite{PERF-2014-03}.
Jets with $\pt>\SI{20}{\GeV}$ and $\abseta<2.5$ are defined as signal jets, and used further in the analysis.

Signal jets containing $b$-flavoured hadrons
are identified with the DL1r \btag algorithm~\cite{FTAG-2018-01,ATL-PHYS-PUB-2017-013}
with an average identification efficiency of 70\% in simulated $t\bar t$ events.
The rejection factor is measured to be approximately 300 for jets initiated by light quarks
and gluons and approximately 9 for jets initiated by charm quarks~\cite{Aad:2019aic}.

Electron candidates are reconstructed as tracks in the ID matched to energy clusters in the EM calorimeter, within $|\eta| <  2.47$~\cite{EGAM-2018-01}.
The analysis considers only candidate electrons with $\pt > \SI{10}{\GeV}$ and not in the transition region between the barrel and endcap calorimeters ($1.37  < |\eta| <  1.52$).
The electron identification is based on a multivariate likelihood-based discriminant that uses the shower shapes in the EM calorimeter and the associated track properties measured in the ID.
The electron candidates must satisfy the `Medium' identification criteria described in Ref.~\cite{EGAM-2018-01},
while the signal electrons must satisfy the `Tight' identification for better rejection of non-prompt or misidentified electrons.
The electron identification efficiency varies with increasing $\pt$ in $Z \to ee$ events,
from 65\% at $\pt = \SI{10}{\GeV}$ to 88\% at $\SI{100}{\GeV}$ for the {Tight} operating point,
and from 75\% at $\SI{20}{\GeV}$ to 94\% at $\SI{100}{\GeV}$ for the {Medium} operating point.
 
For candidate and signal electrons, the longitudinal impact parameter of the electron track, $z_0$,
is required to satisfy $|z_{0}\sin\theta| < \SI{0.5}{mm}$, where $\theta$ is the polar angle of the track.
For signal electrons, the transverse impact parameter divided by its uncertainty, $|d_{0}|/\sigma(d_{0})$, is required to be at most five.
 
For all signal electrons there must be no association with a vertex from a reconstructed photon conversion~\cite{EGAM-2018-01} in the detector material. 
To further reduce the photon conversion background, additional requirements are applied to the signal electrons~\cite{Aad:2020klt}:
i) the candidate must not have a reconstructed displaced vertex with a conversion radius $r > \SI{20}{mm}$ whose reconstruction uses the track associated with the electron,
ii) the invariant mass of the system formed by the track associated with the electron and the closest track (in $\Delta R$) at the primary vertex or a conversion vertex
is required to be larger than $\SI{100}{\MeV}$.
This photon conversion veto has an average efficiency of 99\% for prompt electrons
while providing a rejection factor of 4 for electrons from photon conversion.
 
In the \ll category, signal electrons with wrongly reconstructed charge (charge-flip)
are suppressed using a boosted decision tree (BDT) discriminant exploiting additional tracks in the vicinity of the electron and track-to-cluster matching variables~\cite{EGAM-2018-01}.
A rejection factor of around 9 for electrons with a wrong charge assignment is achieved, while selecting
properly measured electrons with an efficiency of 98\%, in simulated $Z \to ee$ events selected with the Tight identification and isolation operating points~\cite{EGAM-2018-01}.

Muon candidates are reconstructed in the region $\abseta<2.5$ from MS tracks matching ID tracks.
Only muons with $\pt>\SI{10}{\GeV}$ satisfying the `Medium' quality requirements defined in Ref.~\cite{MUON-2018-03} are considered.
The muon reconstruction efficiency is approximately 98\% in simulated $Z \to \mu\mu$ events.
The same longitudinal impact parameter selection as for candidate and signal electrons is applied, while $|d_{0}|/\sigma(d_{0})$ is required to be at most three.

For signal electrons and muons the identification criteria are complemented by an isolation requirement,
which is based on the energy in a cone around the lepton candidate calculated using either reconstructed tracks or energy clusters.
Non-prompt electrons and muons from the decays of $b$- and $c$-flavoured hadrons are further rejected
using a BDT discriminant based on isolation and secondary vertex information,
referred to as the non-prompt-lepton veto~\cite{Aaboud:2017jvq}.
The efficiency of the combined isolation and non-prompt-lepton veto is on average above 80\% for prompt leptons with $\pt>\SI{30}{\GeV}$ in simulated diboson events.
Finally, all signal leptons are required to have $\pt>\SI{15}{\GeV}$.

A sequential overlap removal procedure to resolve ambiguities
between candidate jets and candidate leptons is carried out before the signal selection as follows.
First, candidate electrons sharing their track with a muon candidate are removed.
Furthermore, any non-$b$-jet candidate lying within an angular distance $\Delta R = 0.2$ of a candidate electron is discarded;
non-$b$-jets within $\Delta R = 0.4$ of candidate muons are
removed if the number of tracks associated with the jet is less than three.
Finally, any lepton candidate remaining within a
distance $\Delta R = \min\{0.4, 0.04 + \SI{10}{\GeV}/\pt(\ell)\}$ of any surviving jet candidate is discarded since they likely arise from decays of $b$- or $c$-flavoured hadrons.

Similar to the electron candidates, the photon candidates are reconstructed from calorimeter energy clusters and identified using the `Tight' criteria~\cite{EGAM-2018-01}.
They are required to be in the region $\abseta<2.37$ and have $\pt > \SI{145}{\gev}$.
Signal photons must satisfy the `Tight' calorimeter-based isolation requirements~\cite{EGAM-2018-01}, and are used to validate the background estimation technique detailed in Section~\ref{sec:bkd_njets}.

The missing transverse momentum, with magnitude \met, is defined
as the negative vector sum of the transverse momenta of all identified objects
(muon, electron and jet candidates) and an additional soft term~\cite{PERF-2016-07,ATLAS-CONF-2018-023}.
The soft term is added to recover the contributions from other low-\pt objects,
and is constructed from all tracks that are matched to the primary vertex but are not
associated with any other object.
A dedicated overlap removal procedure, based on removing duplicated energy contributions, is applied.
The \met variable is used to define control regions enriched in certain types of background, as discussed in Section~\ref{subsec:bkd_qcd}, and as input for the multivariate discriminant.


\section{Event selection and analysis strategy}
\label{sec:evtSelectionStrategy}
Two complementary analysis strategies are defined, namely the `jet counting analysis' and the electroweak analysis, labelled `EWK analysis'.
While the first approach is designed to be very  generic and offers a large variety of signal interpretations for strong production models,
the second approach is specifically tailored to reach sensitivity for electroweakino production.
In both analyses, events are assigned to one of two categories according to their lepton content, and further categorized into regions based on jet multiplicity and $b$-jet multiplicity.
This categorization provides a set of regions that are sensitive to decays from all the possible signal models considered in this search, amplifying the ability of the search to discriminate signal from background.
The EWK analysis is an extension of the jet counting analysis, where a third variable, a neural network (NN) discriminant,
is introduced in some of the jet and $b$-jet multiplicity regions in order to improve the separation of signal from background.
Additional kinematic selections are also applied at the preselection level, tailored to the electroweakino signals.
 
Events were selected for read-out using single-lepton triggers that require the electron or muon to
satisfy identification criteria similar to those used in the offline reconstruction and isolation requirements~\cite{TRIG-2018-01,TRIG-2018-05}.
For the analysis selection, at least four jets and at least one signal electron or muon, matched to the trigger lepton, are required in the event.
The highest-\pt lepton in the event has to pass the signal requirements and satisfy $\pt>\SI{27}{\GeV}$, in order to be above the trigger threshold.
Two disjoint event categories are defined according to the lepton content:
a same-charge dilepton selection (\ll), and all other events with at least one lepton (\l).
Events are placed in the \ll category if they contain exactly two signal leptons with same electric charge, and no additional candidate leptons.
In order to reduce backgrounds containing a $Z$ boson decaying into electrons, where one electron has its charge misidentified, events with two electrons have to satisfy a $|m_{ee} - m_Z| > \SI{10}{\gev}$ requirement.
Events with at least three signal leptons, with one same-flavour pair satisfying $|m_{\ell\ell} - m_Z| < \SI{10}{\gev}$ and exactly zero $b$-jets, are included as a separate subregion of the \ll category, in order to be used for background estimation.
All events passing further selections, but which do not enter the \ll category, are assigned to the \l category, including events containing more than one candidate lepton.
The regions with seven or fewer jets and zero $b$-jets in the \l category are further divided into three subregions.
The first subregion is defined by selecting events with two same-flavour candidate leptons fulfilling an invariant-mass requirement,
$|m_{\ell\ell} - m_Z| < \SI{10}{\gev}$.
The remaining events are divided into two subregions according to the electric charge of the highest-\pT\ lepton.
This division in subregions provides additional information that allows more accurate constraints to be placed on the $W$+jets and $Z$+jets backgrounds.

The jet counting analysis is carried out with five
jet-\pT thresholds to provide sensitivity to a broad range of possible signals. These thresholds are applied to all jets in the event and are at $\pT = 20$, $40$, $60$, $80$, and $100~\GeV$.
The optimal jet \pt\ threshold depends on the model and particle masses being tested.
The jet multiplicity is binned from a minimum of four jets to a maximum number ($N_{\text{last}}$) that depends on the \pT\ threshold and the lepton category.
In the \l category the last bin corresponds to 15 or more jets for the $20$~\GeV\ threshold, and 12, 11, 10, and 8 or more jets for the other thresholds in increasing order. In the \ll category it corresponds to 10, 8, 7, 7, and 6 or more jets respectively.
The highest jet-multiplicity bin for each \pT\ threshold is inclusive of larger jet multiplicities.
For each jet multiplicity bin, there are five exclusive bins in the $b$-jet multiplicity (four exclusive bins from zero to three $b$-jets, with an additional inclusive four-or-more bin).
The regions defined in the jet counting analysis are summarised in Table~\ref{tab:regions}.
For a given jet \pt\ threshold all regions are orthogonal and are analysed simultaneously. However, regions defined for different jet \pt\ thresholds can overlap.
The number of bins used in the search ranges from 110 when considering the $20~\GeV$ jet threshold, including the different subregions with zero $b$-jets, to 51 bins when considering the $100~\GeV$ jet threshold.
In this article, the notation $N^{\text{process}}_{j,b}$ is used to denote the number of events
predicted by the background fit model, with $j$ jets and $b$ $b$-jets for a given process, e.g.\ $N^{\ttbar\text{+jets}}_{j,b}$ for $\ttbar$+jets events.
The quantity $N^{\text{process}}_{j}$, referred to as a jet slice, is the number of events with $j$ jets for the considered physics process, and it is inclusive in the number of $b$-jets.

\begin{table}[t!]
\begin{center}
\caption{
Summary of regions considered in the jet counting analysis. The notation N$b$ is used to indicate a requirement on the $b$-jet multiplicity.
The highest jet multiplicity considered ($N_{\text{last}}$) depends on the jet \pT\ threshold and the lepton category.
In the \l category it corresponds to 15, 12, 11, 10, and 8 jets for the different jet \pt\ thresholds in increasing order. In the \ll category it corresponds to 10, 8, 7, 7, and 6 jets respectively.
}
\vspace{0.2cm}
\def\arraystretch{1.4}
\begin{tabular}{| c | c | c |}
\hline
Lepton category & Jet multiplicity & Analysis regions \\
\hline
\multirow{ 2}{*}{\l category} & 4\dots7 jets & $0b~\ell^-$, $0b~\ell^+$, $0b~m_{\ell\ell}$, $1b$,  $2b$,  $3b$,  $\geq 4b$ \\
& 8\dots$\geq N_{\text{last}}^{\l}$ jets & $0b$, $1b$,  $2b$,  $3b$,  $\geq 4b$ \\
\ll category & 4\dots$\geq N_{\text{last}}^{\ll}$ jets & $0b~3\ell$, $0b$, $1b$,  $2b$,  $3b$,  $\geq 4b$ \\
\hline
\end{tabular}
\label{tab:regions}
\end{center}
\end{table}

In order to improve the sensitivity of the search to the model with electroweakino production, the EWK analysis is introduced, which extends the jet counting analysis at the 20 \gev\ jet-\pT threshold.
In the \l category only, a separate NN discriminant is trained in each jet slice with eight or fewer jets, to discriminate the higgsino signal from the \ttbar\ background.
The full distribution of the NN output, binned in four even-width bins with approximately equal signal fraction, is fitted in each of the regions with at least one $b$-jet.
The NN training is performed with the constraint that the NN output distribution of the \ttbar\ background be invariant with respect to the $b$-jet multiplicity.
This property is later exploited in order to estimate the background from data, as described in Section~\ref{sec:Bkg}.
The invariance of the NN output with respect to the $b$-jet multiplicity is achieved with distance-correlation training~\cite{szekely2007,Kasieczka:2020yyl}.
The NNs are trained on a mixture of higgsino samples as the signal, and \ttbar\ as the only background.
 
The NN discriminant is constructed from a combination of low-level and high-level inputs.
The low-level variables considered are the jet and leading-lepton momenta, the individual pseudo-continuous $b$-jet score~\cite{FTAG-2018-01} of all jets, and the \met\ magnitude and direction.
The high-level inputs correspond to the jet and $b$-jet multiplicity of the event, minimum distance between the leading lepton and any jet, scalar \pt\ sum of all jets (\HT), scalar \pt\ sum of all $b$-jets, \minmaxmass (defined below), invariant mass of the three-jet system with highest system \pt, and invariant mass of the 3j + $\ell$ + $\mathbf{p}^{\text{miss}}$ system with highest system \pt (assuming that the $z$-component of the missing momentum  $\mathbf{p}^{\text{miss}}$ is zero). The two invariant mass variables attempt to reconstruct $\chinoonepm \rightarrow bbs$ and $\ninoonetwo \rightarrow tbs$ decays respectively.
The \minmaxmass variable is defined as follows.
All the jets in the event are split into two groups, where both groups have to contain at least one jet.
All possible combinations are tested, including those where the number of jets in each group is very different.
For each grouping, the higher of the masses of the two groups is selected, and then the minimum across all possible groupings is taken.
The \minmaxmass\ distribution has an endpoint for signal events at $m(\ninoone)$ that is reached in events where all partons were reconstructed.
For most events, the value is lower, since the lepton and \met\ components are ignored.
Backgrounds, however, do not show such an endpoint.
In addition, the shape of this variable has only a weak dependence on the number of $b$-jets,
which helps the NN to achieve separation while not introducing sensitivity to the $b$-jet multiplicity.
 
\begin{figure}[t!]
\centering
\includegraphics[width=.49\textwidth]{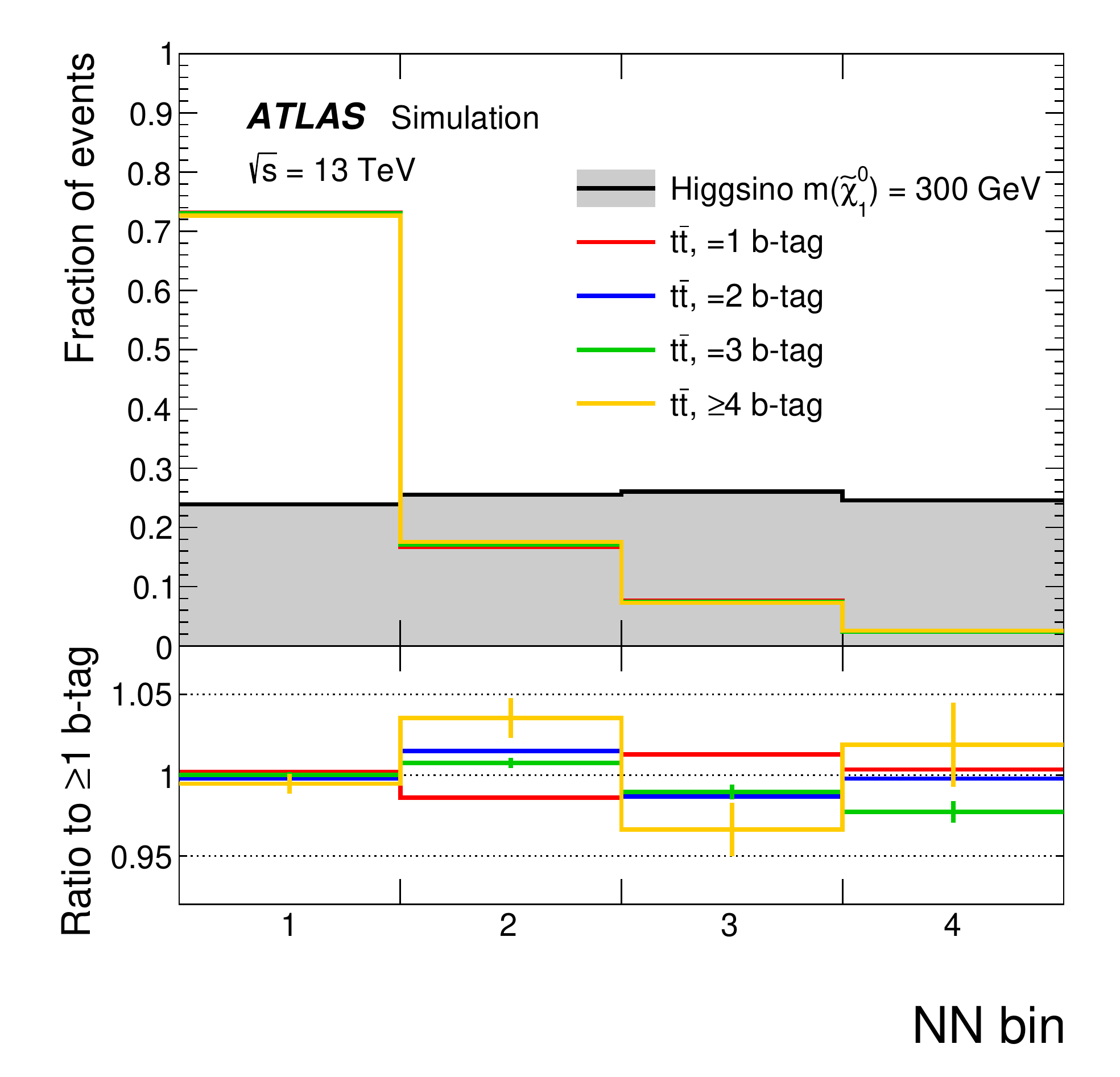}
\caption{
Output distribution of the NN discriminant in the six-jet slice, evaluated over a signal sample and \ttbar\ background split into the different $b$-jet regions. The bottom panel shows the ratio of the NN output distribution for \ttbar\ background in each $b$-jet region to the distribution in an inclusive region with at least one $b$-jet. The NN output distribution is invariant with respect to the number of $b$-jets in the selection, with differences per bin below 4\%.
}
\label{fig:NNinvariance}
\end{figure}

The inputs are connected to a single output node via two fully connected hidden layers of 100 neurons.
The NNs are trained using PyTorch~\cite{NEURIPS2019_9015} and the Adam optimizer~\cite{kingma2017adam}.
Events in the training dataset are sampled according to the inverse of the $b$-jet fraction (defined as the fraction of events in a given $b$-jet bin with respect to the total number of events in the jet slice) in order to flatten the $b$-jet spectrum. In order to achieve invariance of the NN output with respect to the \bjet multiplicity, the loss function of the training contains a term that penalizes a high distance correlation between the output and the $b$-jet multiplicity~\cite{Kasieczka:2020yyl}. A hyperparameter $\lambda$ controls the weight of this penalty term, with a value $\lambda=15$ which was optimized to achieve the highest sensitivity to the signal, accounting for both the separation and the systematic uncertainties derived from non-invariance of the NN, as described in Section~\ref{sec:syst_unc}. The invariance and separation achieved is shown in Figure~\ref{fig:NNinvariance}.
After training, the variables ranked highest in importance (using the integrated gradients method described in Ref.~\cite{sundararajan2017axiomatic}) are
\HT, the individual pseudo-continuous $b$-jet score of all jets,
number of $b$-jets, invariant mass of the 3j + $\ell$ + $\mathbf{p}^{\text{miss}}$ system with highest system \pt, and \minmaxmass.
The $b$-jet multiplicity is highly ranked despite the NN output being independent of it since it can be used to offset the effect from variables that are correlated with the number of $b$-jets such as \HT.

In the \ll category, signal events are produced via the leptonic decays of two top quarks.
However, the dominant backgrounds contain only one leptonic top decay, while the second lepton is a misidentified or non-prompt lepton, or originates from a $W$ boson that is not produced in a top decay (\ttW).
This property is exploited by introducing an additional requirement of $\minmaxmlj < \SI{155}{\gev}$, where the observable \minmaxmlj is defined as
${\minmaxmlj = \min_{a,b} \left\{ \max\left( m(\ell_0,\text{jet}_a), m(\ell_1,\text{jet}_b)\right)\right\}}$, with $\text{jet}_a \neq \text{jet}_b$,
for all possible permutations of $\text{jet}_a$ and $\text{jet}_b$ taken from the four leading jets. No $b$-tag information is used in the selection of the jets, to avoid differences in the variable across the different $b$-tag regions.
The signal has an endpoint in this variable at ${m^{\ell j} = \sqrt{m_{\text{top}}^2 - m_{W}^2} \approx \SI{152}{\gev}}$,
while background events tend to have larger values.

\begin{table}[t!]
\begin{center}
\caption{
The discovery signal regions used in jet counting and NN analyses, in the search for a generic BSM signal.
For every jet \pt\ threshold, four signal regions are defined in the jet counting analysis, leading to a total of 20 discovery signal regions in the jet counting analysis.
Two additional discovery signal regions are defined in the EWK analysis.}
\vspace{0.2cm}
\def\arraystretch{1.4}
\begin{tabular}{| c | c | c | c |}
\hline
\multicolumn{4}{|c|}{Jet counting analysis discovery SRs} \\
\hline
\multirow{ 2}{*}{Jet \pt\ threshold} & Number of jets & Number of jets & \multirow{ 2}{*}{Number of $b$-jets} \\
& \l category & \ll category &  \\
\hline
20 \gev  & $\geq 15$ &  $\geq 10$~~ & $= 0$, $\geq 3$ \\
40 \gev  & $\geq 12$ &  $\geq 8$  & $= 0$, $\geq 3$ \\
60 \gev  & $\geq 11$ &  $\geq 7$  & $= 0$, $\geq 3$ \\
80 \gev  & $\geq 10$ &  $\geq 7$  & $= 0$, $\geq 3$ \\
100 \gev{}~~ & ~~$\geq 8 $ &  $\geq 6$  & $= 0$, $\geq 3$ \\
\hline
\hline
\multicolumn{4}{|c|}{EWK analysis discovery SRs} \\
\hline
\multirow{ 2}{*}{Jet \pt\ threshold} & Lepton category & \multirow{ 2}{*}{Number of jets} & \multirow{ 2}{*}{Number of $b$-jets} \\
& and selection & &  \\
\hline
20 \gev  & \l, NN bin 4 &  $= 6$ & $\geq 4$ \\
20 \gev  & \ll, $\minmaxmlj<155$ \gev &  $= 6$ & $\geq 3$ \\
\hline
\end{tabular}
\label{tab:SRs}
\end{center}
\end{table}

In order to probe a specific BSM model, all the regions in both lepton categories are simultaneously fit to data to constrain the model, in what is called a model-dependent fit.
Separate fits are performed for each analysis and jet \pt threshold, and the configuration providing the best expected sensitivity is used to probe the model.
In the search for a generic BSM signal,
dedicated discovery signal regions (SRs) are defined which could be populated by a signal, and where the SM contribution is expected to be small.
The background in these SRs is estimated from
a fit excluding the SR being tested, in what is called a model-independent fit.
The discovery SR definitions used in the jet counting analysis are shown in Table~\ref{tab:SRs}.
 
Two additional discovery SRs are defined targeting a possible electroweakino signal making use of the EWK analysis.
The first SR is defined in the \l category, with exactly six jets with $\pt > \SI{20}{\GeV}$, at least four $b$-jets, and a selection on the NN discriminant.
The NN selection corresponds to a signal efficiency of 25\% on the higgsino model with a mass of 300 \gev\ and a background rejection of 40 for the \ttbar\ background, which corresponds to bin four in Figure~\ref{fig:NNinvariance}.
The second SR is defined in the \ll category, with exactly six jets with $\pt > \SI{20}{\GeV}$, and at least three $b$-jets.
The discovery SR definitions used in the EWK analysis are also shown in Table~\ref{tab:SRs}.

The dominant background processes are $t\bar{t}$+jets and $W/Z$+jets in the \l category, and \ttW, \ttbar with a misidentified lepton, and diboson production in the \ll category.
The estimation of the dominant backgrounds is carried out using a combined fit to the jet and $b$-jet multiplicity bins described above.
For these backgrounds, the normalization per jet slice is derived using parameterized extrapolations from lower jet multiplicities. The $b$-jet multiplicity shape per jet slice is taken from MC simulation
for the $W/Z$+jets and diboson backgrounds, whereas for background processes involving top quarks it is predicted from the data using a parameterized extrapolation based on observables at medium jet multiplicities.
A separate likelihood fit is carried out for each jet \pT\ threshold, with the fit parameters of the background model determined separately in each fit.
The assumptions used in the parameterization are validated using data and MC simulation.

\section{Background estimation}
\label{sec:Bkg}
The dominant background in the \l category arises from $W/Z$+jets production in the zero $b$-jet regions,
and top-quark pair production in the regions with at least one $b$-jet.
In the \ll category the dominant background in the zero $b$-jet regions originates from diboson production with fully leptonic decays, in particular $WZ$ where one lepton from the $Z$ boson decay is lost.
In the regions with at least one $b$-jet the main backgrounds are the associated production of a top-quark pair and a $W$ boson,
dileptonic \ttbar\ where an electron has its charge misidentified,
and semileptonic \ttbar\ with a jet misidentified as a lepton, or with a non-prompt lepton.
These three background components are merged and labelled as \ttX, and estimated simultaneously.
 
The theoretical modelling of all these backgrounds at high jet multiplicity suffers
from large uncertainties, so they are estimated from the data by extrapolating the jet and $b$-jet multiplicity distributions extracted at moderate jet multiplicities to the high jet multiplicities of the search regions.

\subsection{Jet multiplicity prediction}
\label{sec:bkd_njets}
 
A data-driven approach is used to estimate the contribution of the main backgrounds in each jet multiplicity slice.
The estimate of the normalization relies on assuming a functional form to describe the evolution of the number of background events for process $X$ as a function of the jet multiplicity,
$r^X(j) \equiv N^{X}_{j+1}/N^{X}_{j}$.
 
Above a certain number of jets, $r(j)$ is assumed to be constant, implying a fixed probability of additional jet radiation, referred to as
`staircase scaling'~\cite{Ellis:1985vn,Berends:196850,Giele1990,Gerwick2012}. This behaviour has been observed in $W/Z$+jets by the ATLAS~\cite{STDM-2012-04,STDM-2016-01} and CMS~\cite{Chatrchyan2012}
collaborations. For lower jet multiplicities, a different scaling is expected with $r(j) = k / (j + 1)$ where $k$ is a constant, referred to as `Poisson scaling'~\cite{Gerwick2012}.
The transition point between these scaling behaviours depends on the jet kinematic selections.
 
For the kinematic phase space relevant for this search, a combination of the two scalings is found to describe the data in dedicated validation regions (described later in this section),
as well as in simulated MC samples with an integrated luminosity much larger than that of the data.
This combined scaling is parameterized as
\begin{equation*}
r^X(j) = c^X_0 +  c^X_1 / (j + c^X_2),
\end{equation*}
where $c^X_0$, $c^X_1$ and $c^X_2$ are process-dependent constants that are extracted from the data. The $c^X_2$ parameter is introduced to take into account the ambiguity in the counting of jets originating from the decay products of the process $X$ and the additional jets. The parameter is fixed to $c^X_2=1$ in the estimation of $W$+jets, $Z$+jets, and fully leptonic diboson events, as there is no ambiguity in the counting of jets for these processes. However, $c^X_2$ is a free parameter in the estimation of backgrounds containing top quarks, where the jet counting ambiguity remains.
 
Studies using simulated event samples, both at generator level and after event reconstruction, demonstrate that the flexibility of this parameterization is also able to absorb reconstruction effects
related to the decrease in event reconstruction efficiency with increasing jet multiplicity, which are mainly due to the lepton--jet overlap and lepton isolation requirements.
 
The number of background events from process $X$ in the $j$ jets slice is then parameterized as follows:
\begin{equation*}
N^{X}_{j} = N^{X}_{4} \cdot \prod\limits_{j'=4}^{j'=j-1} r^X(j') ,
\end{equation*}
 
where $N^{X}_{4}$ is a free parameter for the absolute normalization in four-jet events.
Since the last jet-multiplicity bin used in the analysis is inclusive in the number of jets, the model is used to predict this by iterating to higher jet multiplicities and summing the contribution for each jet multiplicity above the maximum used in the analysis.
The four parameters per process, i.e.\ $N^{X}_{4}$, $c^X_0$,  $c^X_1$, and $c^X_2$ (if not fixed to one), are allowed to float in the fit,
and are therefore extracted from the data along with the other background contributions.
Studies in data and MC simulation indicate that the $c^X_0$ and $c^X_1$ parameters for $W$+jets and $Z$+jets are statistically compatible,
and are therefore combined into common parameters $c^\text{$W/Z$+jets}_0$ and $c^\text{$W/Z$+jets}_1$.
The normalization parameters $N^{\text{$W$+jets}}_{4}$ and $N^{\text{$Z$+jets}}_{4}$ are kept independent. The $c^X_i$ parameters are independent among the rest of the backgrounds, including \ttbar\ in the \l category and \ttX in the \ll category.
 
The jet-scaling assumption is validated in data, using $\gamma$+jets and dileptonic \ttbar\ events.
The $\gamma$+jets events are selected using a high-\pt\ photon trigger, and a high-\pt\ signal photon is required in the event selection.
The dileptonic \ttbar\ data sample is selected by requiring an electron candidate and a muon
candidate in the event, with at least two jets of which at least one is a $b$-jet, and the small background predicted by MC simulation is subtracted.
The possible signal contamination in this sample is negligible as the selection is inclusive over the number of $b$-jets.
In this sample, the scaling behaviour can be tested for up to 13 jets, which corresponds to 15 jets for a semileptonic $\ttbar$+jets sample.
Simulated $W$+jets, $Z$+jets, semileptonic \ttbar\ (both the nominal
sample and the alternative samples described in Section~\ref{sec:mcSimulation}), and \ttX\ samples are also found to be consistent with the jets-scaling assumption.

\begin{figure}[t!]
\centering
\includegraphics[trim={0 2cm 0 0}, clip=true,width=.49\textwidth]{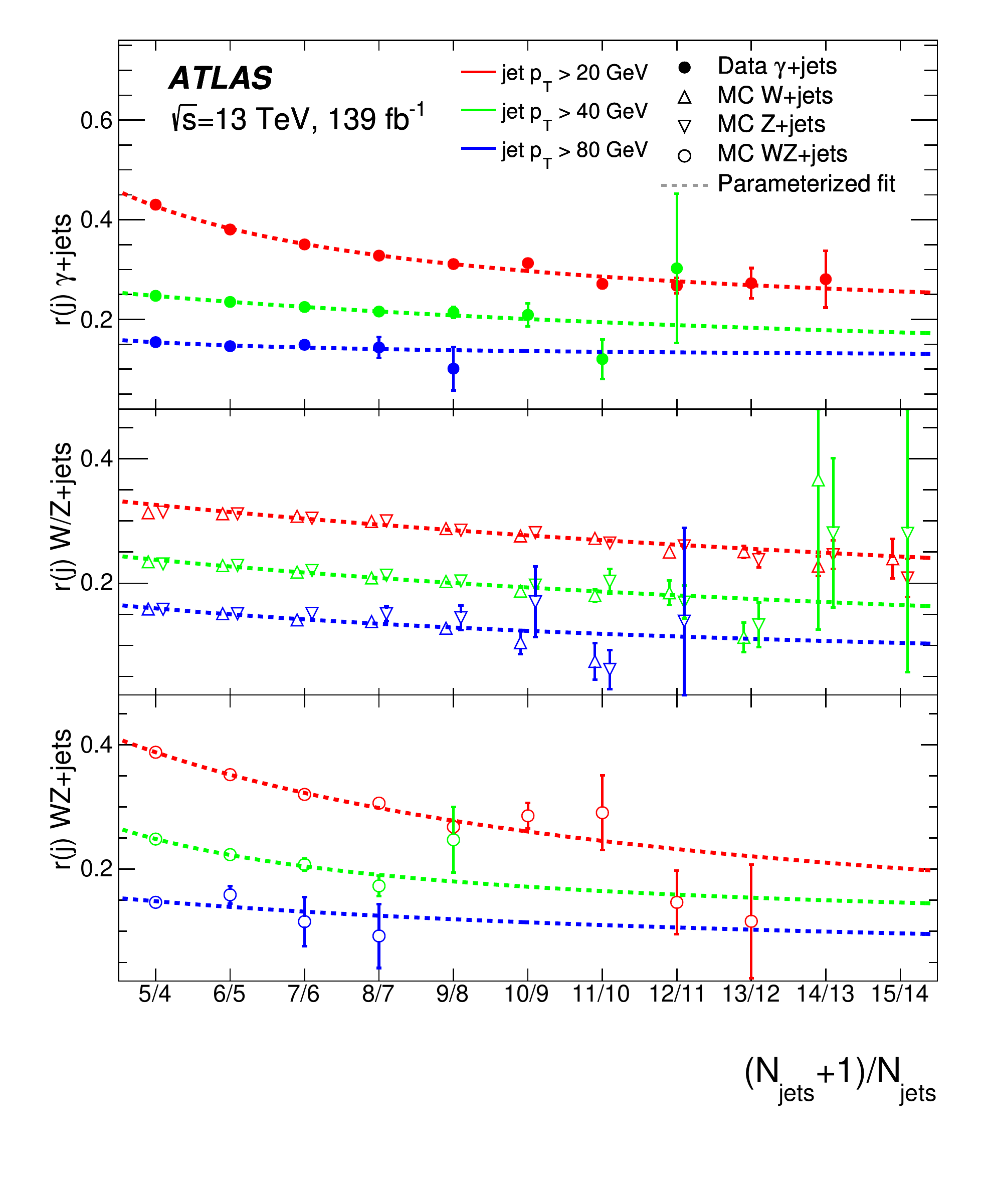}
\includegraphics[trim={0 2cm 0 0}, clip=true,width=.49\textwidth]{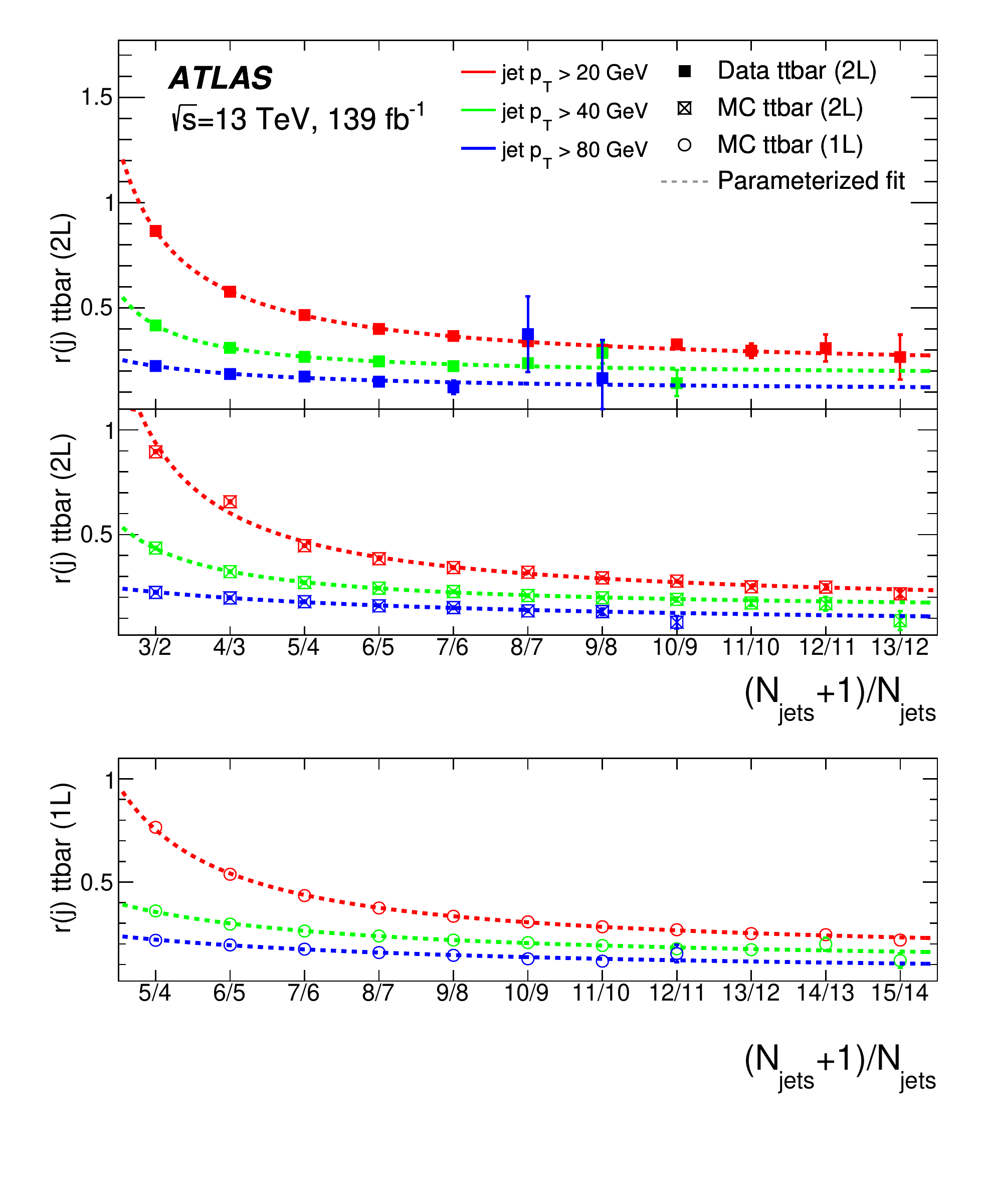}
\caption{
The ratio of the number of events with $(j+1)$ jets to the number with $j$ jets in various event samples (details in the legend), used to validate the jet-scaling parameterization.
In the MC samples of $W$/$Z$/$WZ$+jets the vector bosons are forced to decay to leptons.
Each panel shows the ratio for data or MC simulation with the fitted parameterization overlaid as a dashed line.
The uncertainties shown are statistical.
}
\label{fig:Vscaling_jetn}
\end{figure}
 
Figure~\ref{fig:Vscaling_jetn} shows the $r(j)$ ratio
for various processes used to validate the jet-scaling parameterization. Each panel shows
the $r(j)$ ratio for data or MC simulation with the fitted parameterization overlaid as a line.

\subsection{Prediction of $b$-jet multiplicity}
\label{sec:bkd_bjets}
The number of background events from process $X$ in a given jet and $b$-jet multiplicity region can be expressed as follows:
\begin{equation*}
N^{X}_{j,b} = f^X_{j,b} \cdot N^{X}_{j}
\end{equation*}
where $f^X_{j,b}$ is the fraction of events from process $X$ with $b$ number of $b$-jets in the $j$ jet slice, and satisfies $\sum_{b=0}^4 f^X_{j,b}=1$.
A data-driven model is used to estimate the $b$-jet fraction in background processes containing top quarks.
The basic concept of this model is based  on the extraction of an
initial template of the $b$-jet fraction distribution in events with four jets and the parameterization of the evolution of this template to higher jet multiplicities.
Each jet slice is constrained in the fit as discussed later in this section.
The $b$-jet fractions for $W$+jets, $Z$+jets and diboson backgrounds are taken from MC simulation.
 
The extrapolation of the $b$-jet multiplicity distribution to higher jet multiplicities starts from the assumption that the difference between the $b$-jet multiplicity distribution in events with $j$ and $j+1$ jets arises mainly from the production of additional jets, and can be described by a fixed probability that the
additional jet is a $b$-jet. Given the small mis-tag rate, this probability is dominated by the probability that the additional jet is a heavy-flavour jet which is $b$-tagged. In order to account for
acceptance effects due to the different kinematics in events with high jet multiplicity, the probability of further $b$-tagged jets entering the acceptance is also taken into account.
The extrapolation to one additional jet can be parameterized as:
\begin{equation}
\begin{split}
f_{(j+1),b} & = f_{j,b} \cdot x_0 + f_{j, (b-1)} \cdot x_1 + f_{j, (b-2)} \cdot x_2,
\end{split}
\label{eqn:evolution}
\end{equation}
where the parameters $x_i$ describe the probability of one additional jet to be either not $b$-tagged ($x_0$), $b$-tagged ($x_1$), or $b$-tagged and causing a second jet to be $b$-tagged ($x_2$).
The latter is dominated by cases where the extra jet influences the event kinematics such that a second $b$-jet, previously not $b$-tagged, becomes $b$-tagged. This parameter aims to model the increase in $b$-tagging efficiency with jet \pt.
Given that the $x_i$ parameters describe probabilities, the sum $\sum_{i} x_i$ is normalized to unity.
Terms with a negative number of $b$-tagged jets $(f_{j,b<0})$ are set to zero.
Subsequent application of this parameterization produces
a $b$-jet template for arbitrarily high jet multiplicities.
 
Studies based on MC simulated events with sample sizes corresponding to equivalent luminosities much larger than the collected dataset,
as well as studies using fully efficient generator-level $b$-tagging, indicate the necessity to add a fit parameter that
allows for correlated production of two $b$-jets as may be expected with $b$-jet production from  gluon splitting. This is implemented by changing the evolution described in
Eq.~(\ref{eqn:evolution}) such that any term with $x_1 \cdot x_1$ is replaced by $x_1 \cdot x_1 \cdot \rho_{11}$, where $\rho_{11}$ describes the correlated production of two $b$-jets.
The value of $\rho_{11}$ is a free parameter and is determined in the fit.
 
The initial $b$-jet multiplicity template is extracted from data events with four jets after subtracting all non-$\ttbar$ background processes, and is denoted by $f_{4,b}$ and scaled by the absolute
normalization $N^{\text{$\ttbar$+jets}}_{4}$ in order to obtain the model in the four-jet bin:
\begin{equation*}
N^{\text{$\ttbar$+jets}}_{4,b}  = N^{\text{$\ttbar$+jets}}_{4} \cdot f_{4,b},
\end{equation*}
where the sum of $f_{4,b}$ over the four $b$-jet bins is normalized to unity.
The zero $b$-jet component of the initial $\ttbar$ template, exhibits an anti-correlation with the absolute $W$+jets normalization, which is extracted in the
same region. The division into subregions separated in leading-lepton charge, detailed in Section~\ref{sec:evtSelectionStrategy}, provides a handle to extract the absolute $W$+jets normalization, due to the charge asymmetry in $W^\pm$ production.
The remaining anti-correlation does not affect the total background estimate.
For these regions, the $\ttbar$+jets process is assumed to be charge symmetric and the model is simply split into two halves for these bins.
 
The model described above is based on the assumption that any change in the $b$-jet multiplicity distribution is due to additional jet radiation with a certain probability to lead to $b$-jets.
There is, however, also a small increase in the acceptance for $b$-jets produced in the decay of the \ttbar\ system, when increasing the jet multiplicity, due to the higher jet momentum on average.
The effect amounts to up to 5\% in the one- and two-$b$-jet bins for high jet multiplicities, and is taken into account using a correction to the initial template extracted from simulated $\ttbar$ events.
 
The parameters that model the production of additional $b$-jets ($x_i$, $\rho_{11}$) are correlated in the \l and \ll categories. The initial $b$-jet multiplicity parameters ($f_{4,b}$), and the acceptance correction to the initial template are independent in each lepton category.

\subsection{Neural network template prediction}
\label{sec:bkd_NN}
 
\begin{figure}[t!]
\centering
\includegraphics[width=.49\textwidth]{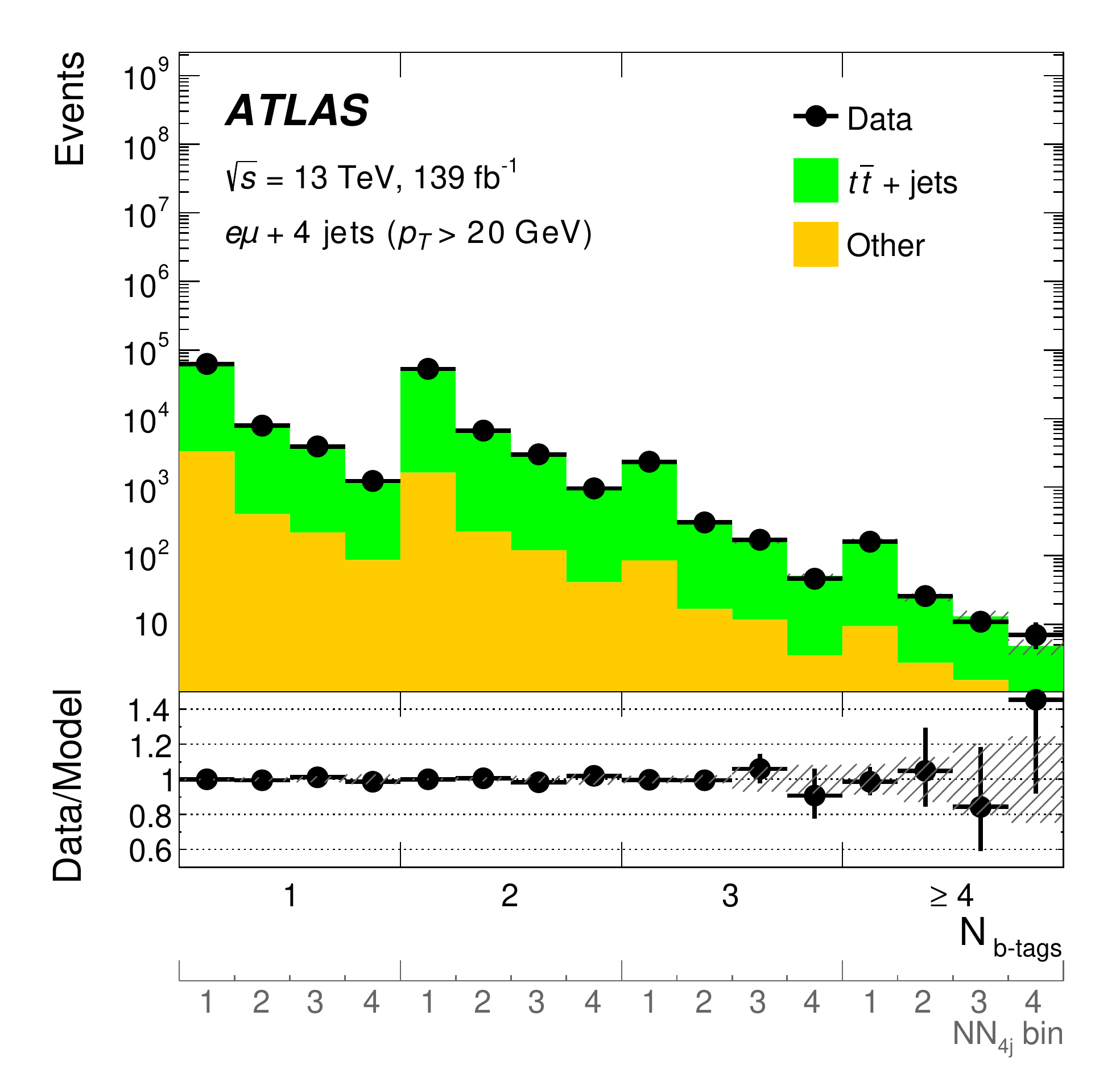}
\includegraphics[width=.49\textwidth]{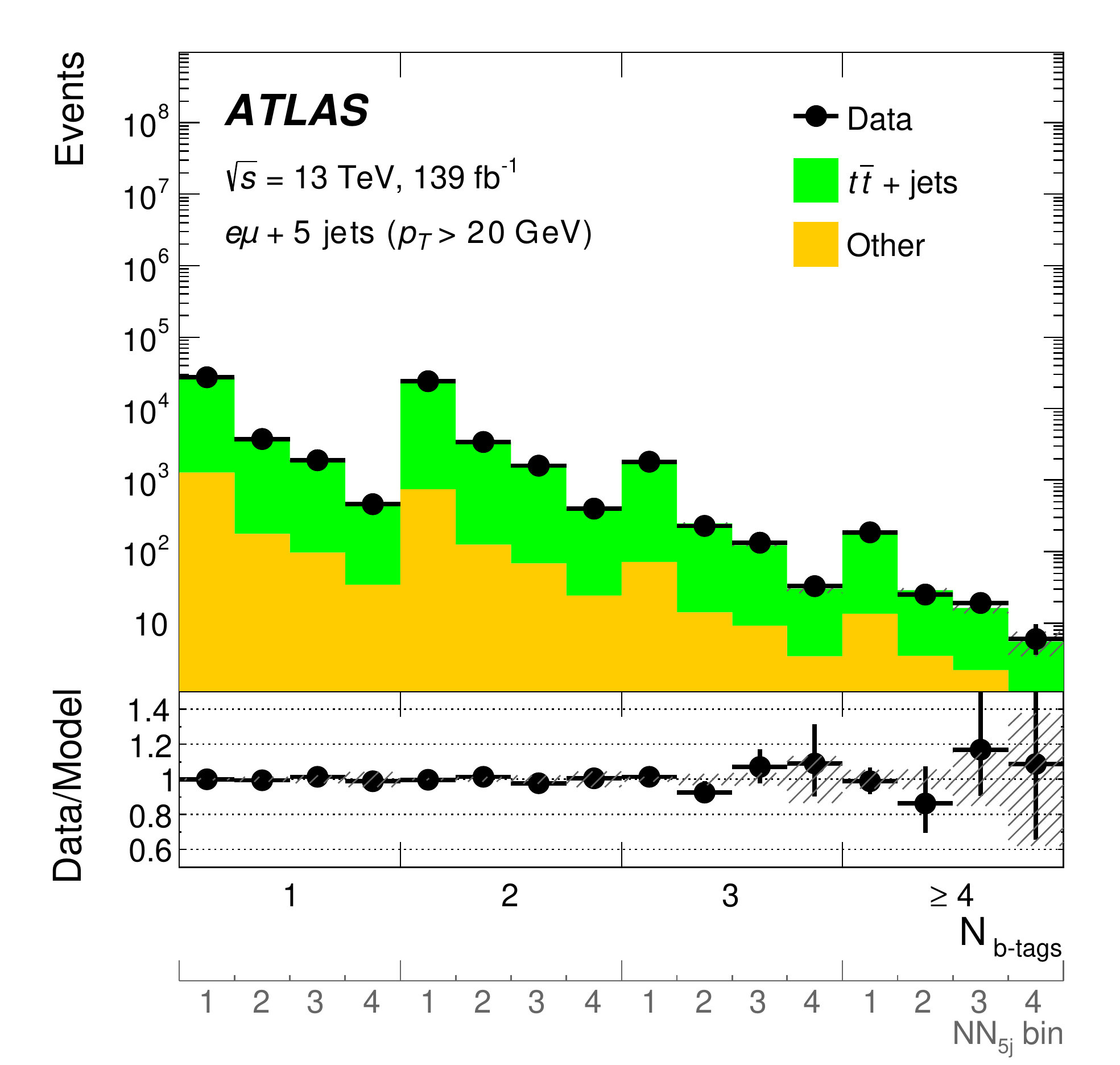}
\caption{Observed data and the corresponding background estimation in regions with one electron, one muon, and four jets (left) or five jets (right).  All uncertainties, which may be correlated across the bins, are included in the error bands (shaded regions). The shape of the NN template for the \ttbar\ background is required to be identical in all $b$-jet regions. Good agreement between the data and estimated background confirms the invariance of the NN output with respect to the $b$-jet region.}
\label{fig:blinded_fitVR}
\end{figure}
 
The NN is only introduced in regions with at least one $b$-jet, where the dominant background is \ttbar\ production. The NN output distribution is obtained from MC simulation for all the subdominant backgrounds.
A data-driven method is developed in order to predict the NN output distribution for \ttbar\ background events,
making use of the invariance of the NN output with respect to the number of $b$-jets in the event. The \ttbar\ background in a given bin of the NN output is parameterized as:
\begin{equation*}
N^{\text{$\ttbar$+jets}}_{j,b,i} = n_{j,i} \cdot N_{j,b}
\end{equation*}
where $n_{j,i}$ is the fraction of \ttbar\ events in bin $i$ of the NN output in the $j$ jet slice, and is independent of the $b$-jet region. The fractions in each jet slice are free parameters and are fitted simultaneously to all $b$-jet regions, constrained by the sum $\sum_{i} n^i_{j}$ being normalized to unity. Given the large statistical power of the one- and two-$b$-jet regions, the fitted NN templates are determined in these regions and not biased by a possible signal entering the high $b$-jet regions.
 
This method relies on the invariance of the NN output with respect to the number of $b$-jets.
This property is validated in large samples of MC simulated \ttbar\ events, including the alternative samples described in Section~\ref{sec:mcSimulation}.
The invariance is also confirmed in data by using a pure sample of dileptonic \ttbar\ events, with a selection requiring one electron, one muon, and at least one $b$-jet.
The signal contamination in this dataset is at most 2\% in the last NN bin of the 4 b-tag region. The only process that is not negligible given this selection is $\ninoone\ninotwo$ production, which has a lower cross-section than the $\chinoonepm\ninoonetwo$ process, that contributes most to the 1-lepton selection.
Figure~\ref{fig:blinded_fitVR} shows that good agreement between data and estimated background is observed in this region, confirming in data the invariance of the NN output with respect to the $b$-jet region.

\subsection{Fake and non-prompt lepton background}
\label{subsec:bkd_qcd}
The contribution from events with a fake or non-prompt (FNP) lepton (such as hadrons misidentified as leptons, leptons originating from the decay of heavy-flavour hadrons, and electrons from photon
conversions), constitutes a minor but non-negligible background.
 
The multi-jet background in the \l category is estimated from the data with a matrix method similar to that described in
Ref.~\cite{paperSS3L}. In this method, two types of lepton identification criteria are defined: \emph{`tight'}, corresponding to the default signal lepton criteria described in Section~\ref{sec:evtReco}, and
\emph{`loose'}, corresponding to candidate  leptons after overlap removal.
The matrix method relates the numbers of observed events in which a \emph{loose} lepton candidate does or does not satisfy the \emph{tight} selection criteria.
The probability for \emph{loose} prompt leptons to satisfy the \emph{tight} selection criteria is
obtained using a $Z\to\ell\ell$ data sample and is modelled as a function of the lepton $\pt$.
The probability for \emph{loose} FNP leptons to satisfy the \emph{tight} selection criteria is determined from a data
control region enriched in non-prompt leptons that requires a \emph{loose} lepton, multiple jets, low \met~\cite{ATL-PHYS-PUB-2015-027,ATL-PHYS-PUB-2015-023}, and low transverse mass.\footnote{The transverse mass of the lepton--\met system is defined as: $\mt=\sqrt{2 \pT^{\ell}\met(1 - \cos(\Delta\phi(\ell,\met))}$).}
This data sample is recorded with prescaled lepton triggers without an isolation requirement.
The efficiencies are measured as a
function of lepton candidate \pt after subtracting the contribution from prompt-lepton processes, and are assumed to be independent of the jet multiplicity.\footnote{To minimize the dependence on the number of jets,
the event selection considers only the highest-\pT\ candidate lepton when checking the more stringent identification and isolation criteria of the signal lepton definitions.}

In the \ll category, the background from FNP leptons in association with a top-quark pair is estimated as part of the \ttX background as described above.
Other contributions from FNP leptons constitute less than 10\% of the total background in the zero-$b$-jet bin, with a negligible contribution in the $b$-jet regions, and are taken from MC simulation.
The estimation is validated in a dedicated validation region requiring zero $b$-jets, two same-flavour leptons satisfying $|m_{\ell\ell} - m_Z| < \SI{10}{\GeV}$,
and an additional candidate lepton failing the signal requirement. This region is dominated by $Z$ boson events containing a FNP lepton, and is used to verify the modelling of FNP leptons in the MC simulation.

\subsection{Minor backgrounds}
\label{subsec:bkd_mc}
The minor background contributions from single-top production, \ttH, and SM four-top-quark production
are estimated using MC simulation. In the \l category the diboson and \ttV\ backgrounds are also estimated from MC simulation, while the estimates are both data-driven (\ttV as part of the \ttX\ background) in the \ll category. In all but the highest jet slices considered, the sum of these backgrounds contributes no more than 10\% of the SM expectation in any of the
$b$-jet bins; for the highest jet slices this can rise to 35\%.
 
\subsection{Fit configuration and validation}
\label{subsec:fit}
Two different fit configurations are used in the search.
When testing a specific BSM model the model-dependent fit set-up is used, where for each jet \pt threshold all the regions in both lepton categories are simultaneously fit to data to constrain the model. The expected signal contribution in all bins is taken into account, including bins with low jet or $b$-jet multiplicity.
The jet counting analysis is used to constrain the models with strong production and the EWK analysis for the models with electroweakino production.
Separate fits are performed for each jet \pt threshold, and the threshold with best expected sensitivity for the given signal mass point is used to set limits.
In the search for a generic BSM signal in a particular SR the model-independent fit set-up is used, where for each jet \pt threshold the simultaneous fit includes all regions except the SR being tested.
Possible signal leakage to the control regions can produce a bias in the background estimation, leading to conservative limits.
Such limits have hence been obtained assuming negligible signal contributions in regions outside of the SR.
Signal processes with final states that the search is targeting generally have negligible leakage into these regions, as is the case for the benchmark models considered.
 
The number of freely floating parameters in the background model is 26 in the jet counting analysis, and 41 in the EWK analysis. The different parameters for each background are summarized in Table~\ref{tab:free_parameters}.
The number of fitted bins varies between 51 and 110 in the jet counting analysis, depending on the jet \pt\ threshold used, and is 170 in the EWK analysis, leading to an over-constrained system in all cases.
 
The fit set-up was extensively tested using MC simulated events, and was demonstrated to give excellent agreement on a background-only dataset, and a
negligible bias in the fitted signal yields. This level of agreement is seen both in the cases where the background-only distributions are fit, and when a
signal is injected into the fitted data.
 
\begin{table}[t!]
\begingroup
\centering
\caption{
Summary of all the free floating parameters in the background model. There are a total of 26 such parameters in the jet counting analysis and 41 in the EWK analysis.
}
\vspace{0.2cm}
\def\arraystretch{1.4}
\setlength{\tabcolsep}{5pt} 
\begin{tabular}{| l | c c c c c | c |}
\hline
Parameters & \ttbar+jets & \ttX & $W$+jets  & $Z$+jets  & $VV$+jets & Constraints \\
\hline
Normalization     & $N^{\ttbar}_{4}$ & $N^{\ttX}_{4}$ & $N^{W}_{4}$ & $N^{Z}_{4}$ & $N^{VV}_{4}$ & -- \\
Jet scaling, {\small $i \in \{0,1,2\}$} & $c^{\ttbar}_i$ & $c^{\ttX}_i$ & \multicolumn{2}{c}{$c^{W/Z}_i$} & $c^{VV}_i$ & $c^{W/Z}_{2}=c^{VV}_{2}=1$ \\
Initial $b$-jet fractions, {\small  $i \in \{0 \dots 4\}$} & $f^{\ttbar}_{4,i}$ & $f^{\ttX}_{4,i}$ & --  & --  & --  & $\sum_{i} f_{4,i}=1$ \\
Extra heavy-flavour jets, {\small  $i \in \{0,1,2\}$} & \multicolumn{2}{c}{$x_i$, $\rho_{11}$} & --  & --  & --  & ~\,$\sum_{i} x_i=1$ \\
NN shape, {\small  $i \in \{1 \dots 4\}$, $j \in \{4 \dots 8\}$}   & $n_{j,i}$ & -- & -- & -- & -- &  $\sum_{i} n_{j,i}=1$ \\
\hline
\end{tabular}
\label{tab:free_parameters}
\endgroup
\end{table}
 

\section{Systematic uncertainties}
\label{sec:syst_unc}
The dominant backgrounds are estimated from the data without the use of MC simulation, and therefore the main systematic uncertainties related to the estimation of these backgrounds arise from the
assumptions made in the construction of the parameterized model. Uncertainties related to the theoretical modelling of the specific processes and due to the modelling of the
detector response in simulated events are only relevant for the minor backgrounds,
which are taken from MC simulation, and for the estimates of the signal yields after selections.
 
For the $W/Z$+jets background estimation, the uncertainty related to the assumed jet scaling is taken from studies of this behaviour in $W$+jets and $Z$+jets MC simulation, as well as in
$\gamma$+jets data control regions chosen to be kinematically similar to the search selection. Deviations from the assumed scaling
behaviour are assigned as a systematic uncertainty, uncorrelated in each jet slice. In the case of no deviation the statistical precision of the validation is assigned.
The uncertainty in the lower jet multiplicities is at the percent level for all jet \pt\ thresholds, and up to 50\% in the highest jet multiplicities, driven by the statistical precision of the method.
The uncertainty related to the parameterization of the jet multiplicity of the \ttbar background is determined with the same strategy,
and is derived from MC simulation closure tests (including alternative MC generators), as well as dileptonic \ttbar control regions in data.
No evidence of a deviation from the assumed scaling behaviour is seen, and the statistical precision of the closure in data is used as an uncertainty.
 
The expected uncertainty of the charge asymmetry in $W$+jets
production is $3\%$--$5\%$ from PDF variations~\cite{Botje:2011sn}, but in the seven-jet region the uncertainty is dominated by the limited number of MC events (up to 10\% for the 80~\GeV\ jet-\pT\ threshold). The
uncertainty in the shape of the $b$-jet multiplicity distribution in $W$+jets, $Z$+jets, and diboson events is derived by comparing different MC generator configurations (e.g.\ varying the renormalization and factorization
scale and the parton-shower model parameters). It is seen to grow as a function of jet multiplicity and is about 50\% for events with five jets, after which the MC statistical uncertainty becomes very large. A conservative
uncertainty of 50\% per additional heavy-flavour quark that is generated is assigned to the fractional contribution from $V(V)$+$b$ and $V(V)$+$c$ events, uncorrelated among the three backgrounds. This uncertainty has a negligible impact on the final result as the background from $W/Z$ boson or diboson production with additional heavy-flavour jets is small compared to that from top-quark pair production.
In addition, the uncertainties related to the $b$-tagging efficiency and mis-tag rate are taken into account in the uncertainty in the $W/Z$+jets $b$-jet template.
 
The $b$-jet fraction estimation method exhibits good closure in studies based on MC simulated events with sample sizes corresponding to integrated luminosities much larger than that of the collected dataset,
as well as studies using fully efficient generator-level $b$-tagging, so no systematic uncertainty related to these studies is assigned. A small uncertainty related to the acceptance correction for the
initial $b$-jet multiplicity template is derived by varying the MC generator configuration for the $t\bar{t}$ sample used to estimate the correction. This leads to a 3\% uncertainty in the correction,
and has no significant effect on the final uncertainty.
 
The prediction of the NN template in \ttbar\ events relies on the invariance of the NN output with respect to the number of $b$-jets in the event.
This assumption is tested in MC simulation (both the nominal sample
and the alternative samples described in Section~\ref{sec:mcSimulation}), and seen to hold within 5\% in the five-jet and six-jet slices,
where the best signal-to-background ratio is expected, and within 10\% in the rest.
The largest deviation from the $b$-jet-inclusive template that is seen per bin across $b$-jet regions is assigned as an uncorrelated uncertainty in each bin and ranges from 1\% to 10\%.
 
The dominant uncertainty in the multi-jet background estimate arises from the number of data events in the control regions.
An uncertainty in the subtraction of electroweak backgrounds from these control regions is estimated at 20\% of the expected yield of these background processes.
Additional uncertainties are assessed to cover the possible dependencies of the prompt and FNP lepton efficiencies~\cite{paperSS3L} on variables other than lepton \pt (for example the dependence on the number of jets in the event). The total uncertainty in the multi-jet background yields is about 50\%.
 
The uncertainty in the expected yields of the minor backgrounds includes theoretical uncertainties in the cross-sections and in the modelling of the kinematics by the MC generator, as well as experimental
uncertainties related to the modelling of the detector response in the simulation. The uncertainties assigned to cover the theoretical estimate of these backgrounds in the relevant regions are 50\%
for diboson in the \l category and single top-quark production, and 30\% for $\ttbar V/H$ production.
An additional uncertainty of 50\% is assigned to the contribution from $\ttV+b$ and $\ttV+c$ events.
These uncertainties are conservative estimates based on the impact seen from renormalization and factorization scale variations, PDF variations, and comparisons with samples with an alternative parton-shower model.
 
Uncertainties in the modelling of \tttt\ are assigned from renormalization and factorization scale variations,
as well as from a comparison with simulated samples with an alternative parton-shower model.
An uncertainty of 100\% is assigned to the cross-section to cover the difference between the predicted and measured cross-sections~\cite{Aad:2020klt}. Using instead the central value and uncertainty from the ATLAS measurement leads to a 1\% decrease in expected sensitivity.
 
The final uncertainty in the background estimate in the SRs is dominated by the uncertainty in the fitted model parameter values, which stems from the statistical uncertainty of the data events in the different jet slices.
Systematic uncertainties do not contribute significantly in the jet counting analysis, and cause only a 1\% loss in sensitivity.
The leading systematic uncertainty in the EWK analysis is related to the NN invariance assumption and causes a 30\% loss in sensitivity, while other systematic uncertainties are subdominant.
 
The uncertainties assigned to the expected signal yields for the SUSY benchmark processes considered include the experimental uncertainties related to the detector modelling, which are dominated by the
modelling of the jet energy scale, and the $b$-tagging efficiencies and mis-tagging rates.
For example, for a signal model with four $b$-quarks, the $b$-tagging uncertainties are $\approx$10\%, and the jet-related uncertainties are typically $\approx$5\%.
The uncertainty in the signal cross-sections used is discussed in Section~\ref{sec:mcSimulation}.
The uncertainty in the signal yields related to the modelling of additional jet radiation is studied by varying the factorization, renormalization, and jet-matching scales as well as the parton-shower tune in the simulation. The corresponding uncertainty is small for most of the signal parameter space, but is as large as 30\% for very light or very heavy LSPs, where the contribution from additional jet radiation is relevant.
The difference between fast and full simulation is evaluated for selected signal points. The jet multiplicity, $b$-jet multiplicity, and NN output distributions are found to be statistically compatible, so no additional uncertainty is considered due to the usage of fast simulation.
 

\section{Results}
\begin{figure}[htbp!]
\centering
\includegraphics[width=.32\textwidth]{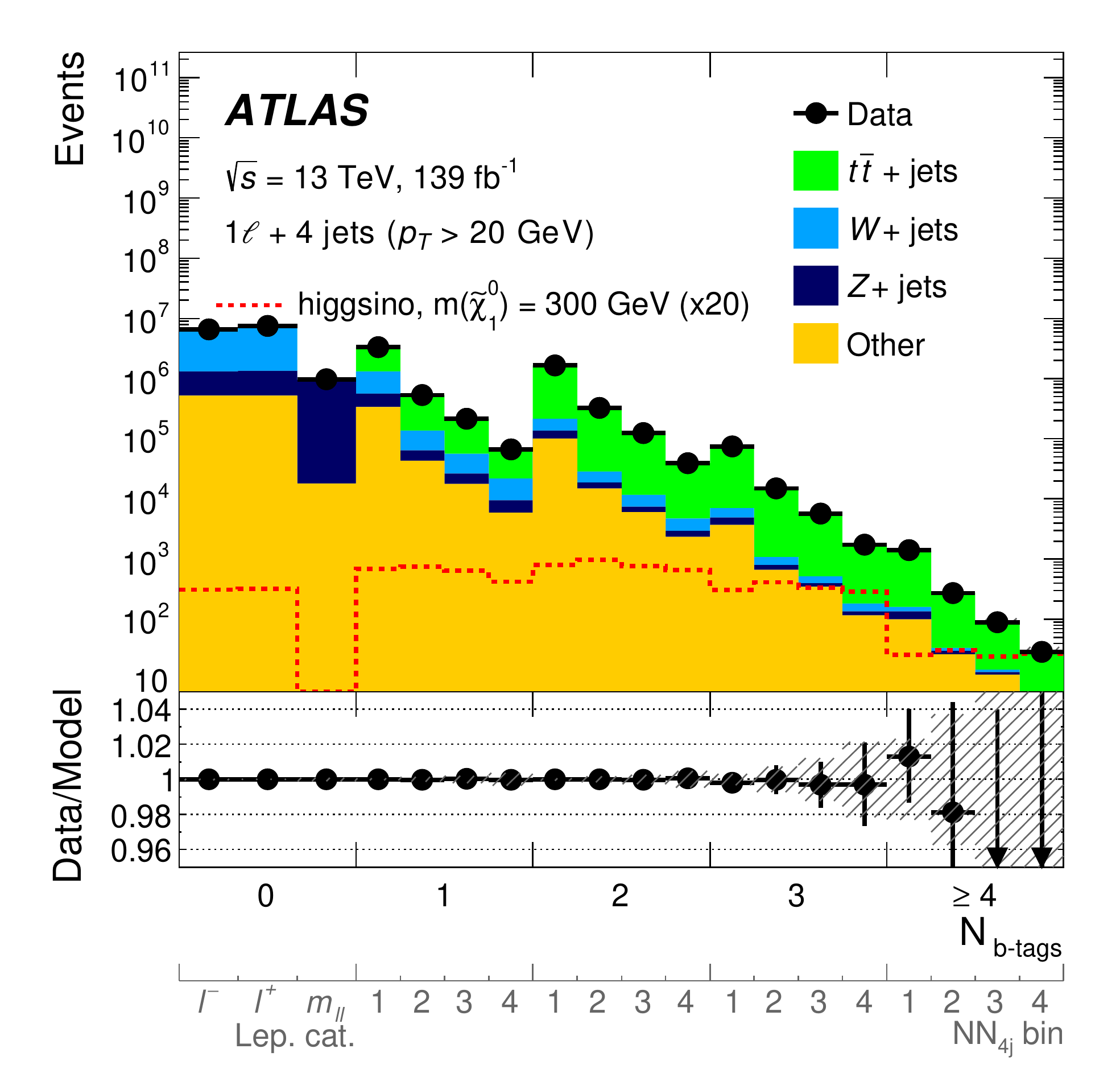}
\includegraphics[width=.32\textwidth]{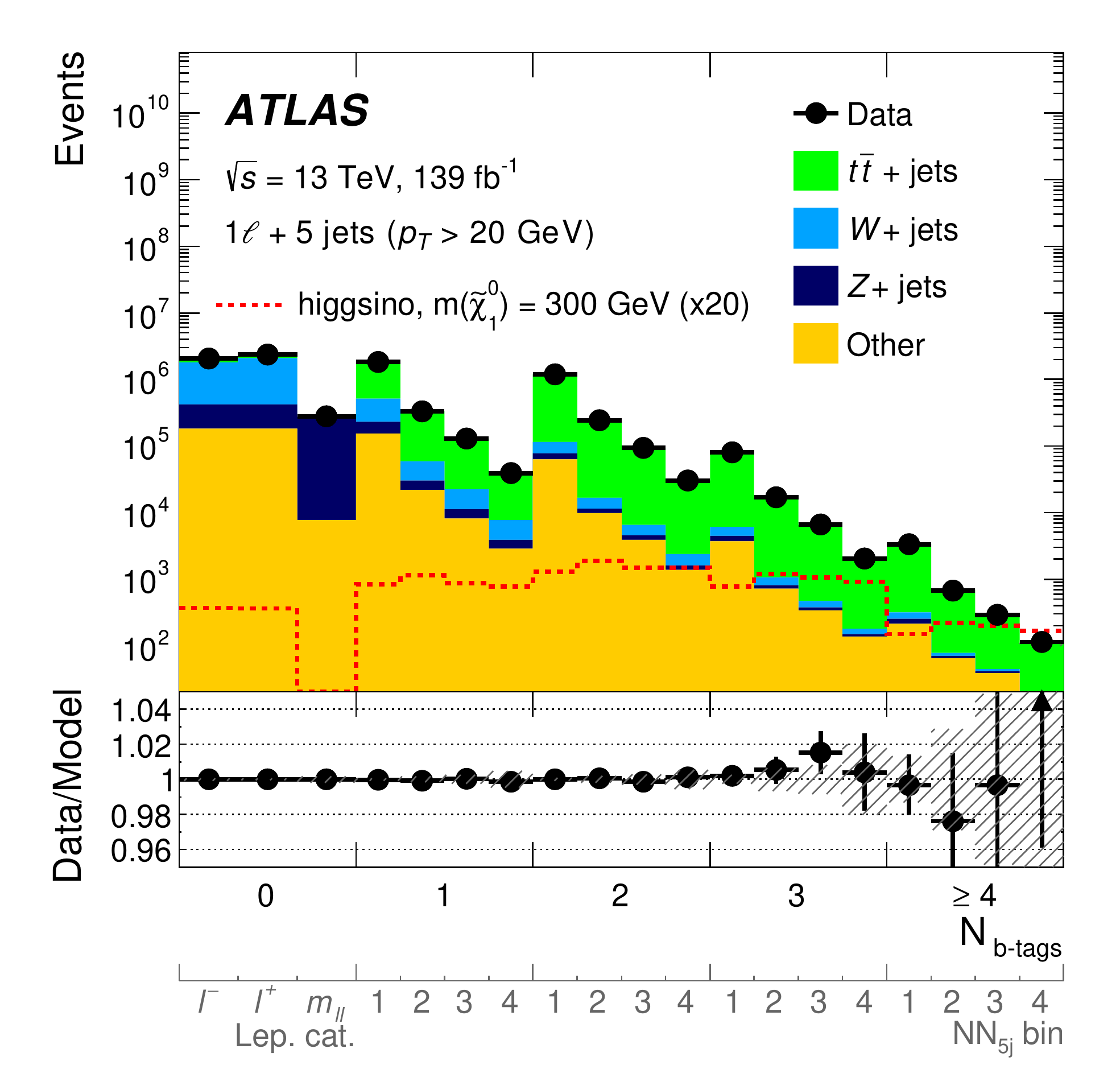}
\includegraphics[width=.32\textwidth]{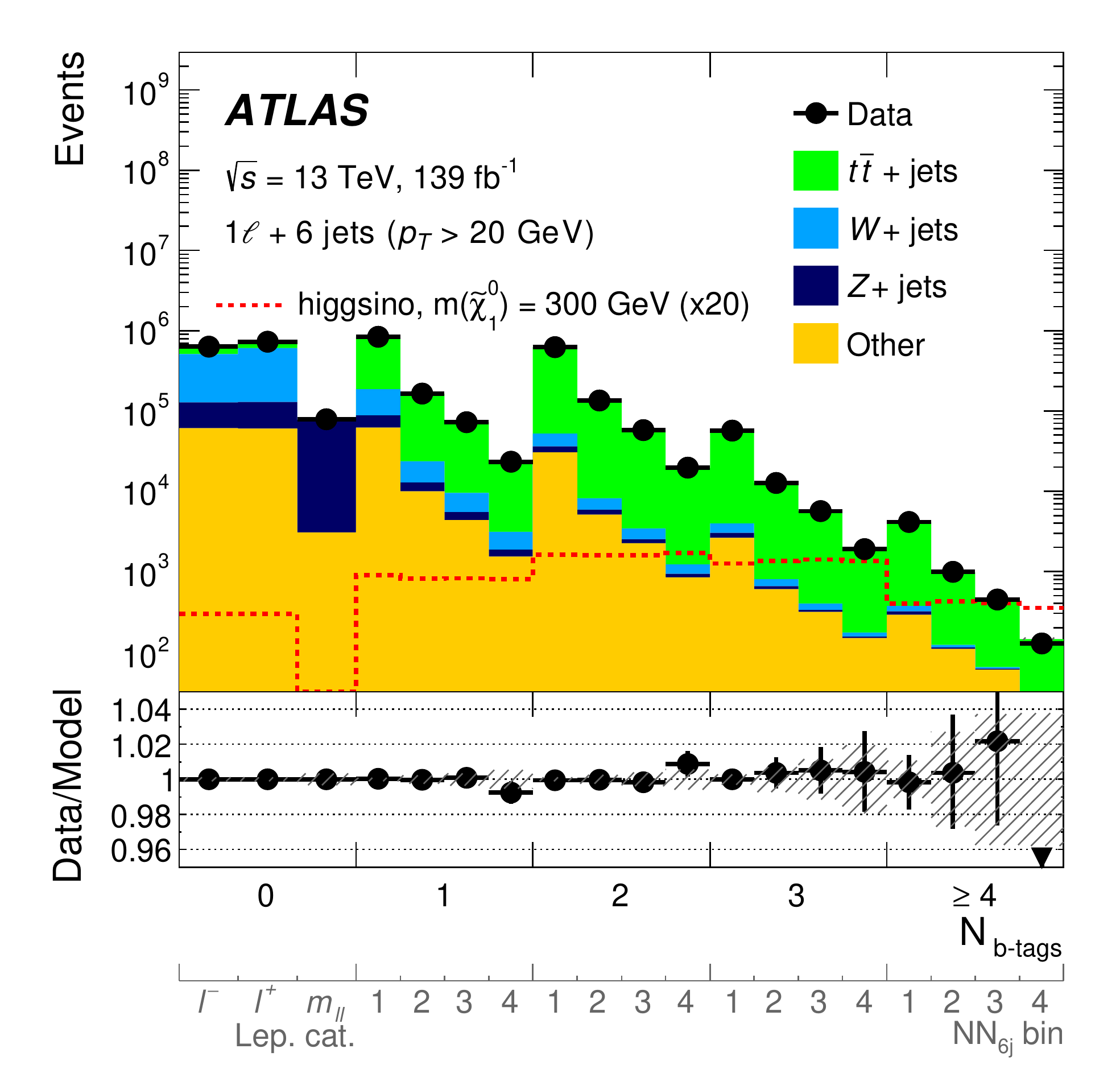}
\includegraphics[width=.32\textwidth]{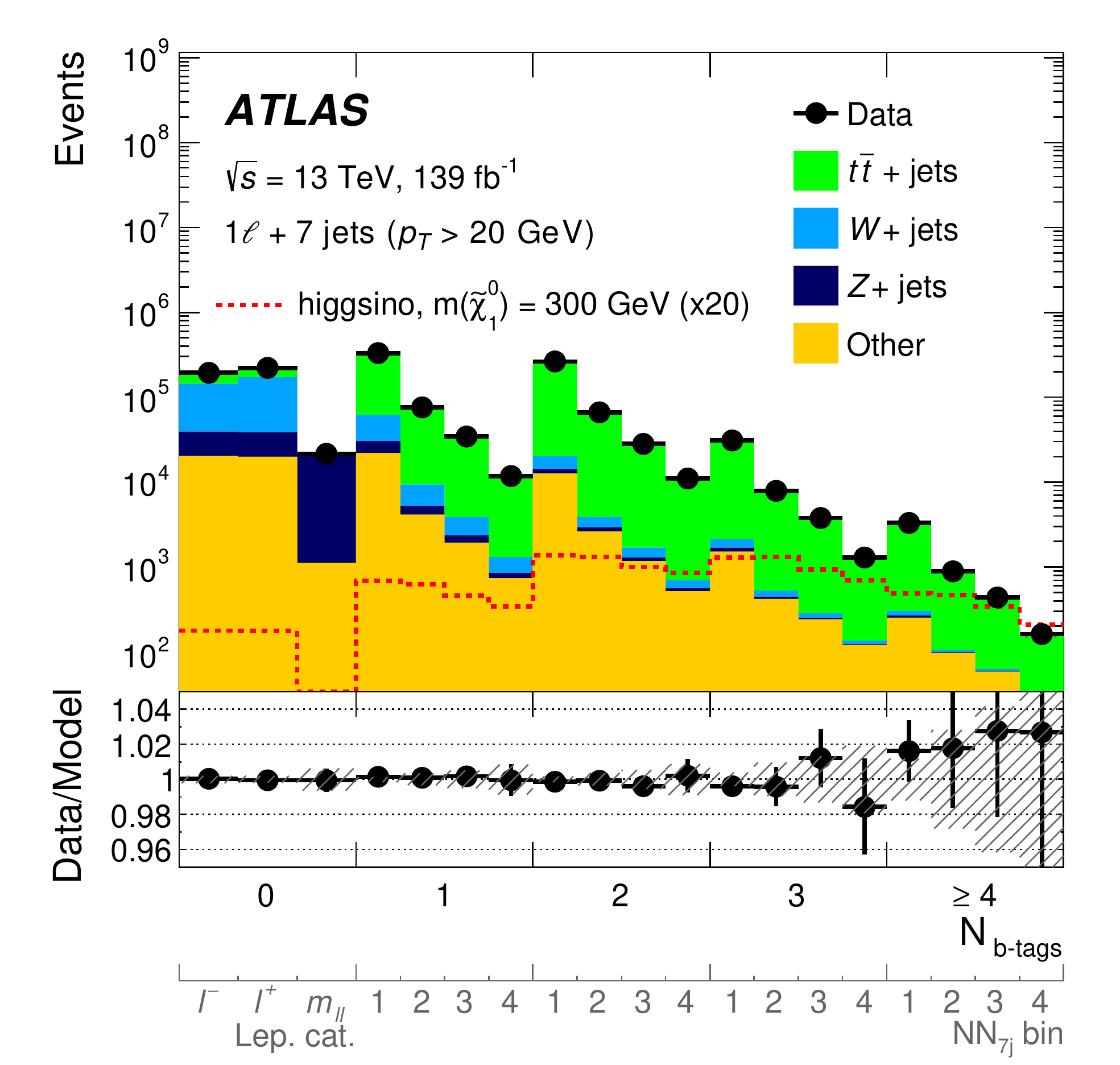}
\includegraphics[width=.32\textwidth]{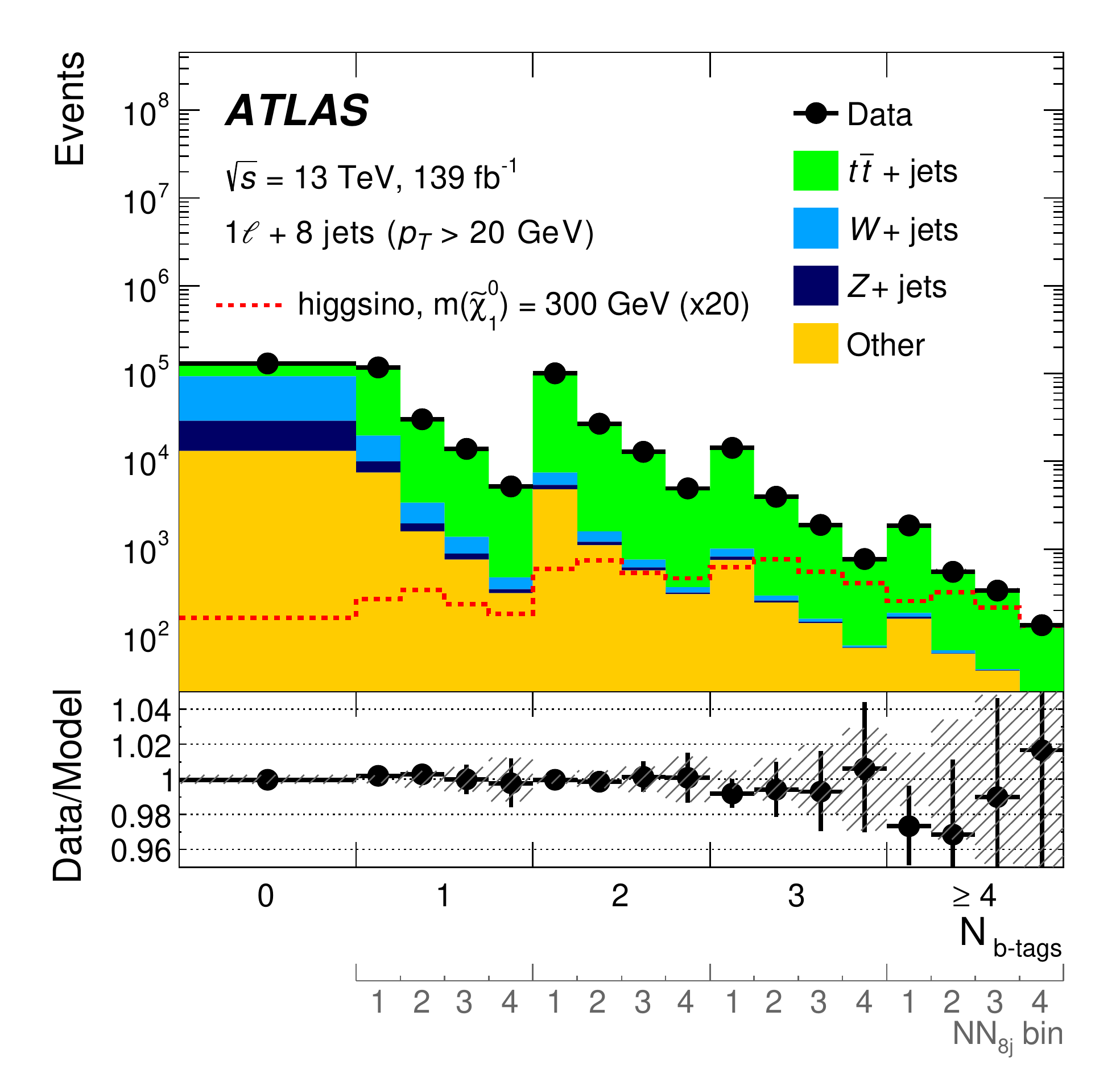}
\includegraphics[width=.32\textwidth]{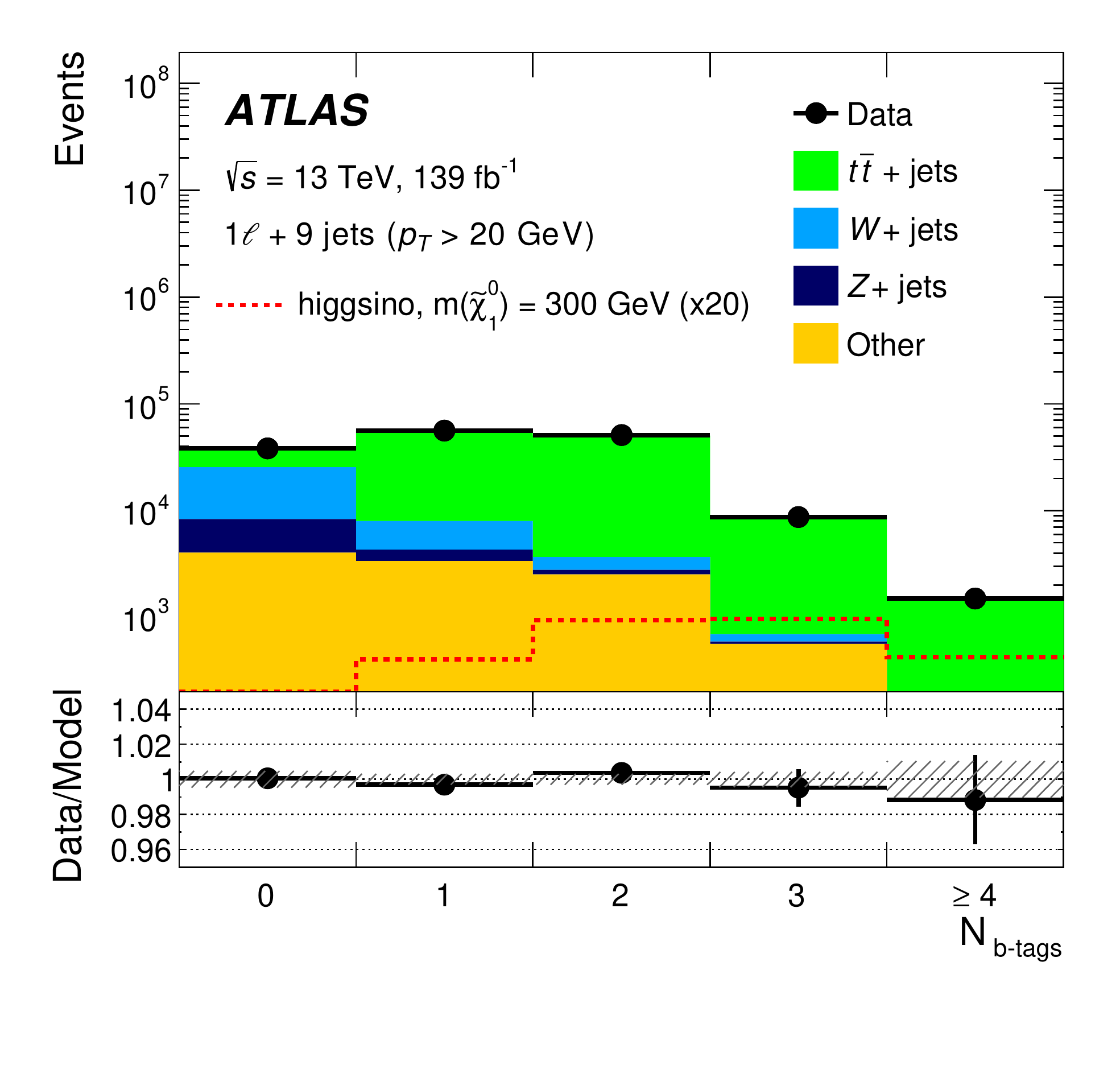}
\includegraphics[width=.32\textwidth]{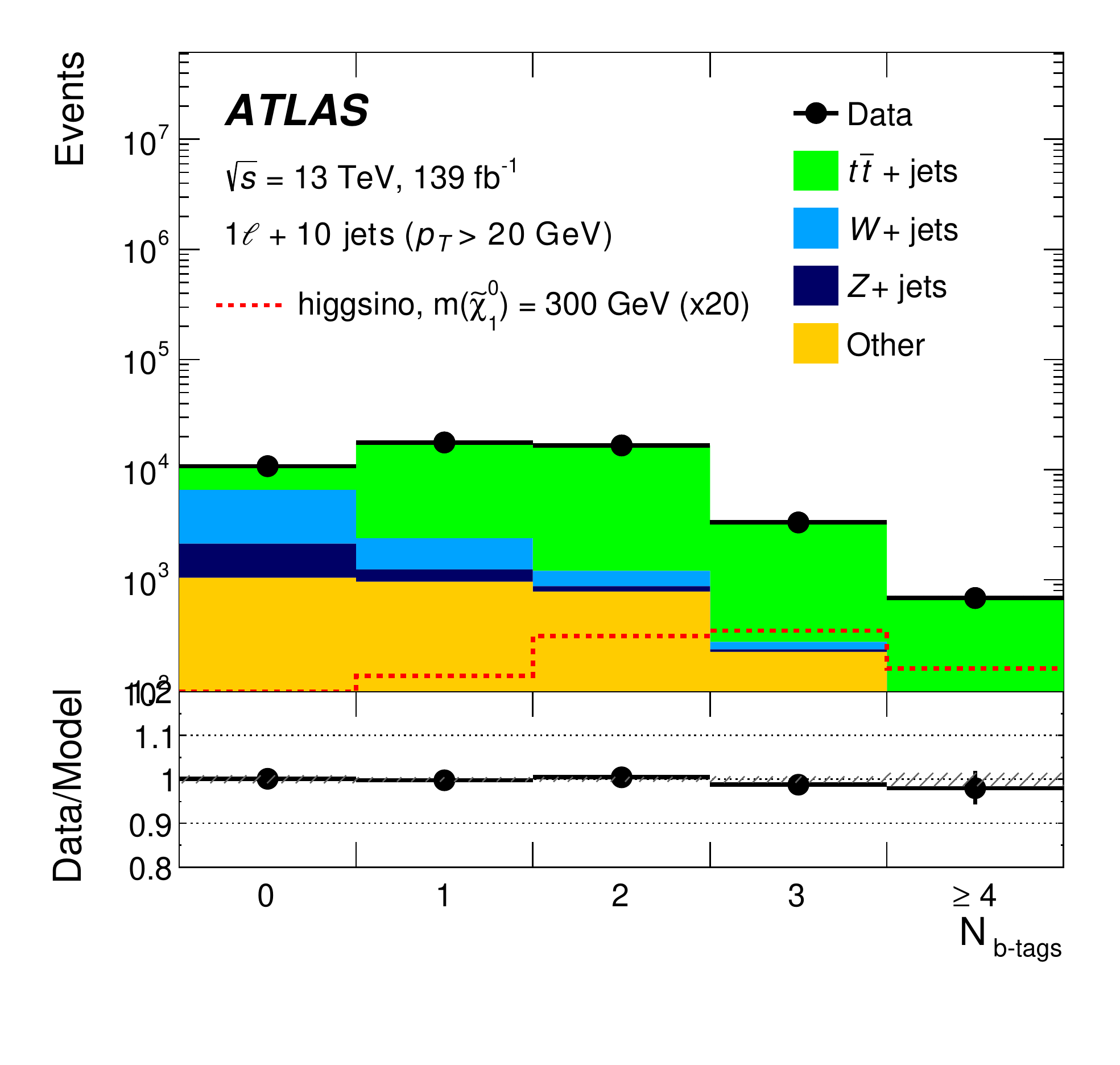}
\includegraphics[width=.32\textwidth]{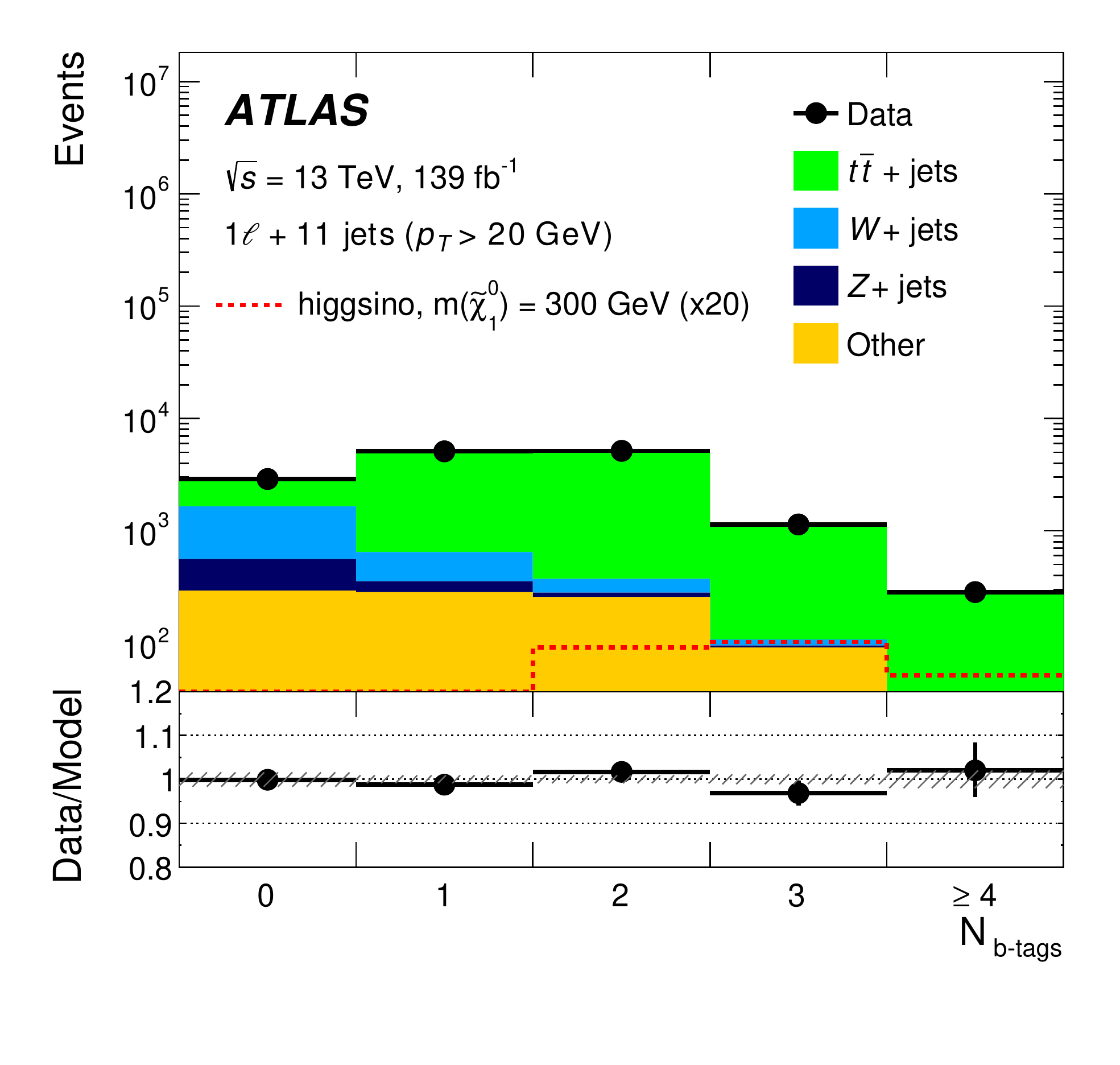}
\includegraphics[width=.32\textwidth]{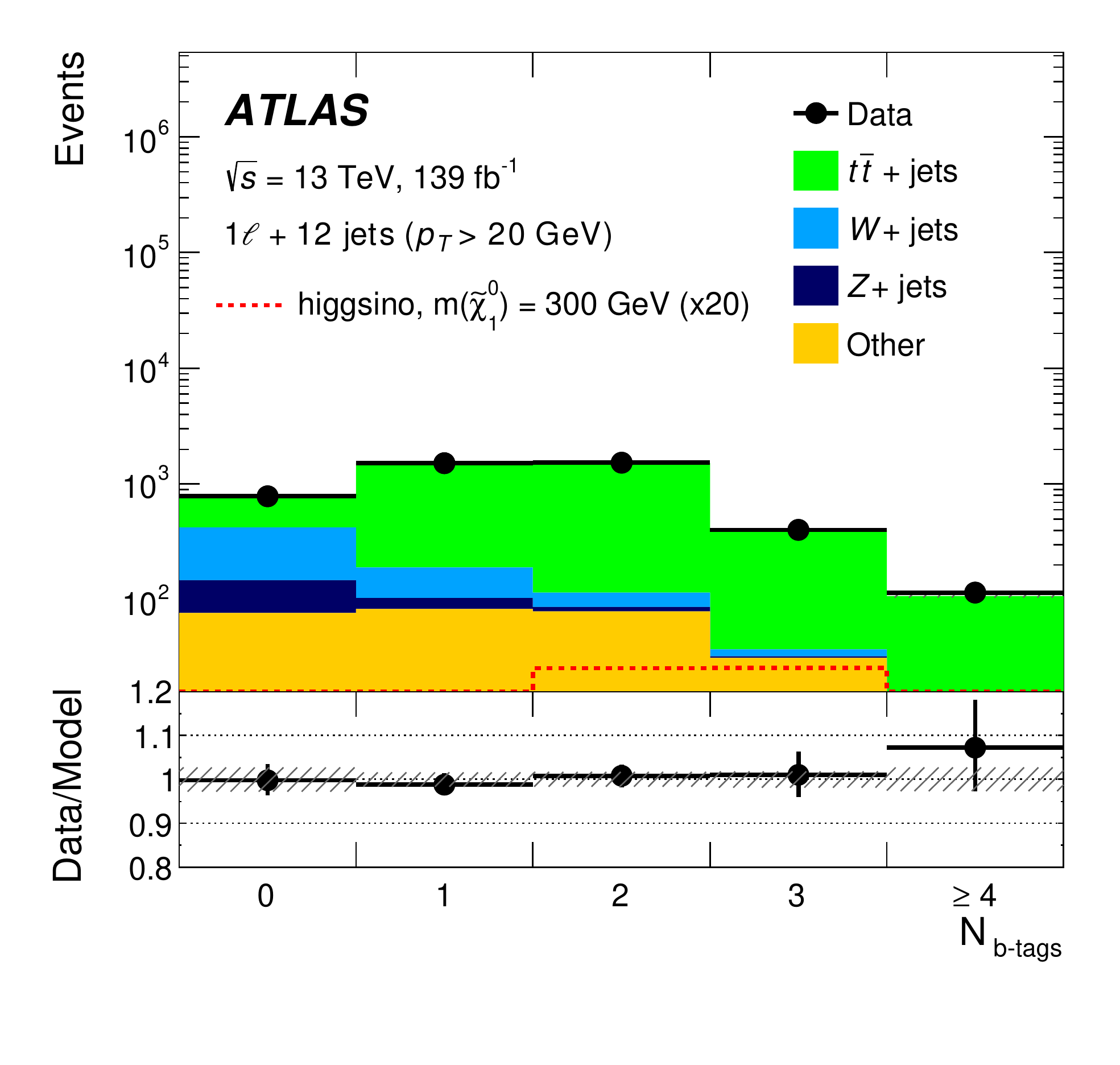}
\includegraphics[width=.32\textwidth]{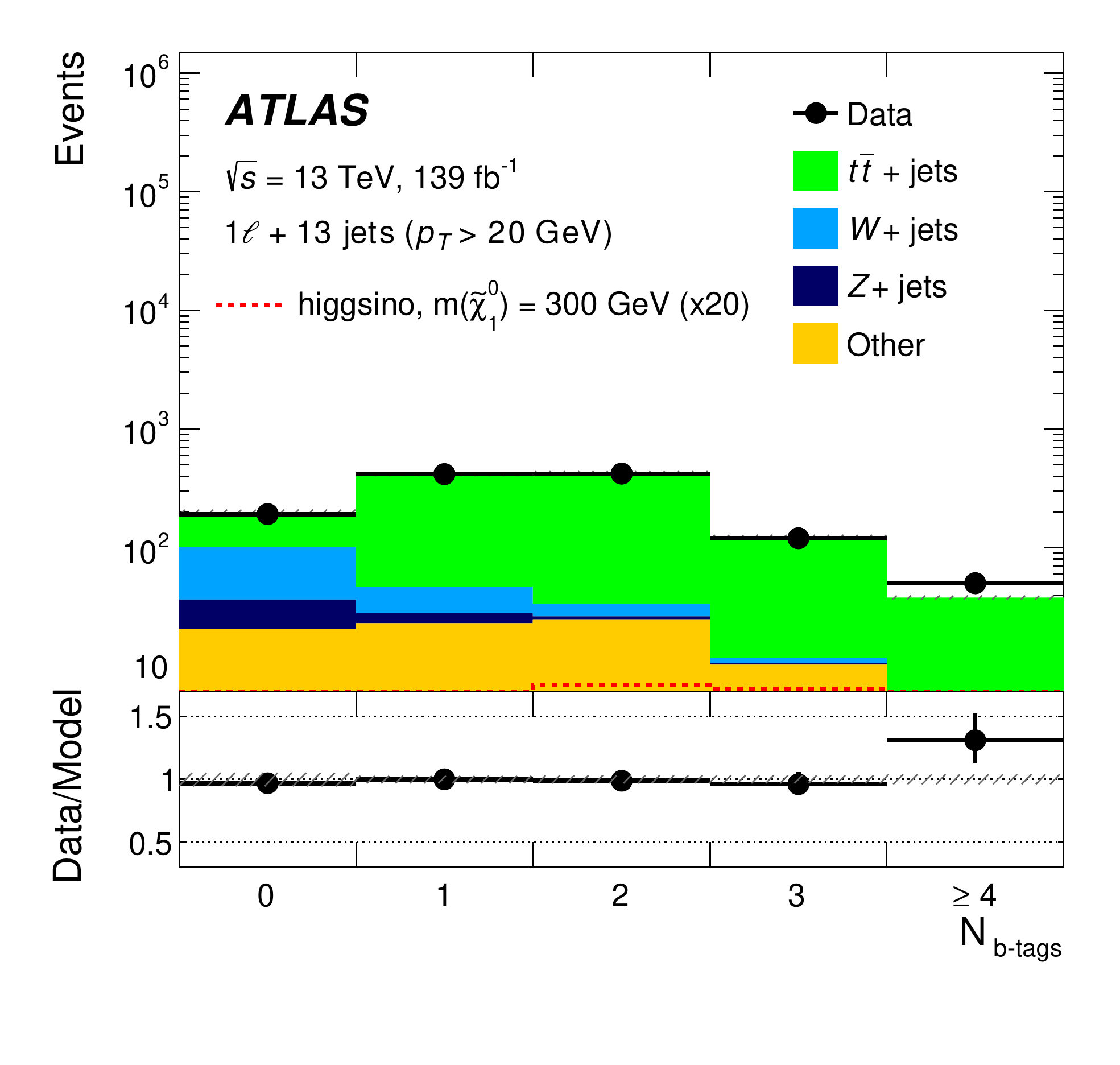}
\includegraphics[width=.32\textwidth]{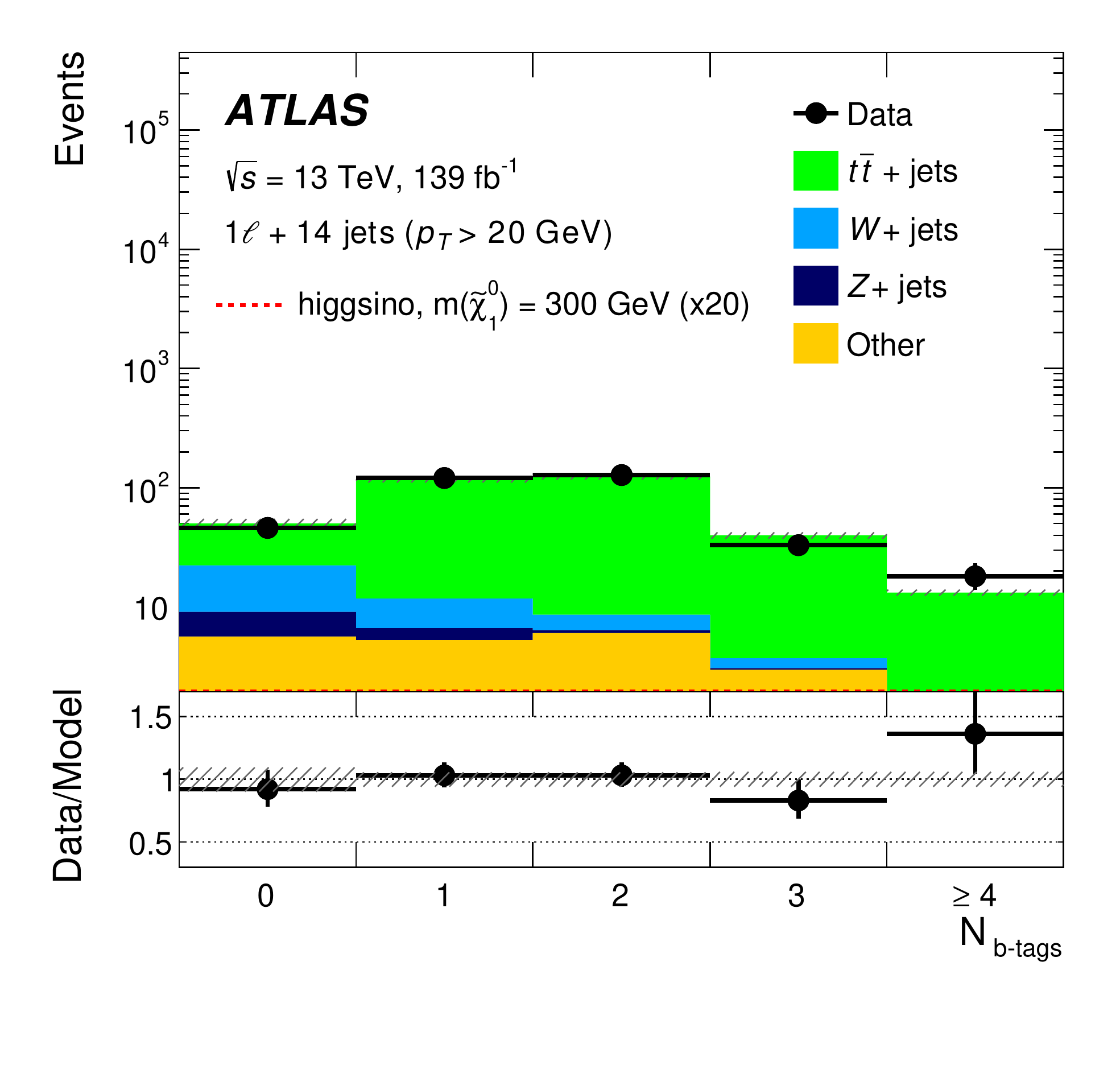}
\includegraphics[width=.32\textwidth]{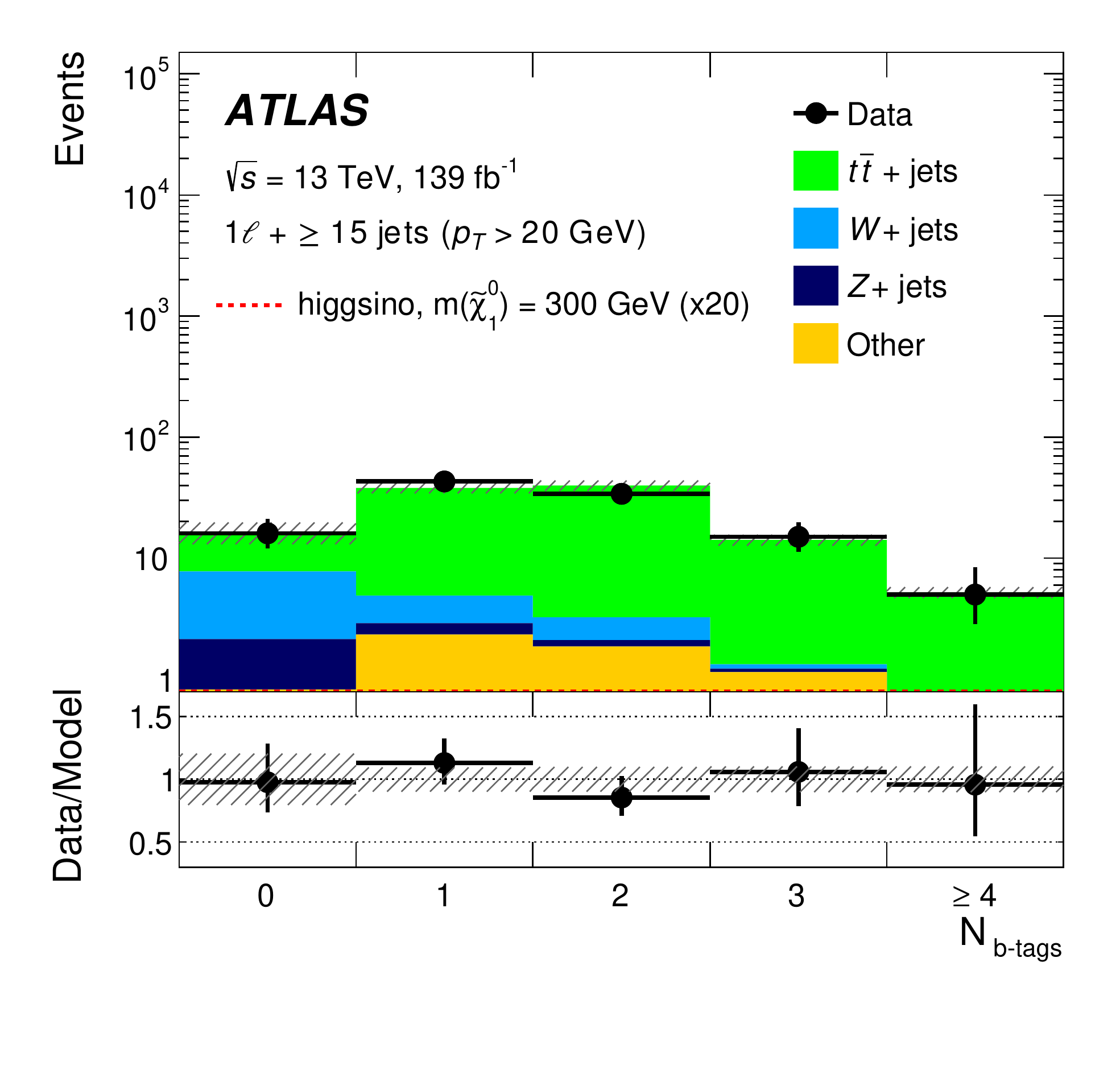} \\
\caption{
The observed data event yields and the corresponding estimates for
the backgrounds in the different $b$-jet multiplicity bins for the \SI{20}{\GeV} jet-\pt\ threshold regions defined for the EWK analysis in the \l category.
The background shown is estimated by including all bins in the fit.
All uncertainties, which may be correlated across the bins, are included in the error bands (shaded regions).
The expected signal distribution for the higgsino LSP $m$(\ninoone)~=~\SI{300}{\GeV} hypothesis normalized to 20 times its expected cross-section is also overlaid.
}
\label{fig:unblinded_1L_20gev_shape}
\end{figure}
 
\begin{figure}[htbp!]
\centering
\includegraphics[width=.32\textwidth]{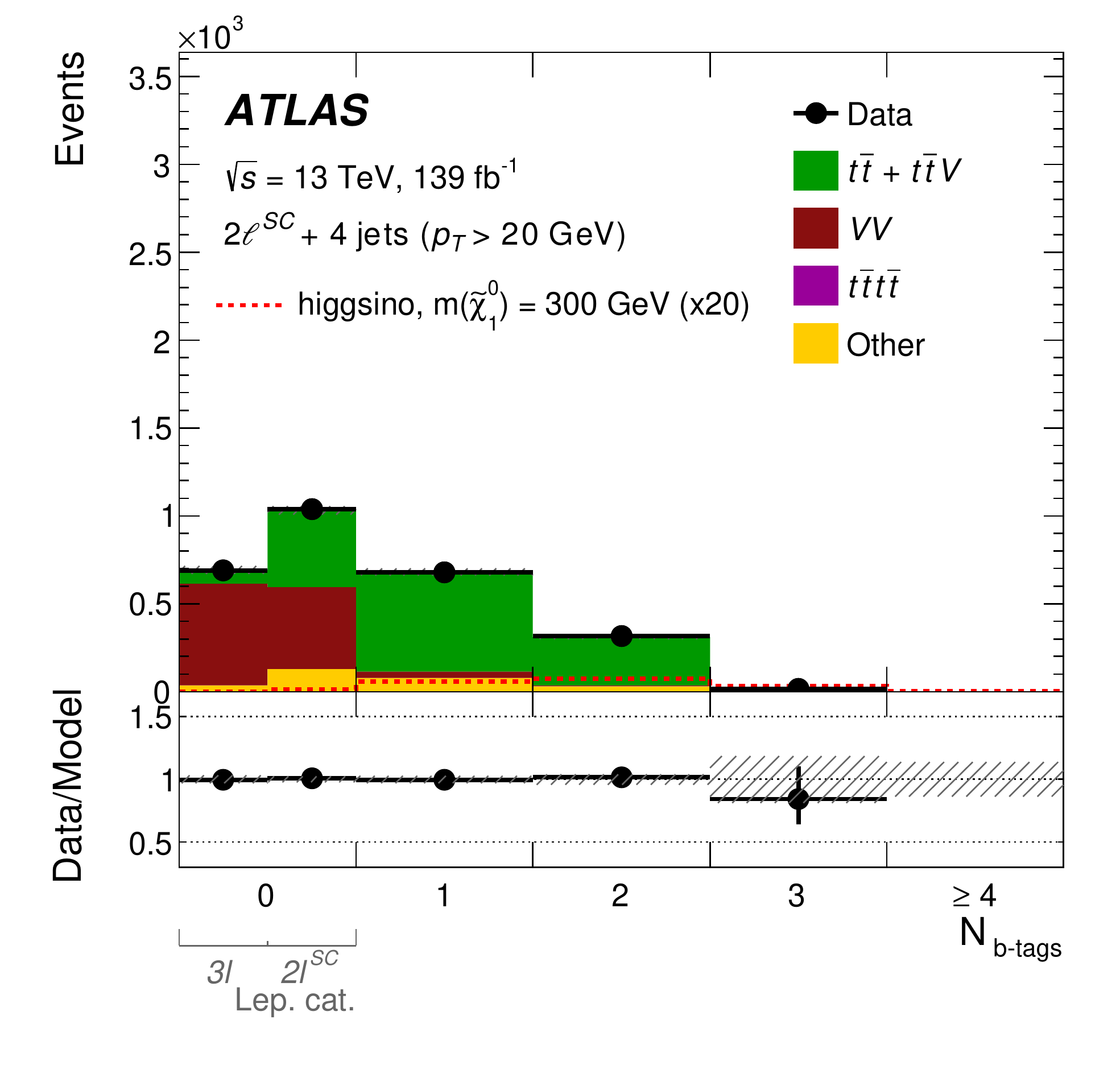}
\includegraphics[width=.32\textwidth]{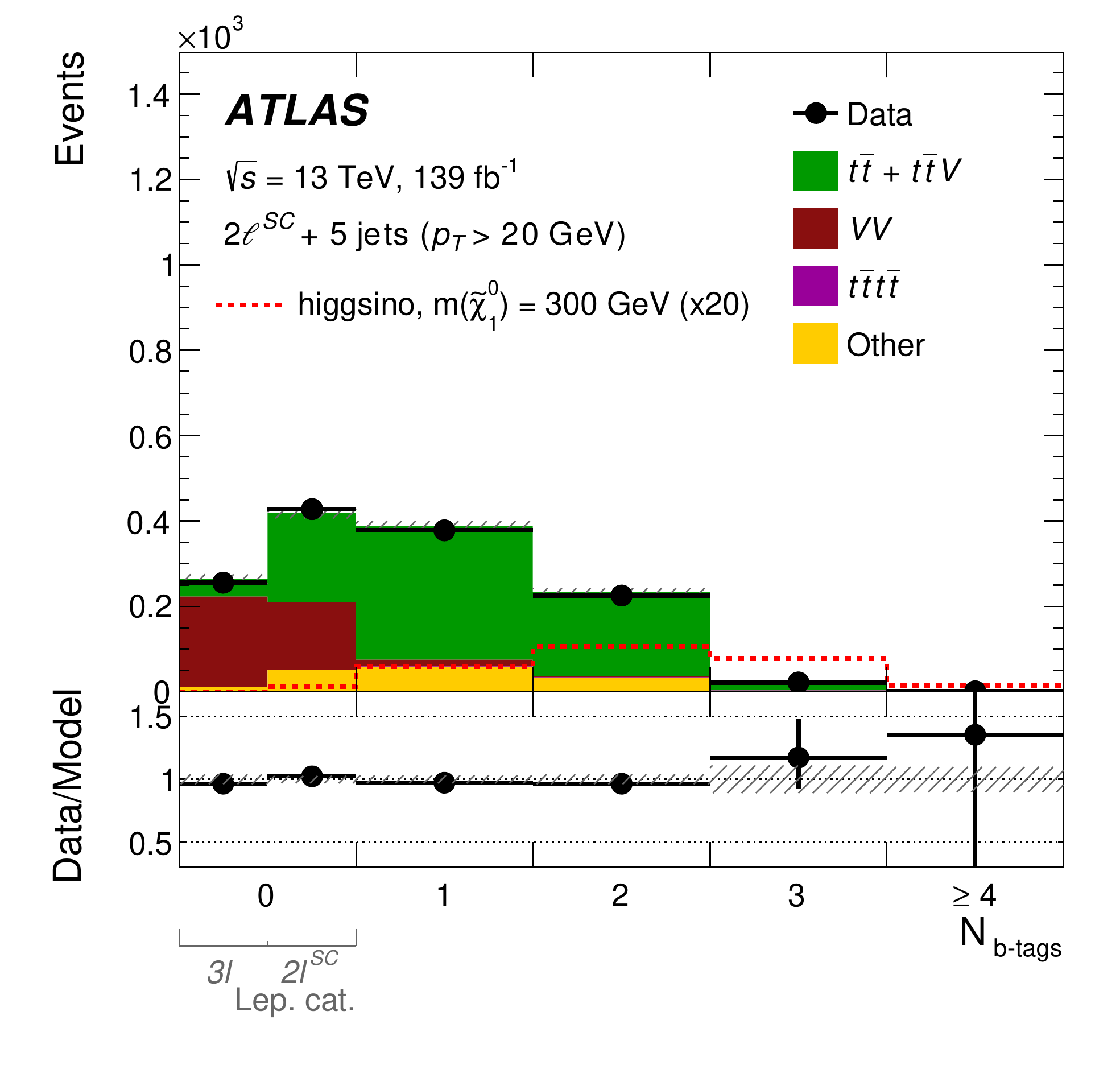}
\includegraphics[width=.32\textwidth]{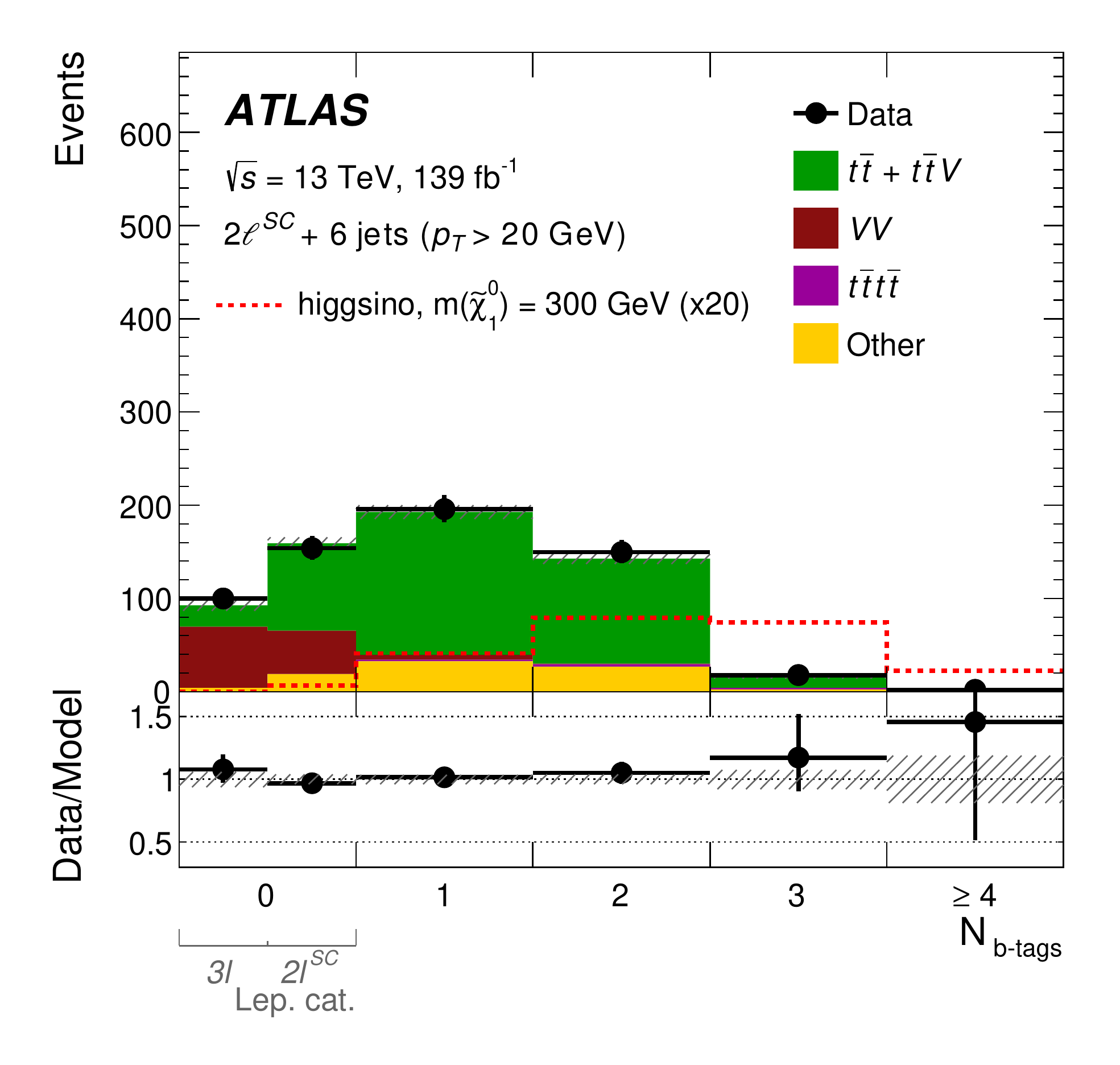}
\includegraphics[width=.32\textwidth]{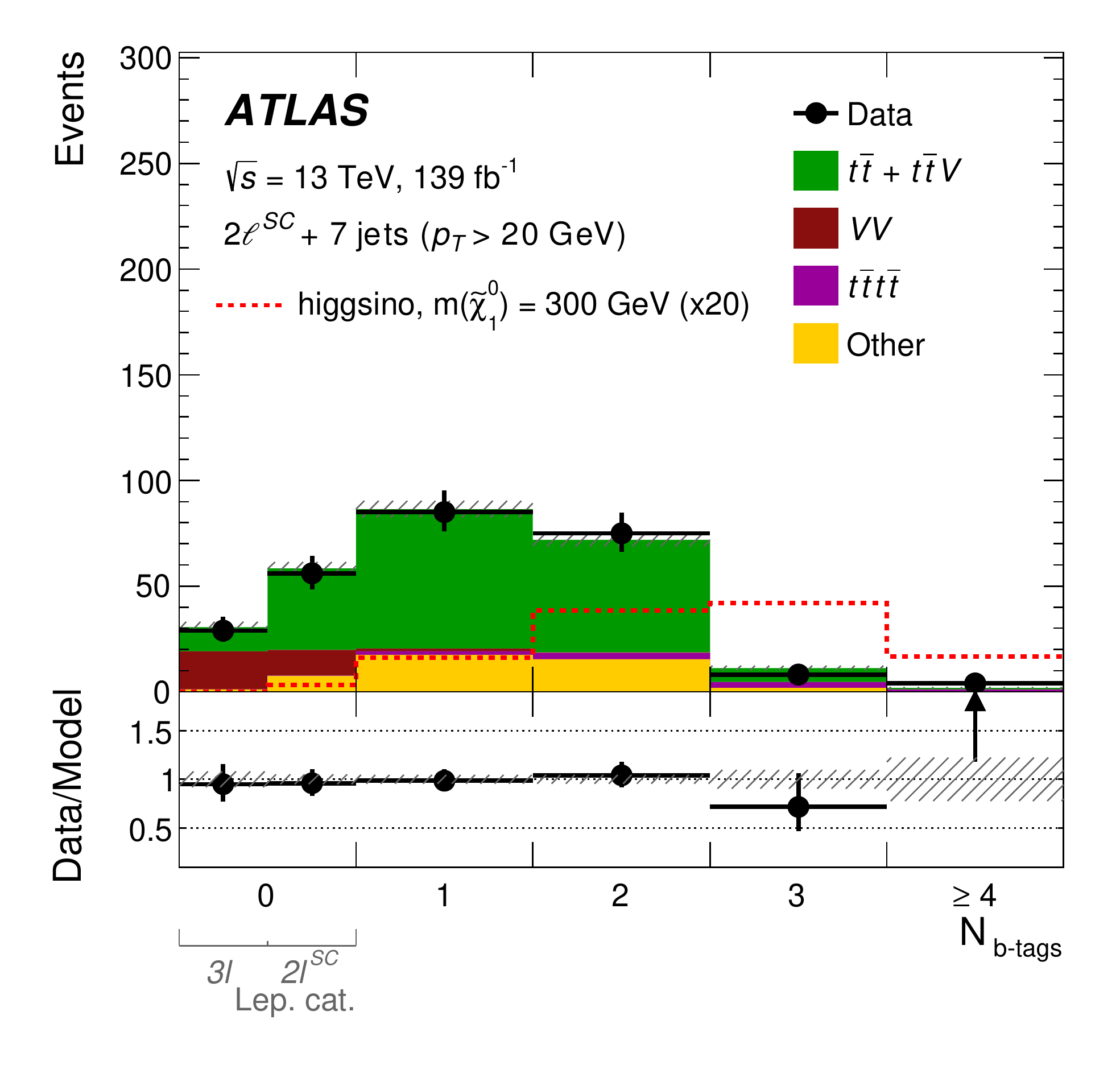}
\includegraphics[width=.32\textwidth]{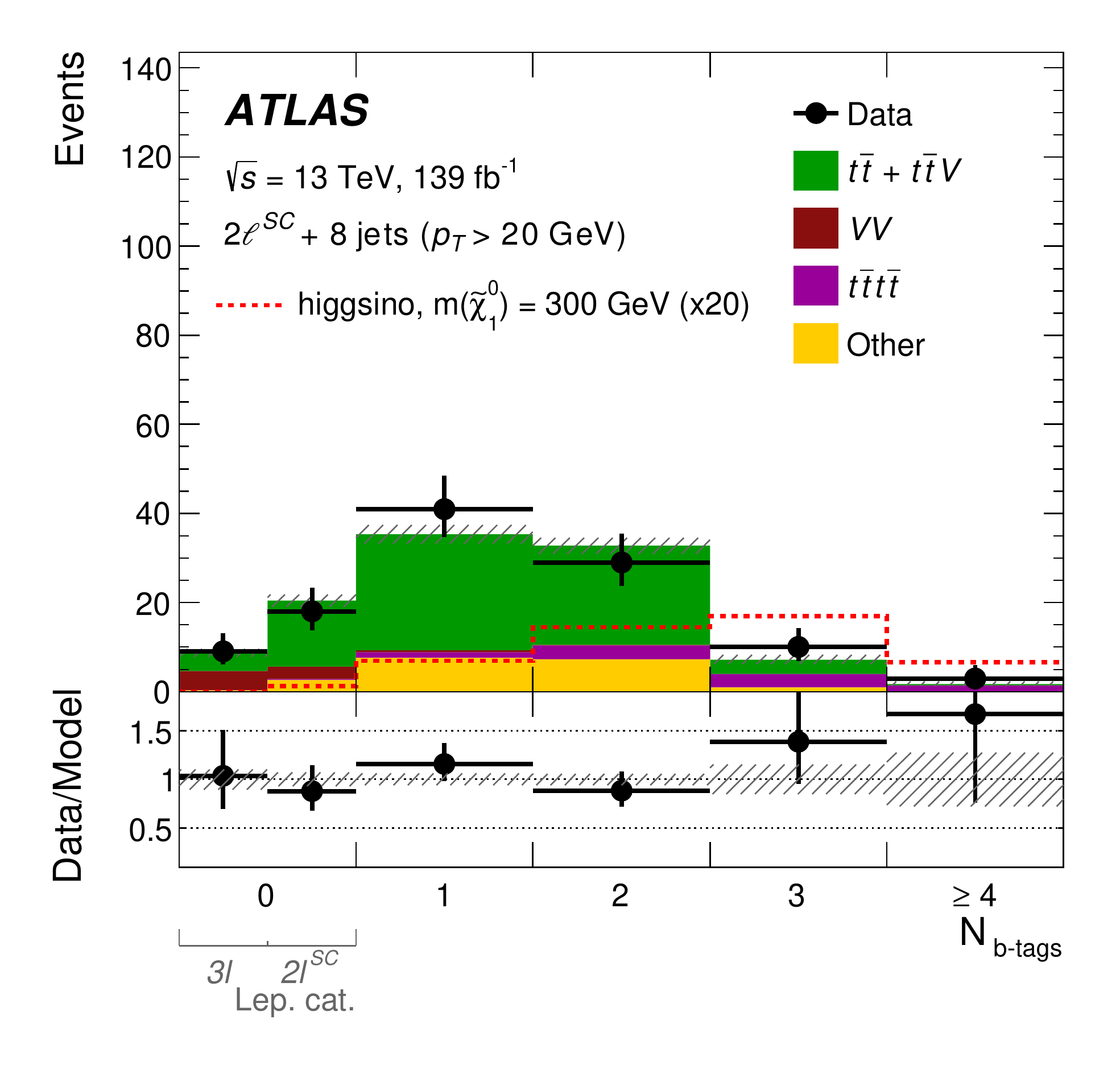}
\includegraphics[width=.32\textwidth]{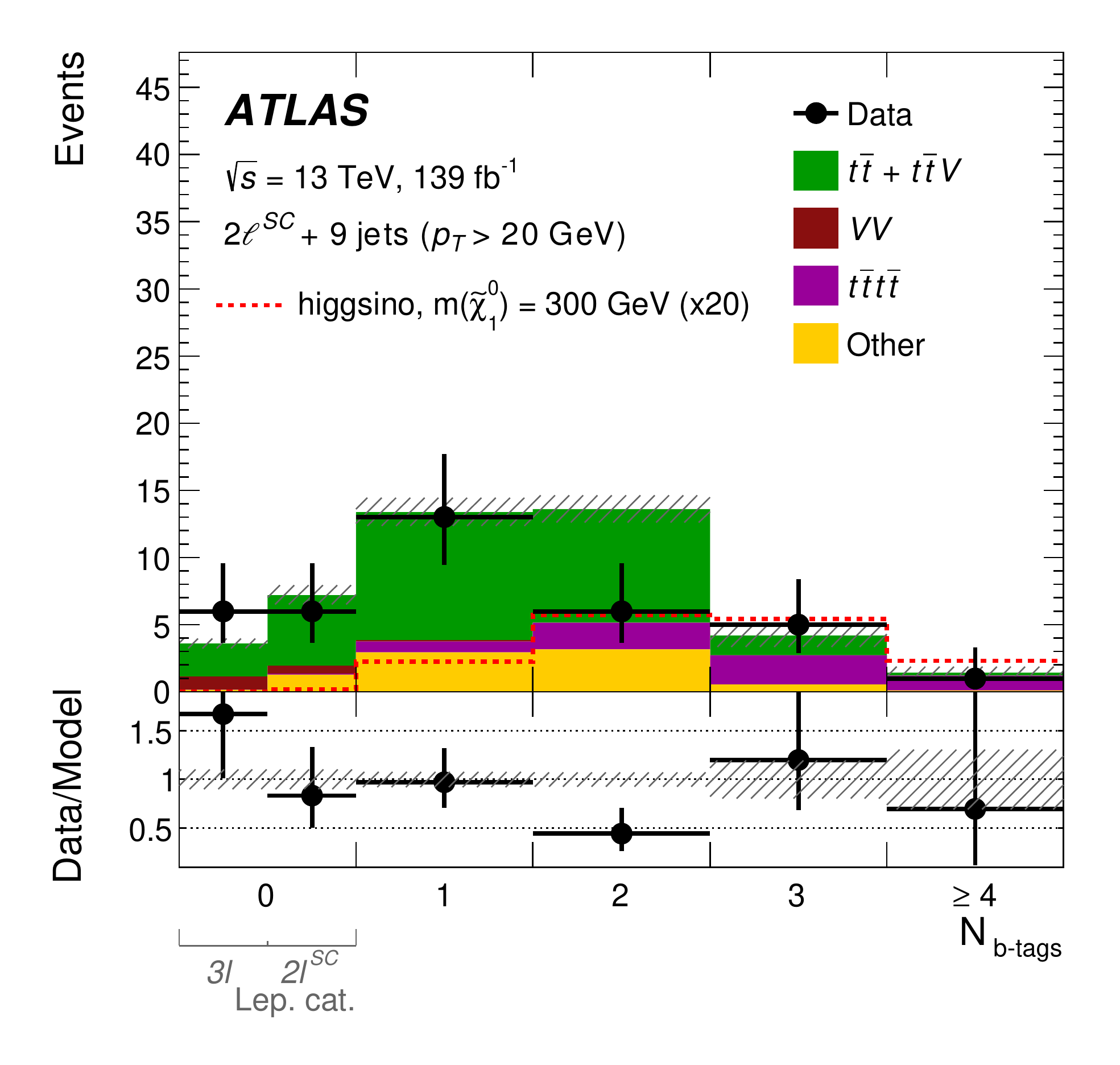}
\includegraphics[width=.32\textwidth]{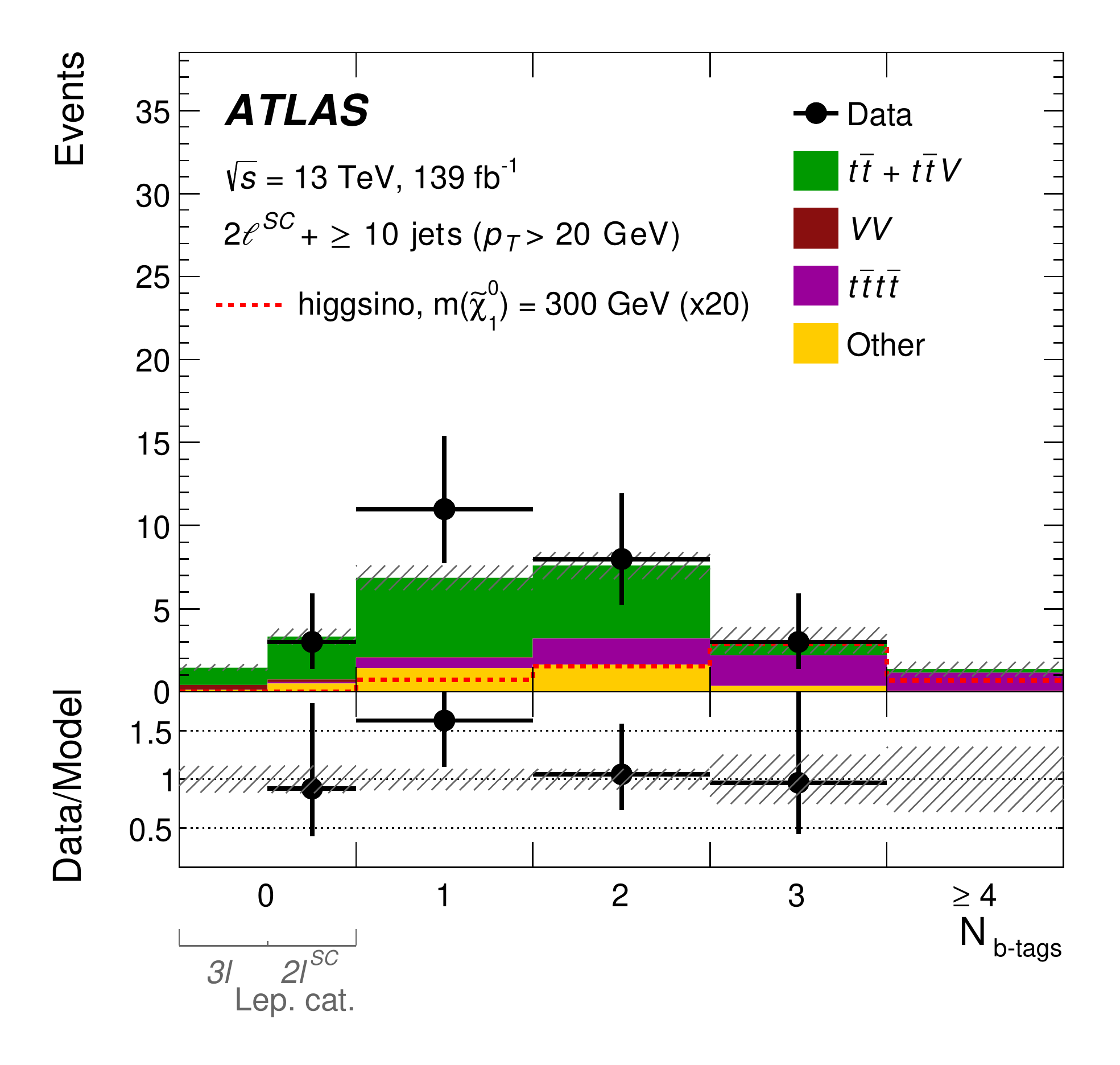} \\
\caption{
The observed data event yields and the corresponding estimates for
the backgrounds in the different $b$-jet multiplicity bins for the \SI{20}{\GeV} jet-\pt\ threshold regions defined for the EWK analysis in the \ll category.
The background shown is estimated by including all bins in the fit.
All uncertainties, which may be correlated across the bins, are included in the error bands (shaded regions).
The expected signal distribution for the higgsino LSP $m$(\ninoone)~=~\SI{300}{\GeV} hypothesis normalized to 20 times its expected cross-section is also overlaid.
}
\label{fig:unblinded_2LSS_20gev_shape}
\end{figure}
 
\begin{figure}[htbp!]
\centering
\includegraphics[width=.49\textwidth,valign=t]{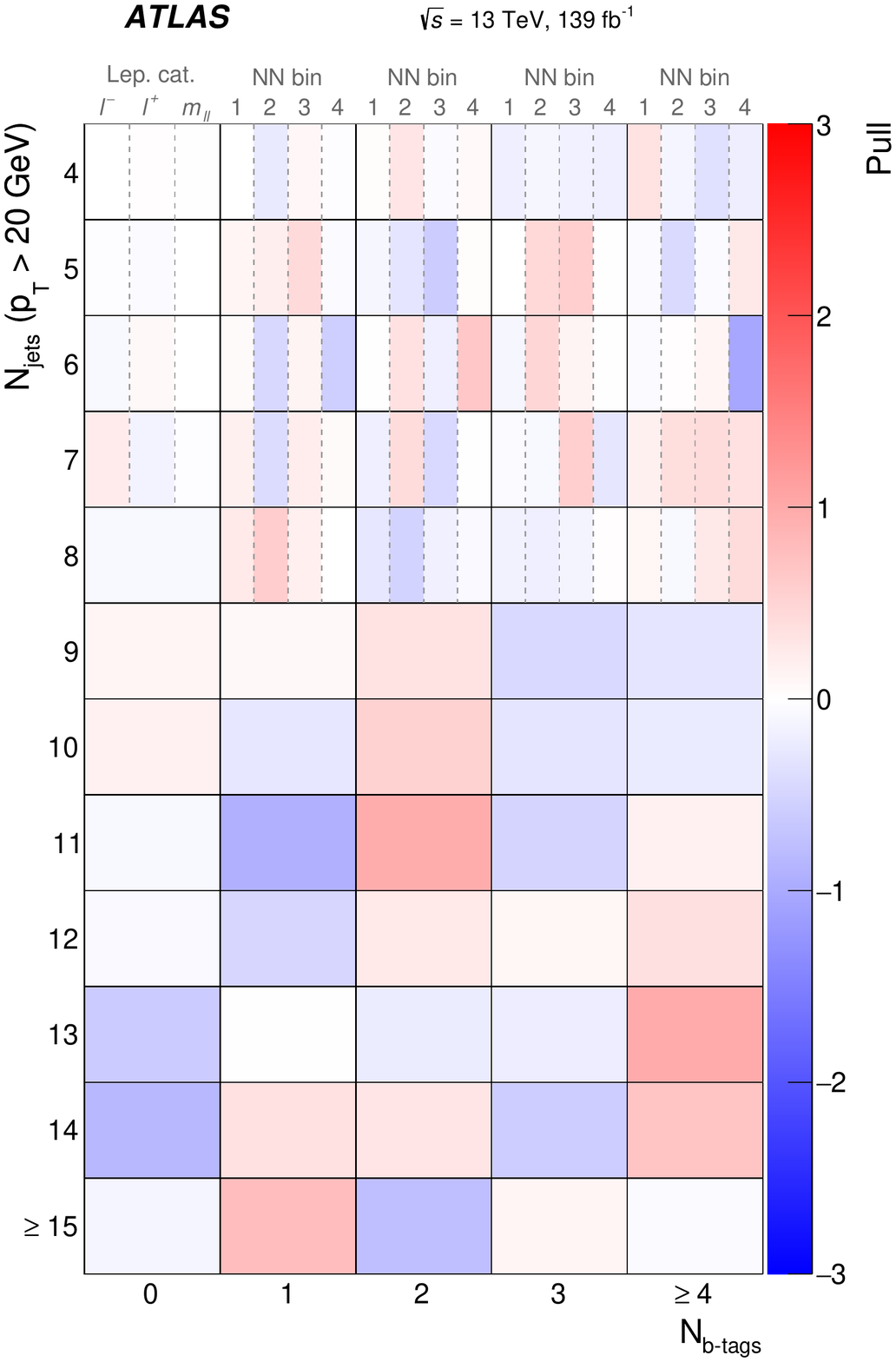}
\includegraphics[width=.49\textwidth,valign=t]{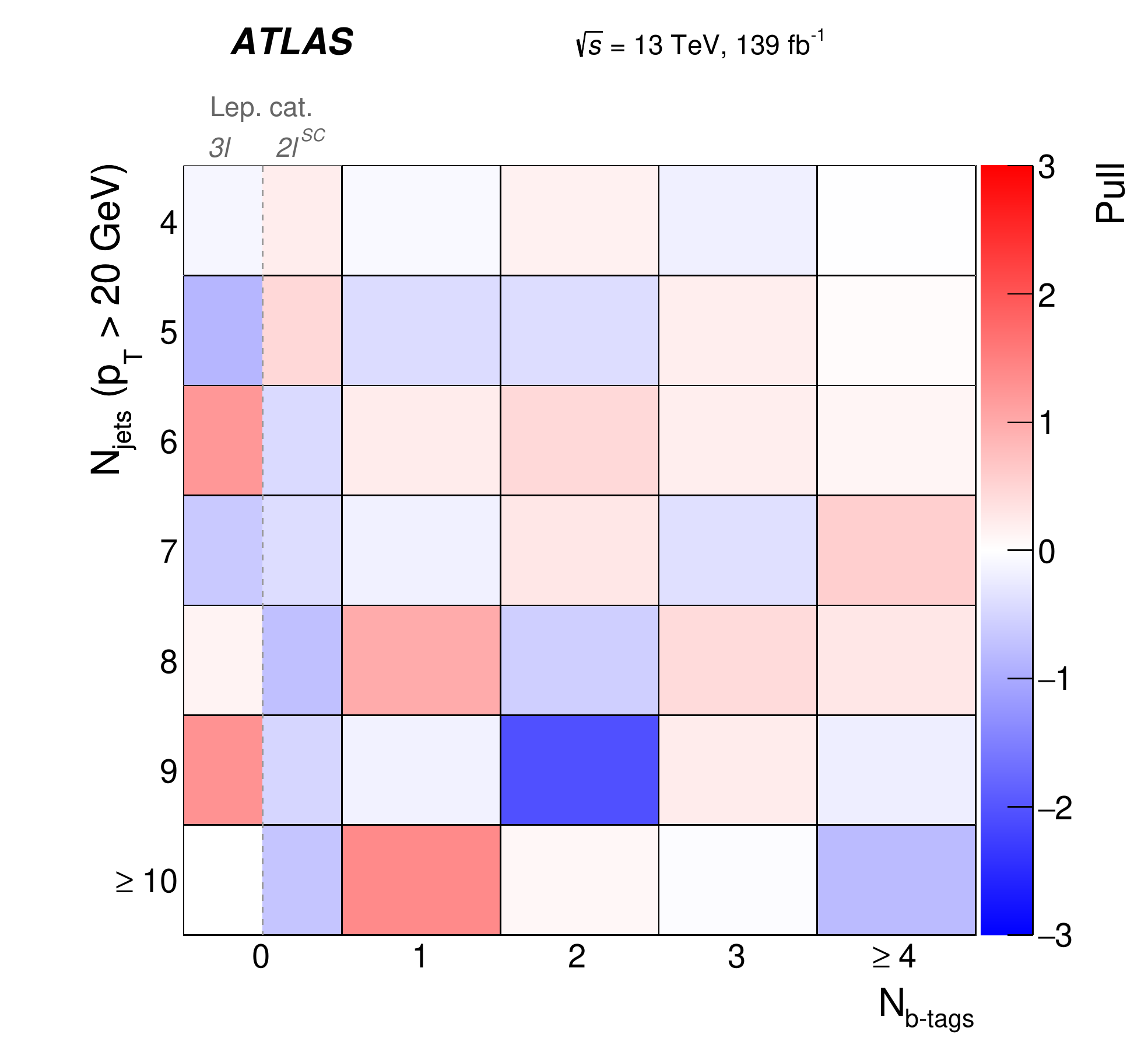}
\caption{
The observed pulls in all the regions considered in the EWK analysis, in the \l category (left) and the \ll category (right). The pull is defined as the difference between the observed number of events and the total expected
number of events determined by the fit divided by the total uncertainty. The total uncertainty is the sum in quadrature of the statistical error of the observed data and the uncertainty in the predicted background.
}
\label{fig:pull_unblinded_20gev_shape}
\end{figure}
 
\begin{figure}[htbp!]
\centering
\includegraphics[clip, trim=0 6cm 0 0, width=.4\textwidth]{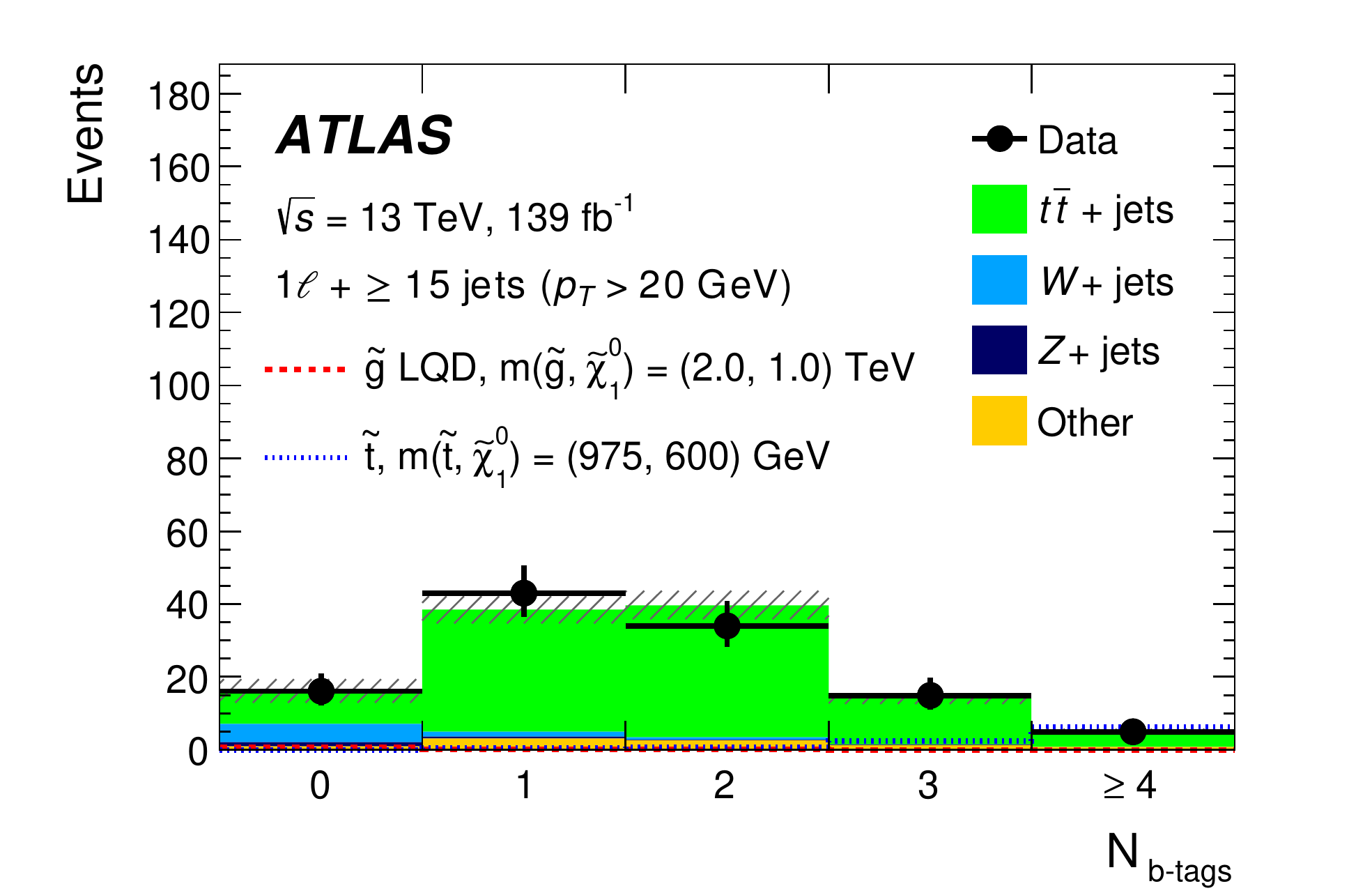}
\includegraphics[clip, trim=0 6cm 0 0, width=.4\textwidth]{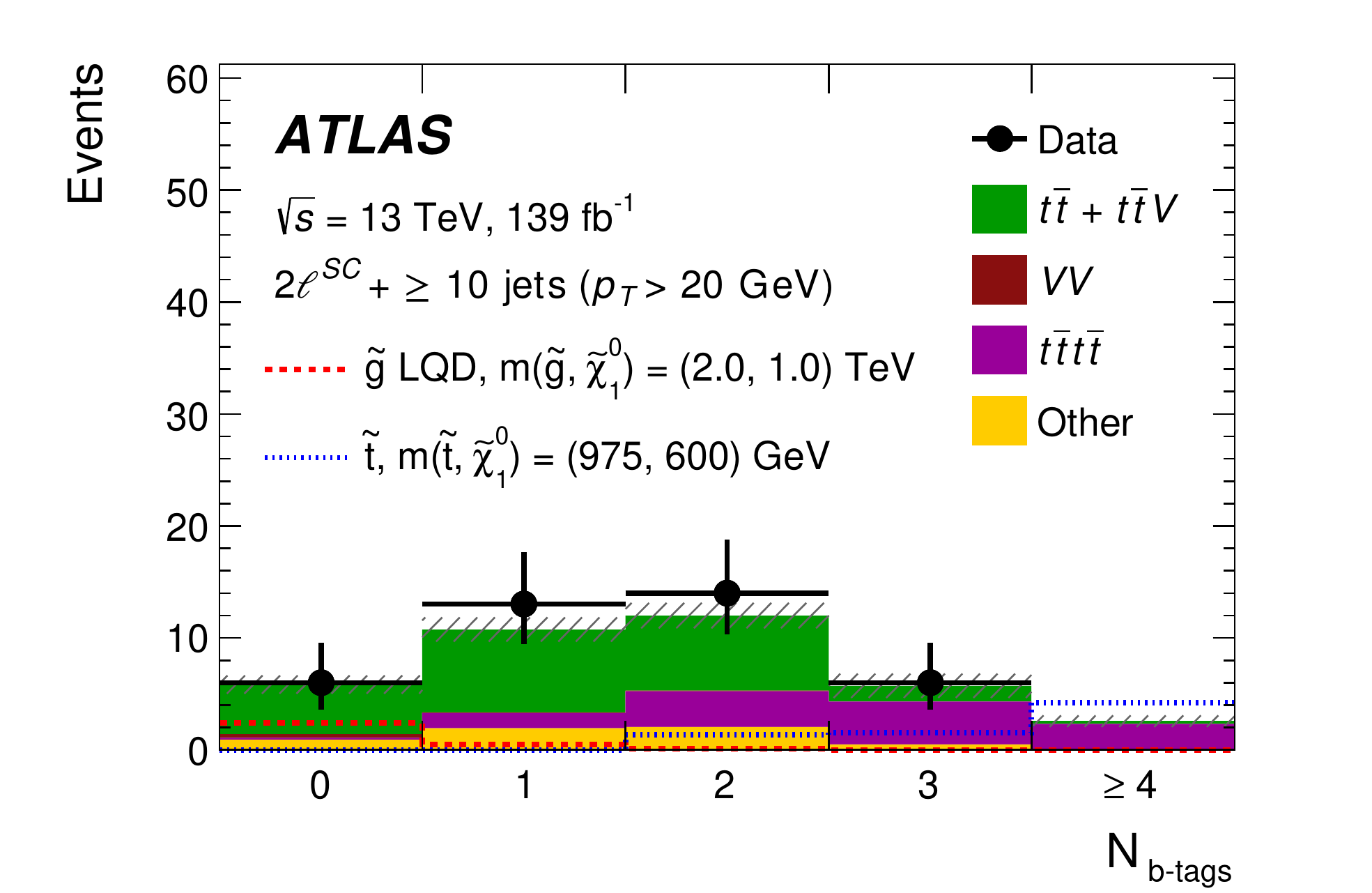} \\
\includegraphics[clip, trim=0 6cm 0 0, width=.4\textwidth]{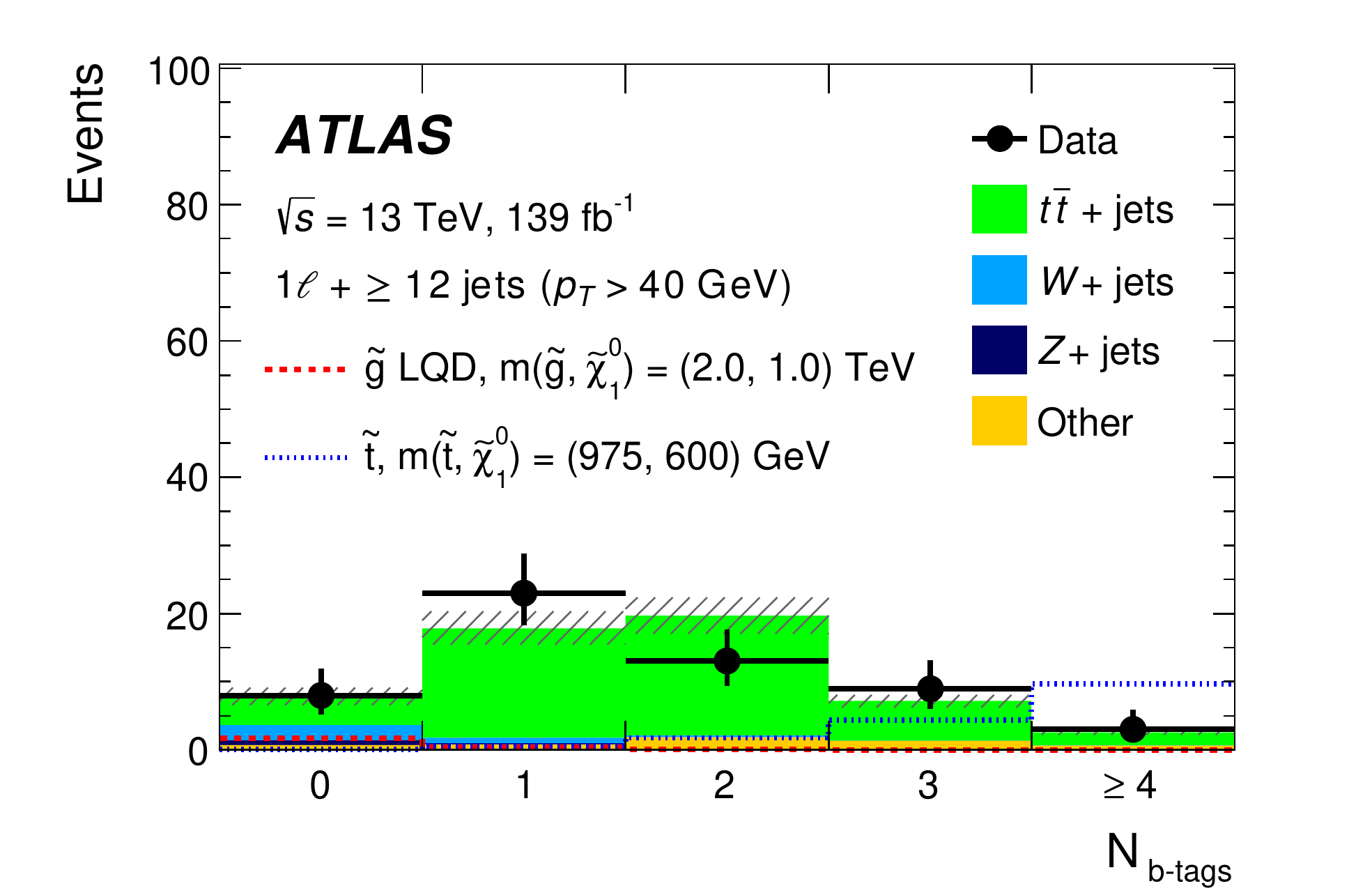}
\includegraphics[clip, trim=0 6cm 0 0, width=.4\textwidth]{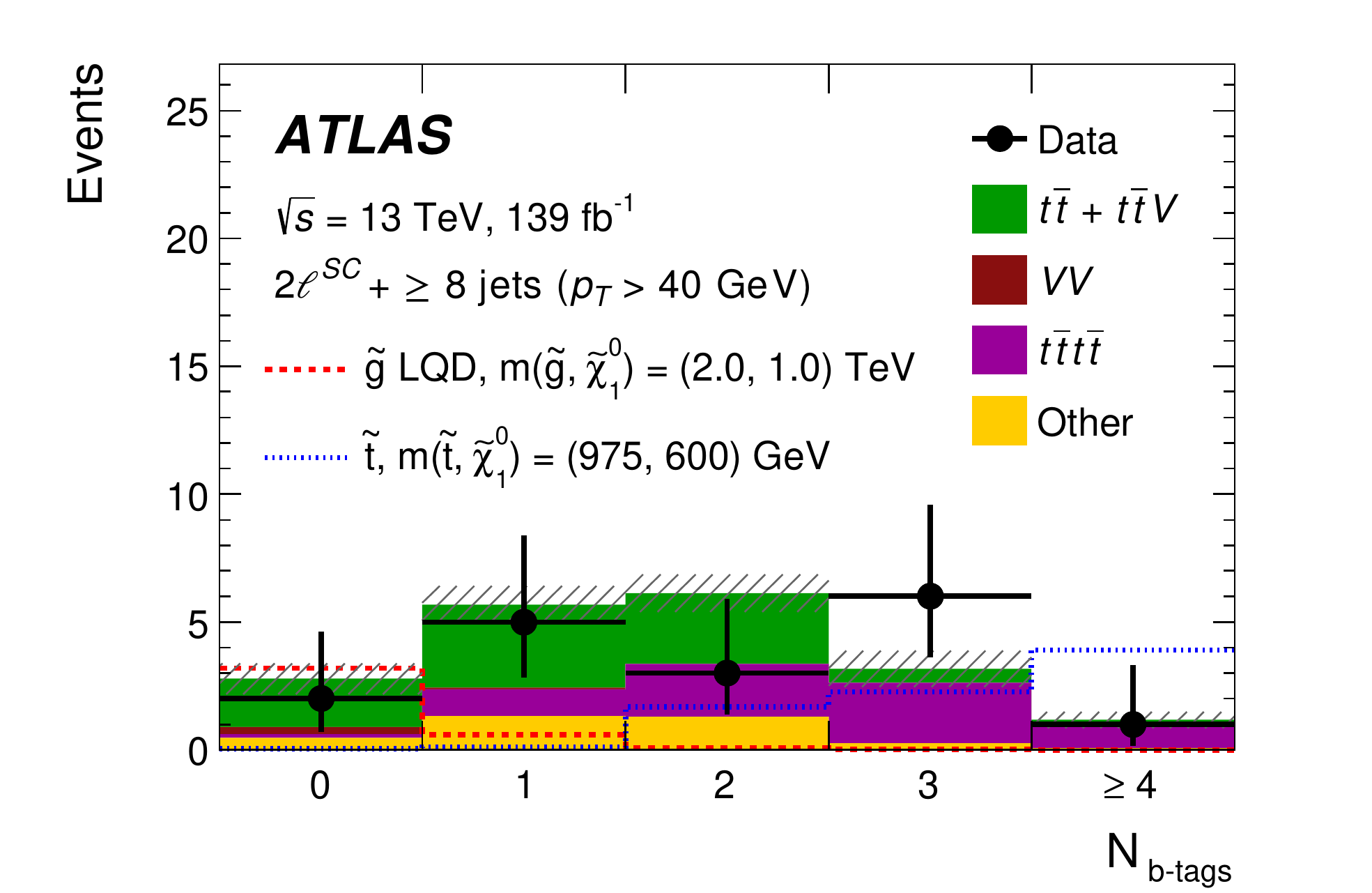} \\
\includegraphics[clip, trim=0 6cm 0 0, width=.4\textwidth]{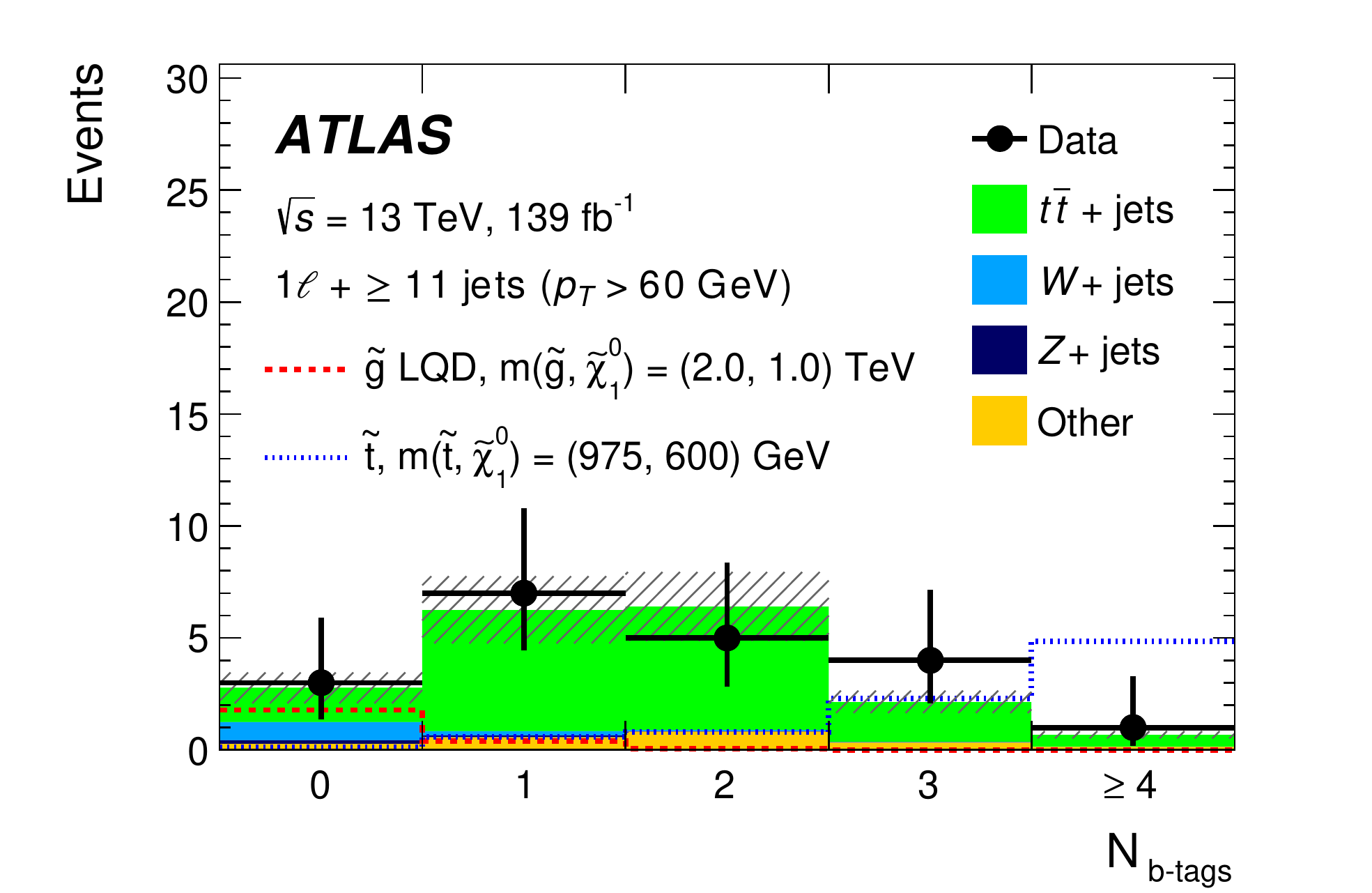}
\includegraphics[clip, trim=0 6cm 0 0, width=.4\textwidth]{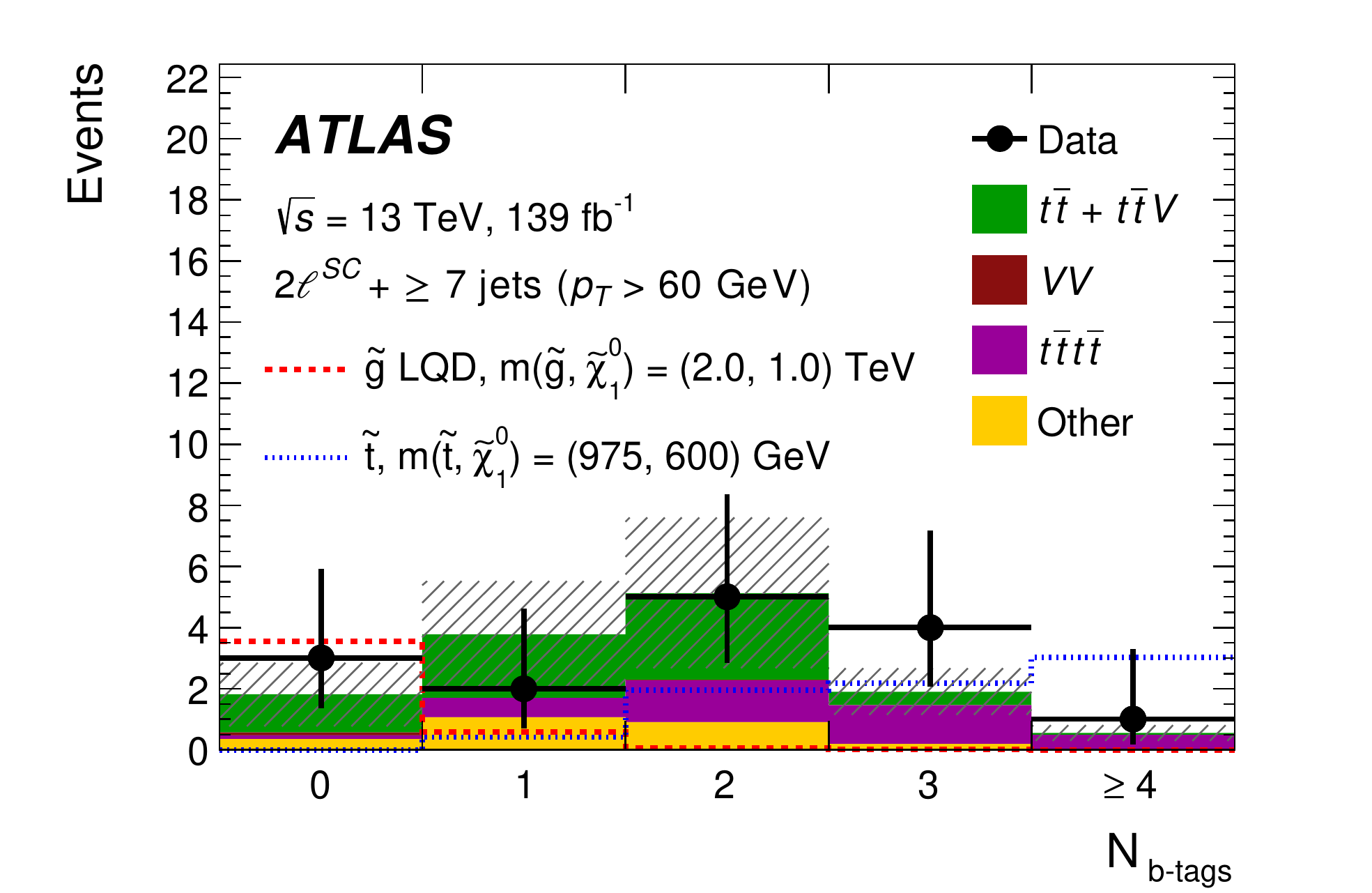} \\
\includegraphics[clip, trim=0 6cm 0 0, width=.4\textwidth]{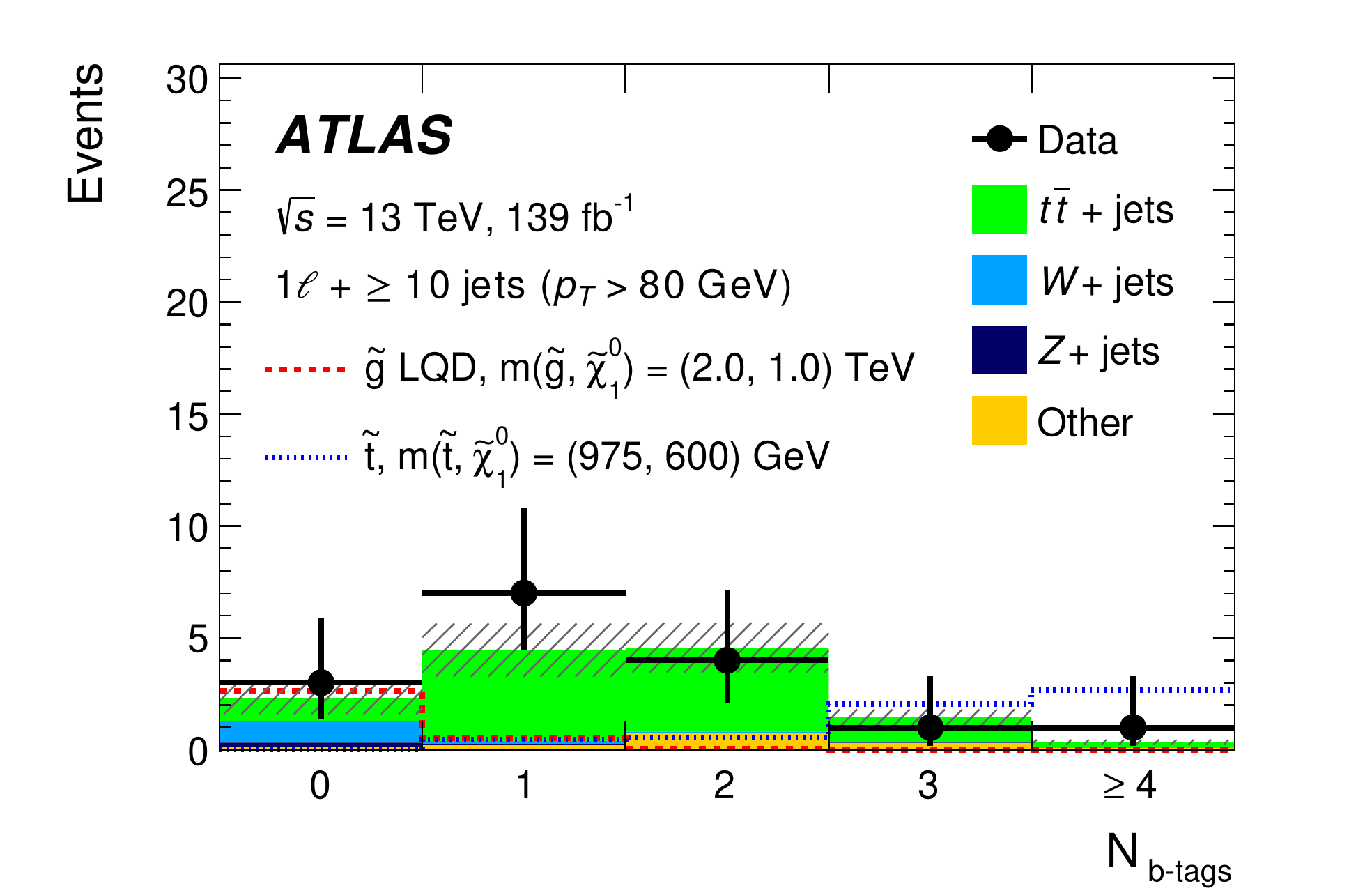}
\includegraphics[clip, trim=0 6cm 0 0, width=.4\textwidth]{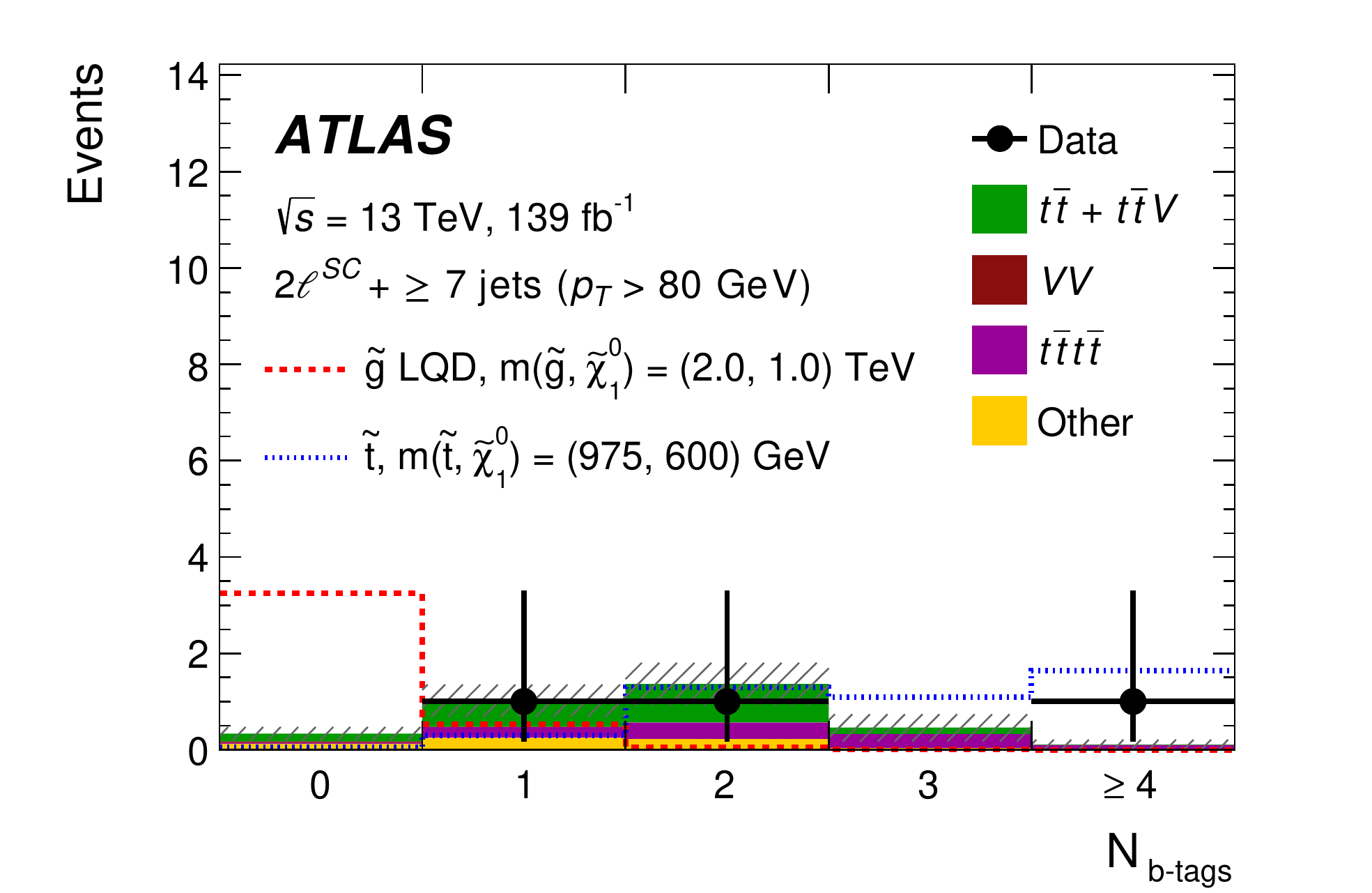} \\
\includegraphics[clip, trim=0 6cm 0 0, width=.4\textwidth]{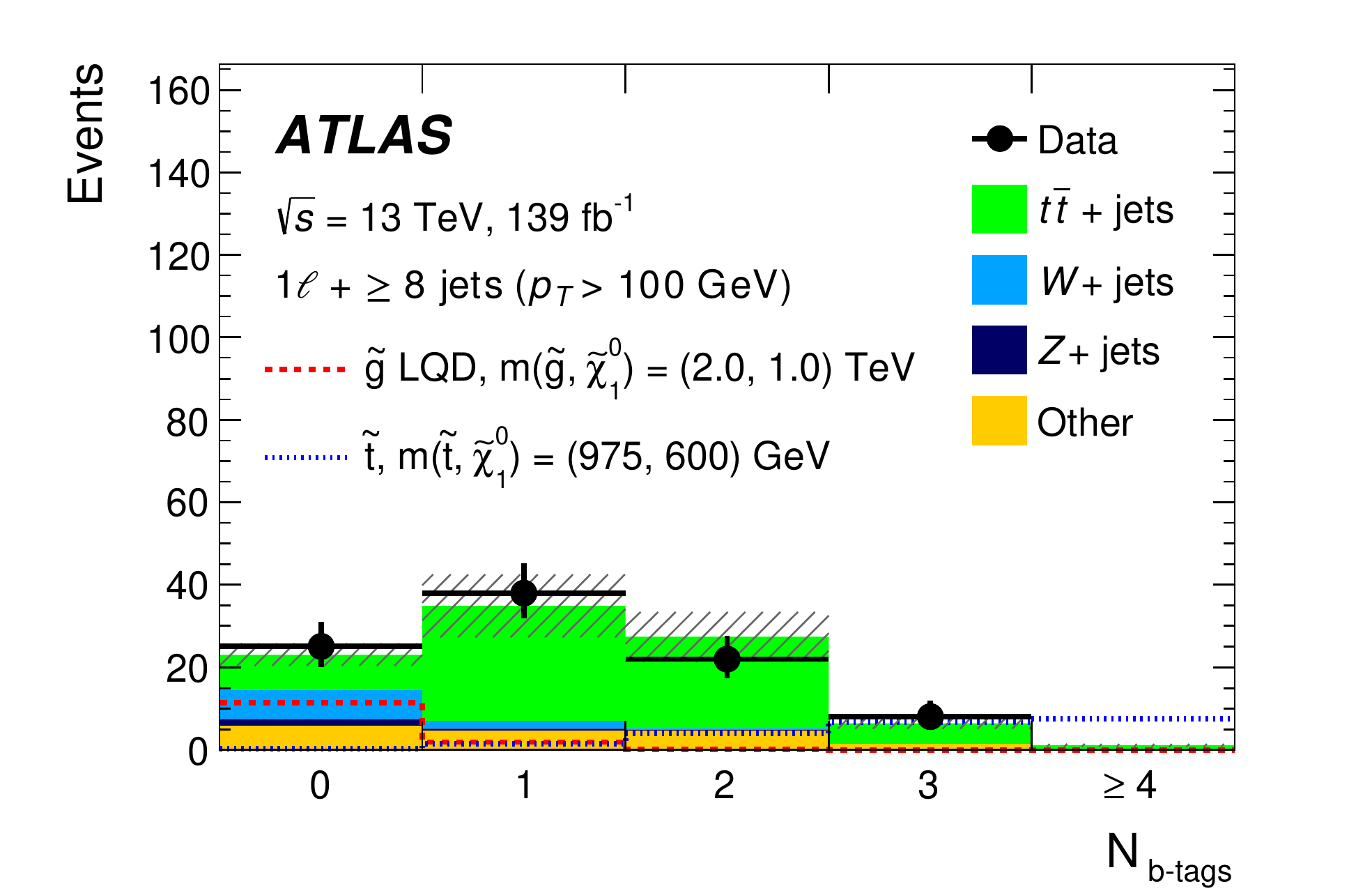}
\includegraphics[clip, trim=0 6cm 0 0, width=.4\textwidth]{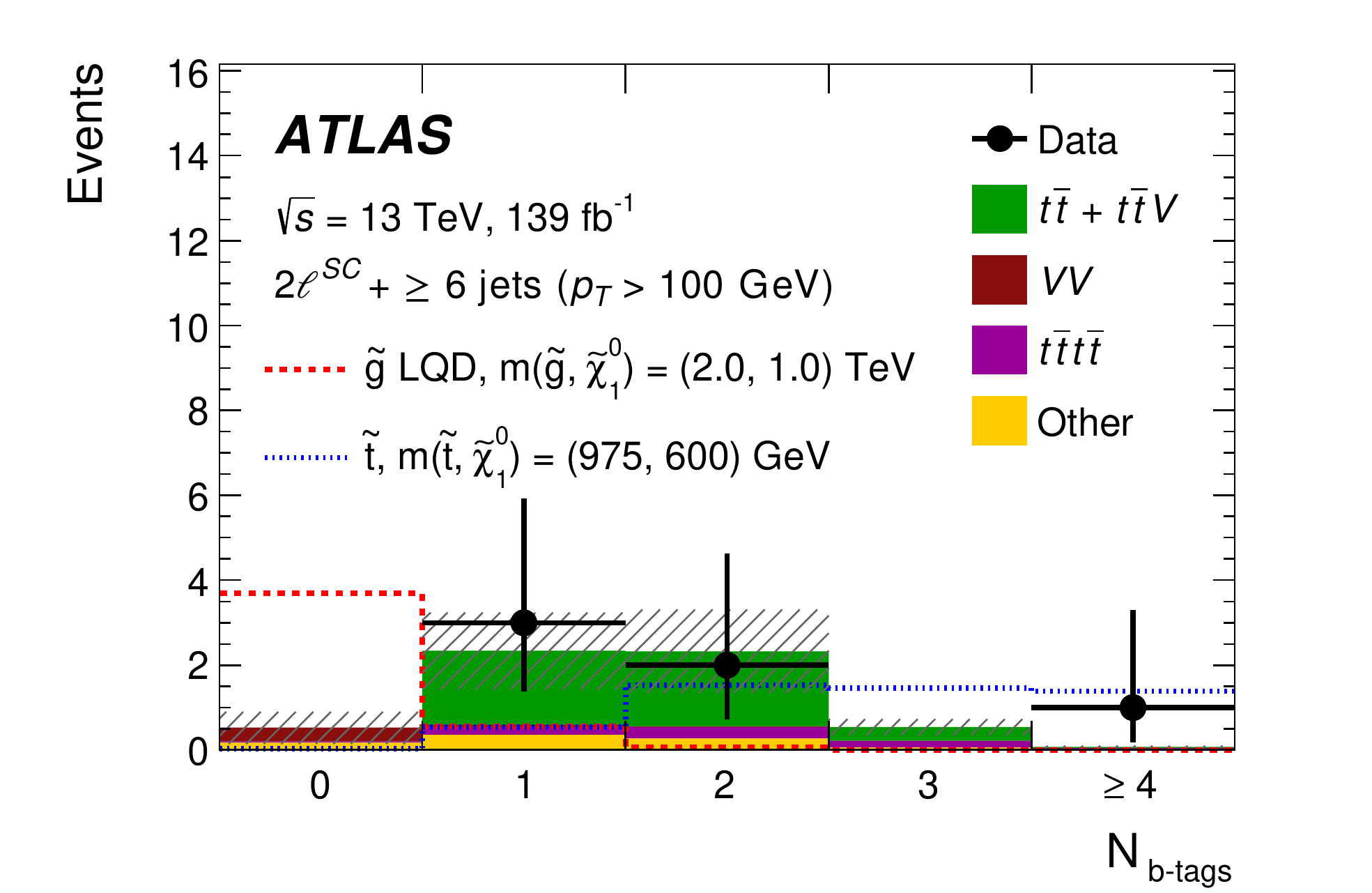} \\
\caption{
The observed data event yields and the corresponding estimates for
the backgrounds in the different last jet-multiplicity bins defined for the \l (left) and \ll (right) categories.
The background shown is estimated by including all bins in the fit.
All uncertainties, which may be correlated across the bins, are included in the error bands (shaded regions).
Hypothetical contributions from representative RPV SUSY scenarios are displayed as dashed and dotted lines.
}
\label{fig:unblinded_jet_counting}
\end{figure}

Results are provided both as model-independent limits on the contribution from BSM physics to the dedicated SRs,
and in the context of the five SUSY benchmark models discussed in Section~\ref{sec:signalModels}.
As described in Section~\ref{subsec:fit}, two different fit configurations are used for the two sets of results.
In both cases, the profile likelihood-ratio test~\cite{Cowan:2010js} is used to establish 95\% confidence intervals using the CL$_\text{s}$ prescription~\cite{Read:2002hq}.
The parameter of interest is the signal strength, defined as the cross-section of the hypothetical contribution from physics beyond the SM in units of the cross-section of the benchmark model.
 
The $b$-jet multiplicity distributions are shown in Figures~\ref{fig:unblinded_1L_20gev_shape} and~\ref{fig:unblinded_2LSS_20gev_shape}
for the EWK analysis defined with a 20~\gev\ jet-\pt threshold, for the \l and \ll categories respectively.
Figure~\ref{fig:pull_unblinded_20gev_shape} summarizes the observed pulls in all analysis regions, defined as the difference between the observed number of events and the total
expected number of events determined by the fit divided by the total uncertainty. The total uncertainty is the sum in quadrature of the statistical error of the observed data and the uncertainty in the predicted background.
The pulls follow a gaussian distribution centered at zero.
Figure~\ref{fig:unblinded_jet_counting} shows the $b$-jet multiplicity distribution for the last jet-multiplicity bin defined for each of the jet \pt\ thresholds, in both the \l and \ll categories, which contains the discovery SRs at zero $b$-jet and high $b$-jet multiplicity.
The likelihood fit is configured using the model-dependent configuration where all bins are input to the fit, and fixing the signal-strength parameter to zero.

\subsection{Model-independent results}
 
\begin{table}[t!]
\centering
\caption{
Data event yields compared with the expected contributions from relevant background sources, in the
discovery signal regions defined for the \l category.
The $p_0$-value, and corresponding significance (Z), as well as the observed and expected 95\% CL model-independent upper limits on the product of cross-section, acceptance and efficiency (in ab) are also shown ($\sigma^\text{excl}$).
In SRs with a deficit of data compared to the background prediction the $p_0$-value is capped at 0.5.
The parameters of the background model are determined in a fit to a reduced set of bins, corresponding to the model-independent fit discussed in the text.
}\def\arraystretch{1.4}
\begin{center}
\begin{tabular}{|r|l|lc|cccc|}
\hline
Jet \pt   & Selection            & Total        &   Data &   $p_0$ &   Z &    $\sigma^\text{excl}$ & $\sigma^\text{excl}$ \\
threshold & $\dagger$=NN bin 4   & background   &        &         &     &           obs. [ab]      &     exp. [ab]            \\
\hline
\multirow{3}{*}{20 GeV}  & $= \phantom{0}6 j$, $\geq 4 b$,$\dagger$    & $\phantom{.}145 \pm 6\phantom{.0}$      &    126 &    0.5\phantom{0}  &            0\phantom{.0}   &  130 & $200^{+80}_{-60}$ \\
& $\geq 15 j$, $= 0 b$                        & $16.3 \pm 5.2$   &     \phantom{0}16 &    0.5\phantom{0}  &            0\phantom{.0}   &  100 & $100^{+40}_{-30}$ \\
& $\geq 15 j$, $\geq 3 b$                     & $17.0 \pm 1.9$   &     \phantom{0}20 &    0.26 &            0.7 &   \phantom{0}92 & $\phantom{0}74^{+29}_{-21}$  \\
\hline
\multirow{2}{*}{40 GeV} & $\geq 12 j$, $= 0 b$                        & $\phantom{0}7.8 \pm 1.5$    &      \phantom{00}8 &    0.47 &            0.1 &    \phantom{0}55 & $\phantom{0}53^{+21}_{-15}$  \\
& $\geq 12 j$, $\geq 3 b$                     & $\phantom{0}8.6 \pm 1.1$    &     \phantom{0}12 &    0.17 &            0.9 &    \phantom{0}76 & $\phantom{0}55^{+21}_{-15}$  \\
\hline
\multirow{2}{*}{60 GeV} & $\geq 11 j$, $= 0 b$                        & $\phantom{0}2.7 \pm 0.8$    &      \phantom{00}3 &    0.44 &            0.2 &    \phantom{0}38 & $\phantom{0}35^{+14}_{-10}$  \\
& $\geq 11 j$, $\geq 3 b$                     & $\phantom{0}2.3 \pm 0.6$    &      \phantom{00}5 &    0.1\phantom{0}  &            1.3 &    \phantom{0}56 & $\phantom{0}33^{+13}_{-9}$   \\
\hline
\multirow{2}{*}{80 GeV} & $\geq 10 j$, $= 0 b$                        & $\phantom{0}2.1 \pm 1.1$    &      \phantom{00}3 &    0.38 &            0.3 &    \phantom{0}42 & $\phantom{0}34^{+13}_{-9}$   \\
& $\geq 10 j$, $\geq 3 b$                     & $\phantom{0}1.7 \pm 1.5$    &      \phantom{00}2 &    0.47 &            0.1 &    \phantom{0}35 & $\phantom{0}33^{+13}_{-9}$   \\
\hline
\multirow{2}{*}{100 GeV}& $\geq \phantom{0}8 j$, $= 0 b$              & $22.7 \pm 1.9$   &     \phantom{0}25 &    0.38 &            0.3 &    \phantom{0}96 & $\phantom{0}85^{+33}_{-24}$  \\
& $\geq \phantom{0}8 j$, $\geq 3 b$           & $\phantom{0}7.5 \pm 1.0$    &      \phantom{00}8 &    0.41 &            0.2 &    \phantom{0}55 & $\phantom{0}51^{+20}_{-14}$  \\
\hline
\end{tabular}

\label{tab:SR_1L}
\end{center}
\end{table}
 
\begin{table}[t!]
\centering
\caption{
Data event yields compared with the expected contributions from relevant background sources, in the
discovery signal regions defined for the \ll category.
The $p_0$-value, and corresponding significance (Z), as well as the observed and expected 95\% CL model-independent upper limits on the product of cross-section, acceptance and efficiency (in ab) are also shown.
In SRs with a deficit of data compared to the background prediction the $p_0$-value is capped at 0.5.
The parameters of the background model are determined in a fit to a reduced set of bins, corresponding to the model-independent fit discussed in the text.
}
\def\arraystretch{1.4}
\begin{center}
\begin{tabular}{|r|l|lc|cccc|}
\hline
Jet \pt   & Selection                         & Total        &   Data &   $p_0$ &   Z &    $\sigma^\text{excl}$  & $\sigma^\text{excl}$ \\
threshold & $\dagger$=$\minmaxmlj<155~\gev$    & background  &        &         &     &     obs. [ab]      &     exp. [ab]            \\
\hline
\multirow{3}{*}{20 GeV}  & $= \phantom{0}6 j$,$\geq 3 b$,$\dagger$     & $16.1 \pm 1.2$     &     20 &    0.21 &            0.8 &       92 & $69^{+27}_{-19}$  \\
& $\geq 10 j$, $= 0 b$                        & $\phantom{0}5.8 \pm 0.8$      &      \phantom{0}6 &    0.46 &            0.1 &       48 & $45^{+18}_{-13}$  \\
& $\geq 10 j$, $\geq 3 b$                     & $\phantom{0}8.2 \pm 1.5$      &      \phantom{0}6 &    0.5\phantom{0}  &            0\phantom{.0}   &       41 & $54^{+21}_{-15}$  \\
\hline
\multirow{2}{*}{40 GeV} & $\geq \phantom{0}8 j$, $= 0 b$               & $\phantom{0}2.8 \pm 0.7$      &      \phantom{0}2 &    0.5\phantom{0}  &            0\phantom{.0}   &       31 & $35^{+14}_{-10}$  \\
& $\geq \phantom{0}8 j$, $\geq 3 b$            & $\phantom{0}3.6 \pm 1.2$      &      \phantom{0}7 &    0.13 &            1.1 &       67 & $39^{+15}_{-11}$  \\
\hline
\multirow{2}{*}{60 GeV} & $\geq \phantom{0}7 j$, $= 0 b$               & $1.71 \pm 0.35$    &      \phantom{0}3 &    0.2\phantom{0}  &            0.8 &       41 & $29^{+12}_{-8}$   \\
& $\geq \phantom{0}7 j$, $\geq 3 b$            & $\phantom{0}2.0 \pm 0.7$      &      \phantom{0}5 &    0.09 &            1.3 &       58 & $32^{+13}_{-9}$   \\
\hline
\multirow{2}{*}{80 GeV} & $\geq \phantom{0}7 j$, $= 0 b$               & $0.34 \pm 0.13$    &      \phantom{0}0 &    0.5\phantom{0}  &            0\phantom{.0}   &       22 & $22^{+9}_{-0}$    \\
& $\geq \phantom{0}7 j$, $\geq 3 b$            & $0.54 \pm 0.20$    &      \phantom{0}1 &    0.34 &            0.4 &       27 & $22^{+9}_{-0}$    \\
\hline
\multirow{2}{*}{100 GeV}& $\geq \phantom{0}6 j$, $= 0 b$               & $\phantom{0}0.5 \pm 0.4$      &      \phantom{0}0 &    0.5\phantom{0}  &            0\phantom{.0}   &       22 & $22^{+9}_{-0}$    \\
& $\geq \phantom{0}6 j$, $\geq 3 b$            & $0.52 \pm 0.22$    &      \phantom{0}1 &    0.28 &            0.6 &       28 & $22^{+9}_{-0}$    \\
\hline
\end{tabular}

\label{tab:SR_2LSS}
\end{center}
\end{table}
 
The observed data event yields and the corresponding estimates for
the backgrounds in the discovery SRs defined for the \l and \ll categories are shown in Tables~\ref{tab:SR_1L} and~\ref{tab:SR_2LSS}.
For each SR a fit is performed to predict the background using the model-independent set-up, which excludes the SR under consideration.
In addition, the discovery $p_0$-values and corresponding gaussian significance (Z) are shown, which measure the compatibility of the observed data with the background-only (zero signal strength) hypothesis relative to fluctuations of the background. Larger values indicate greater relative compatibility.
No significant excess of data over the expected event yields is observed in any of the SRs.
The two largest excesses are observed in the \SI{60}{\gev}, $\geq 11$ jets, $\geq 3$ $b$-jets SR defined for the \l category and
in the 60 \gev, $\geq 7$ jets, $\geq 3$ $b$-jets SR defined for the \ll category, and both correspond to a significance of 1.3 standard deviations.
Upper limits on the product of cross-section, acceptance, and efficiency are set at 95\% CL, ranging from 22 to 200 ab, depending on the SR.

\subsection{Model-dependent results}
\begin{figure}[htb]
\centering
\subcaptionbox{}{\includegraphics[width=.49\textwidth]{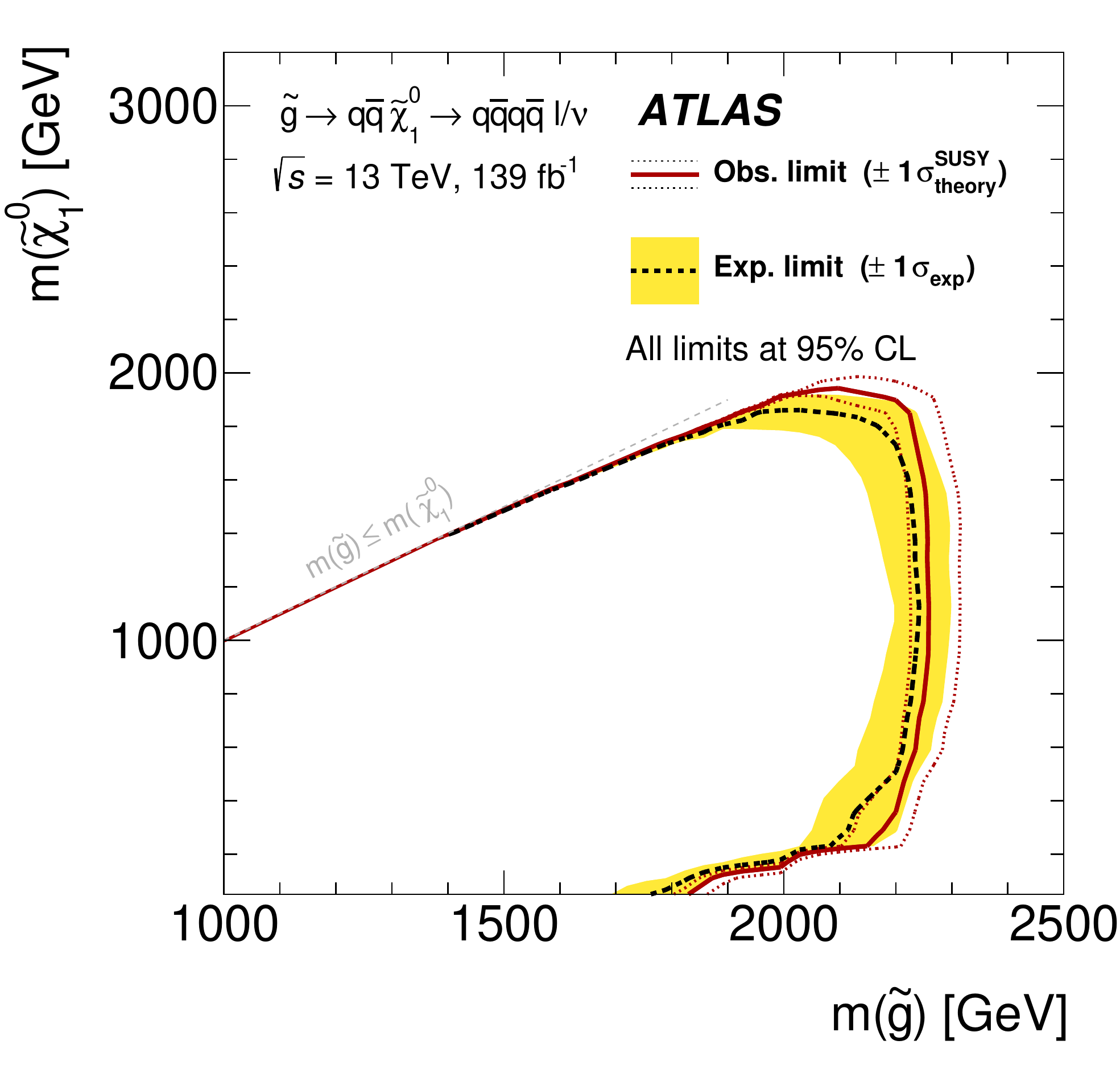}}
\subcaptionbox{}{\includegraphics[width=.49\textwidth]{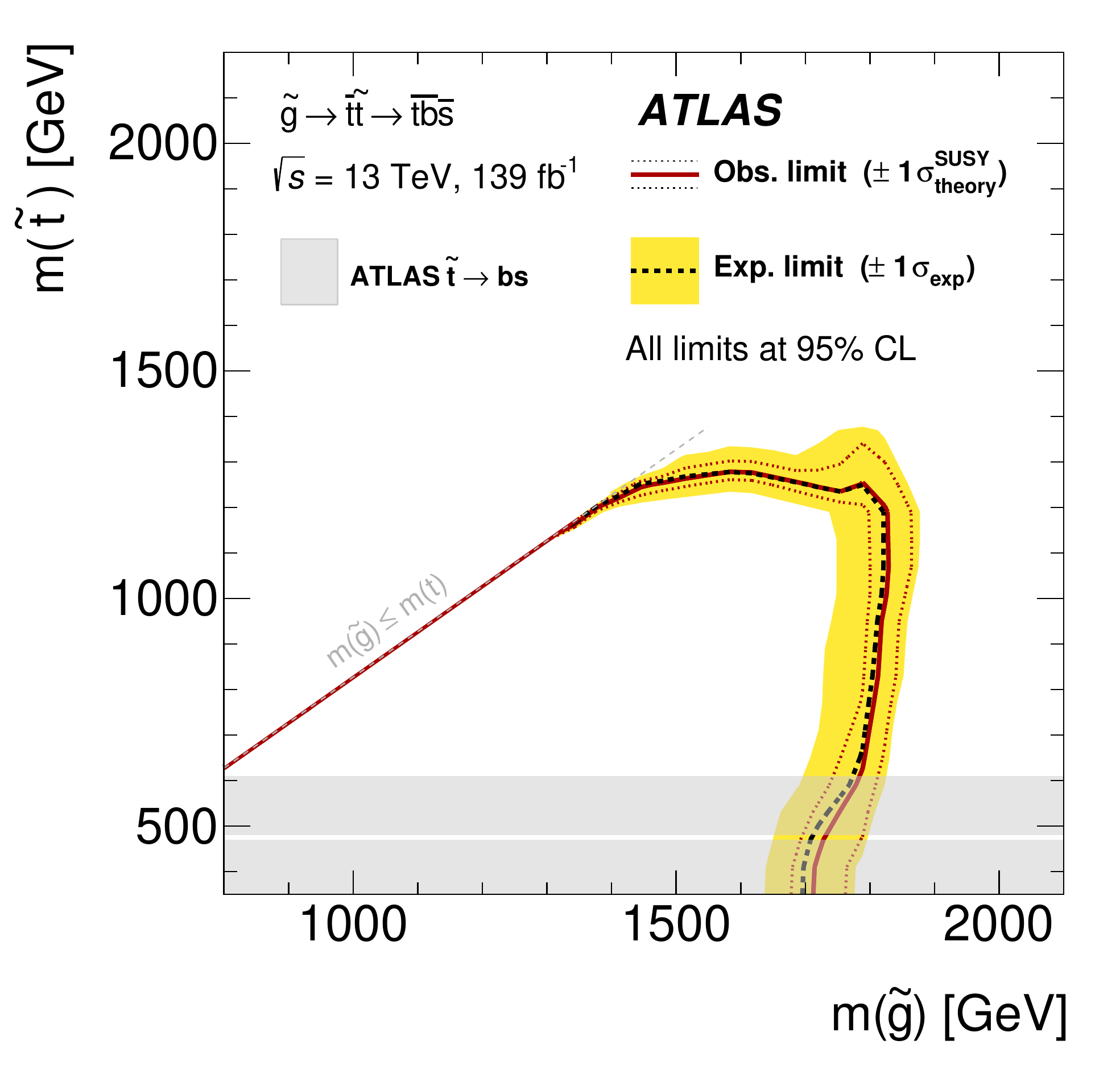}}
\hspace{0.cm}
\subcaptionbox{}{\includegraphics[width=.49\textwidth]{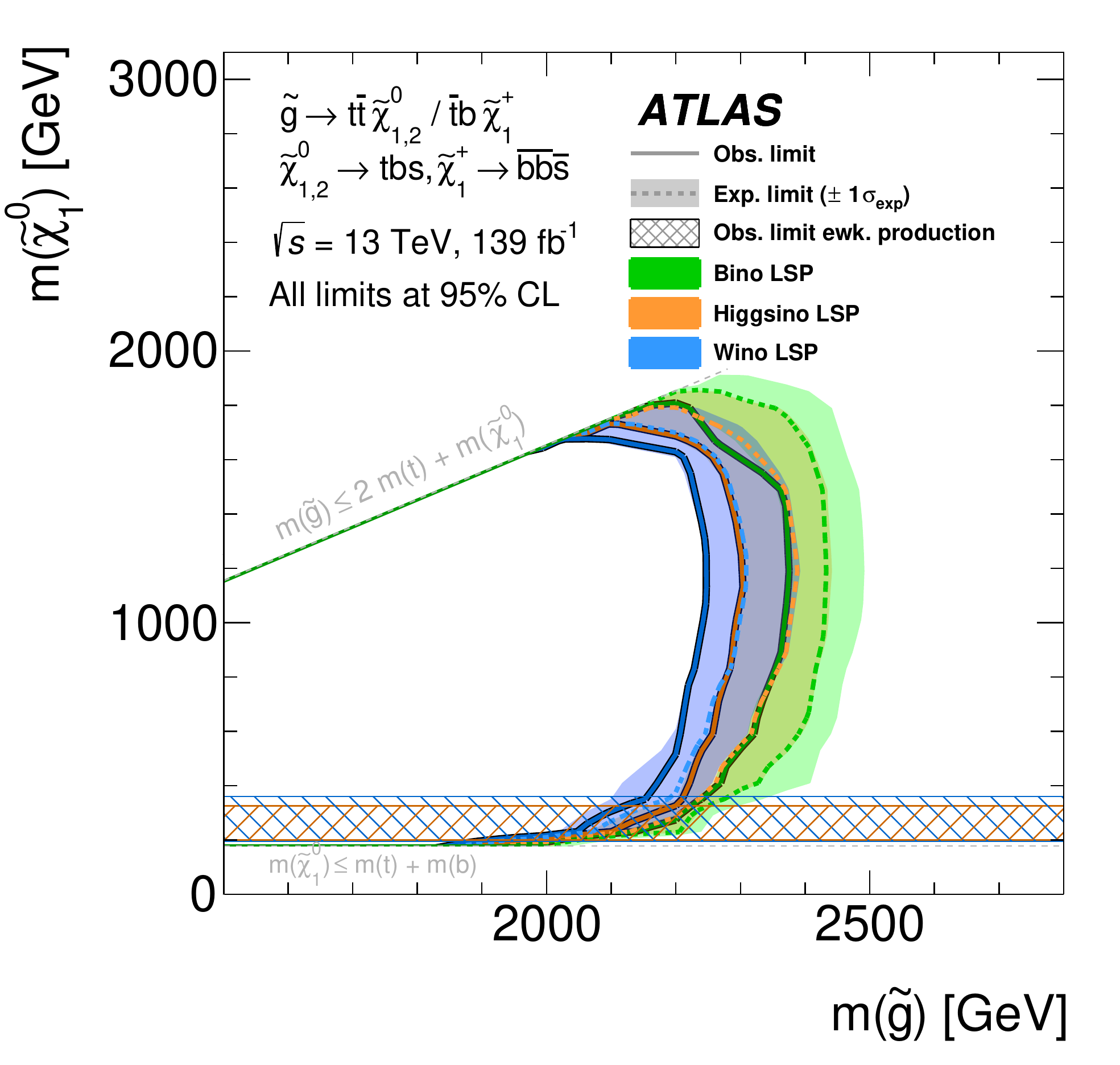}}
\subcaptionbox{}{\includegraphics[width=.49\textwidth]{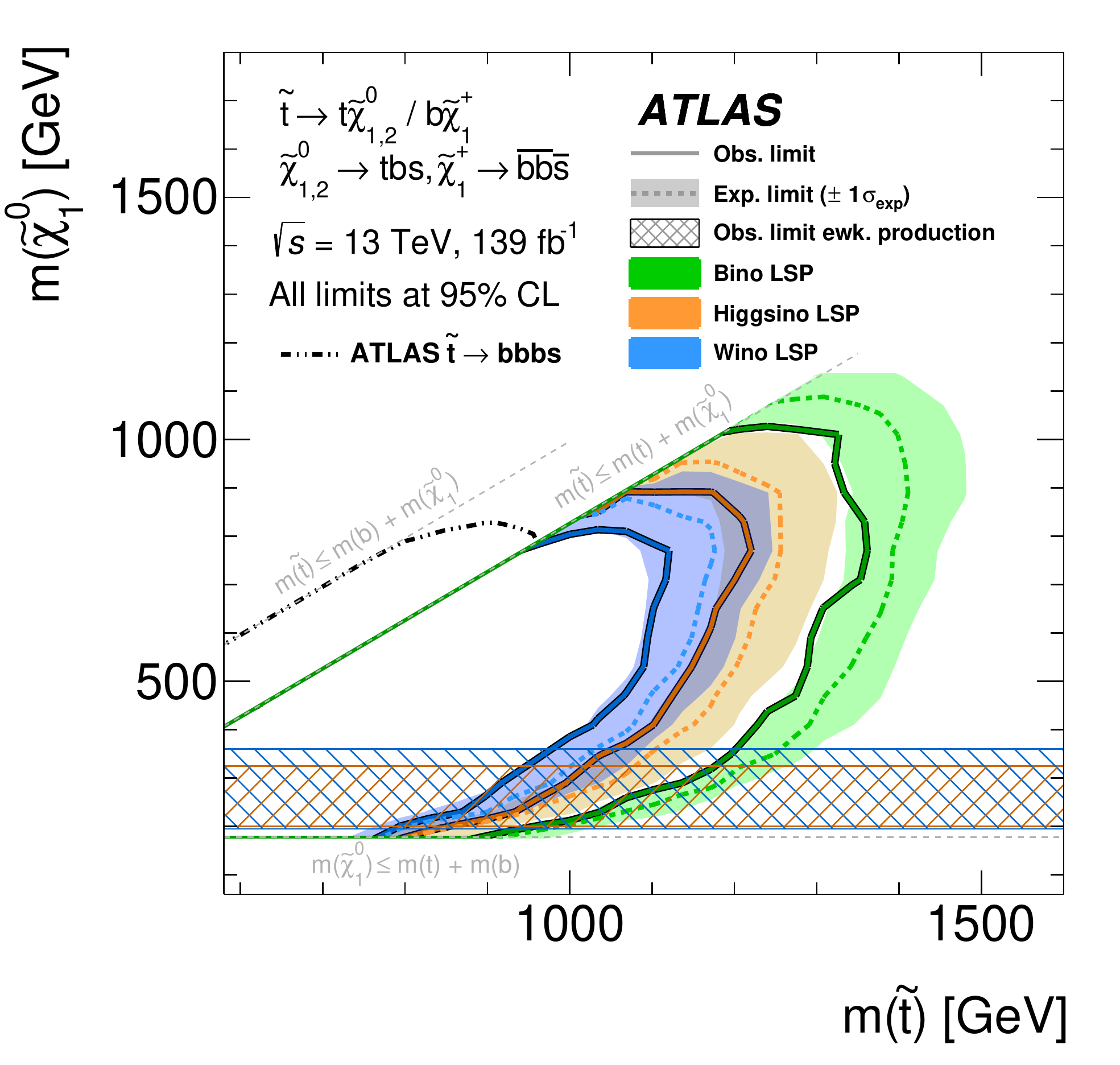}}
\caption{
Observed and expected exclusion contours for the RPV models with strong production.
The results are shown for (a) \GGLQD, (b) \GGRpvlamppThreeThreeOne, (c) \GTT/\GTB\ and (d) stop pair production.
The contours of the band around the expected limit are the $\pm 1 \sigma$ variations, including all uncertainties.
The dotted lines around the observed limit illustrate the change in the
observed limit as the nominal signal cross-section is scaled up and down by the theoretical uncertainty.
All limits are computed at 95\% CL.
The diagonal line indicates the kinematic limit for the decays in each specified scenario.
The limits on direct electroweakino production obtained with the EWK analysis are displayed as horizontal hatched bands in (c) and (d).
When relevant, the limit on the stop mass from Refs.~\cite{SUSY-2016-09,Aad:2020uwr} is also shown. A small range in stop mass between 460 \gev\ and 470 \gev\ is not excluded by the search for $\tilde{t} \rightarrow bs$~\cite{SUSY-2016-09}.
}
\label{fig:exclusion_strong}
\end{figure}
 
\begin{figure}[htb]
\centering
\subcaptionbox{}{\includegraphics[width=.49\textwidth]{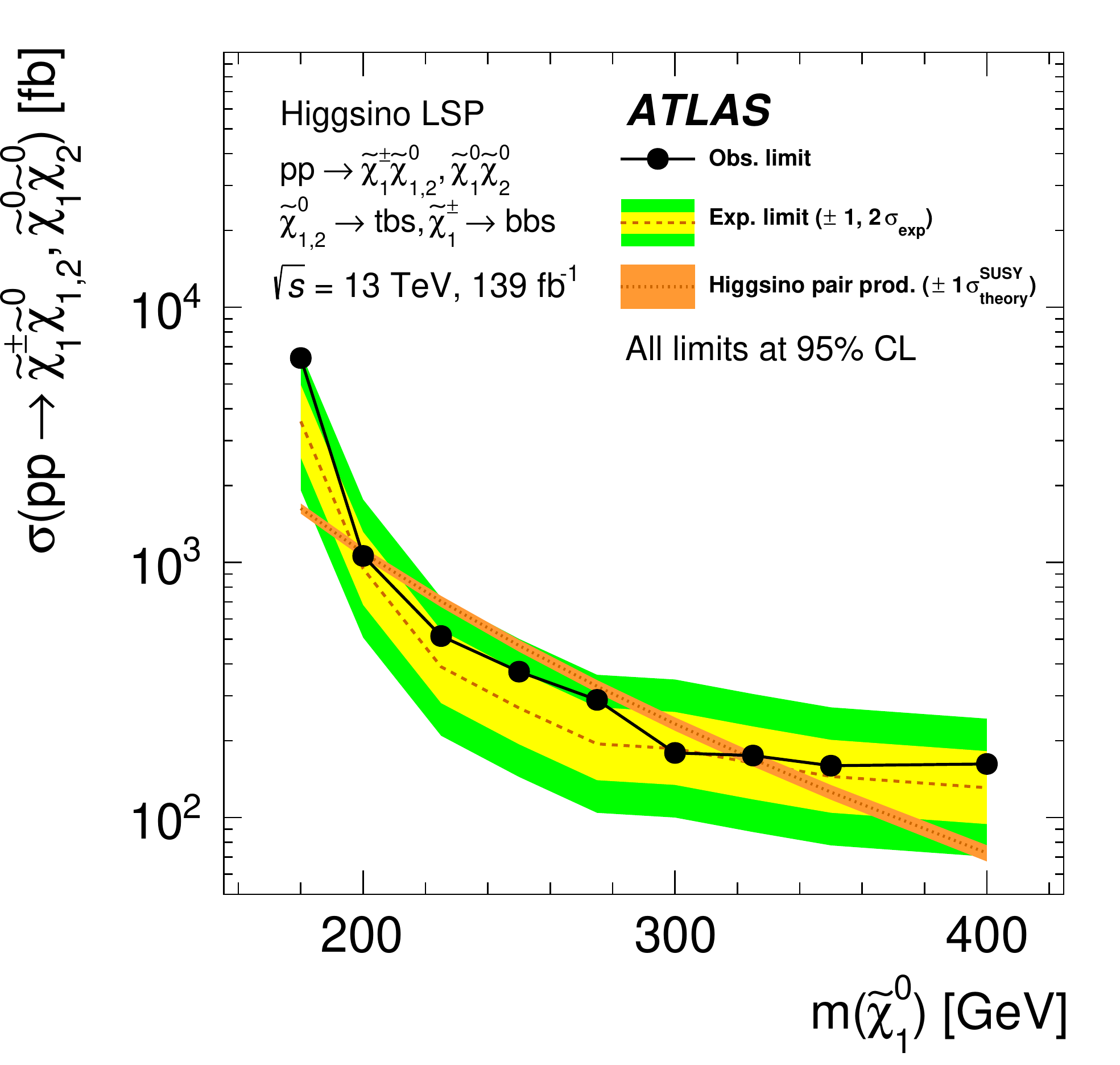}}
\subcaptionbox{}{\includegraphics[width=.49\textwidth]{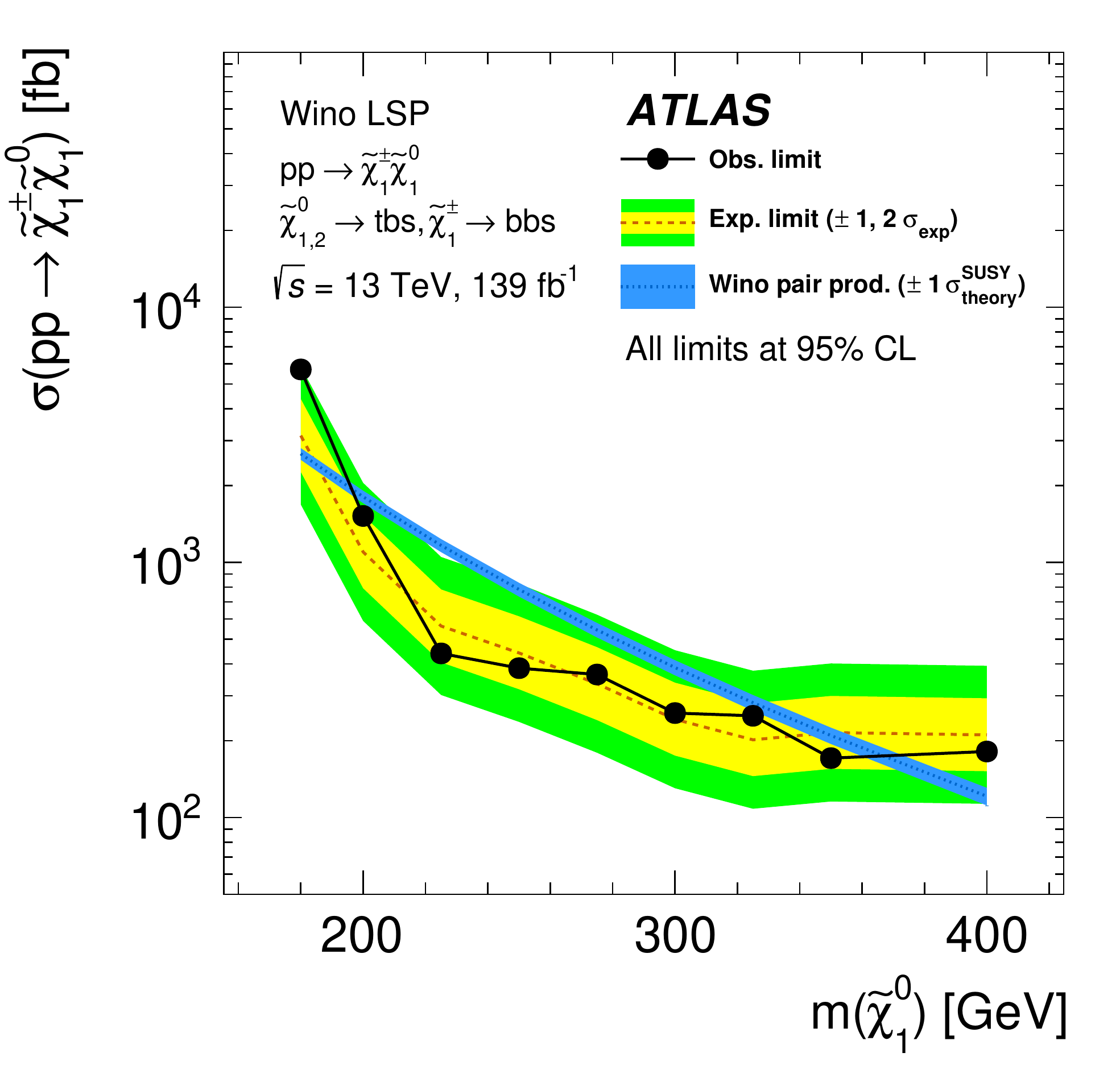}}
\caption{
Observed and expected exclusion contours for the RPV models with electroweakino production models with (a) higgsino and (b) wino LSP hypotheses.
The yellow and green contours of the band around the expected limit are the $\pm 1 \sigma$ and $\pm 2 \sigma$ variations including all uncertainties, respectively.
The theoretical prediction is also shown, with the uncertainties in the prediction shown as a coloured band.
The production of $\chinoonepm\chinoonepm$ is not considered as it decays to a final state with no leptons.
}
\label{fig:exclusion_higgsino}
\end{figure}

For each signal model probed, the fit is configured using the model-dependent configuration.
No significant excess is observed in any of the model-dependent fits.
Figure~\ref{fig:exclusion_strong}
shows the observed and expected exclusion limits for the strong production signal
models featuring gluino and stop pair production, as a function of the gluino mass or stop mass.
Gluino masses up to \SI{2.4}{\TeV} are excluded for high LSP masses, and up to \SI{2}{\TeV} for low LSP masses.
Stop masses are excluded up to 1--\SI{1.3}{\TeV}, depending on the LSP mass.
The sensitivity decreases for low LSP masses due to the high boost of the LSP, resulting in close-by decay products that lead to reconstruction and isolation inefficiencies.
The best sensitivity is achieved for a bino LSP, while the wino LSP exhibits the worst sensitivity due to the lower number of top quarks in the final state.
The exclusion limits for the models with gluino or stop production are stronger than in the previous version of the analysis documented in Ref.~\cite{Aaboud:2017faq},
thanks to the larger dataset and the inclusion of the same-sign lepton category.

Figure~\ref{fig:exclusion_higgsino} shows exclusion limits in the electroweakino pair-production model, versus the LSP mass.
The limits for pure higgsino and wino LSPs are shown separately, taking into account the processes discussed in Section~\ref{sec:signalModels}. The wino signal can only contribute to the \l category, via $\chinoonepm\ninoone$ production, while the higgsino signal is also present in the \ll category through $\ninotwo\ninoone$ production. This leads to differences in the observed limits between the two models.
Depending on the LSP hypothesis, LSP masses between 200 (197)~\gev\ and 320 (365)~\gev\ are excluded for a higgsino (wino) LSP.

The analysis is also sensitive to SM \tttt\ production, which produces a final state similar to the targeted signals.
An additional model-dependent fit is performed where the \tttt\ normalization is a free parameter.
The fitted normalization of the four-top process relative to the Standard Model value is ${\mu_{\tttt} = 2.0^{+0.9}_{-0.7}}$.
Modelling uncertainties due to scale variations and parton-shower variation are taken into account, as is a 20\% cross-section uncertainty for the reference SM prediction of $\sigma_{\tttt}$~\cite{Frederix:2017wme}.
This is in agreement with the measured value in Ref.~\cite{Aad:2020klt} of ${\mu_{\tttt} = 2.0^{+0.8}_{-0.6}}$.
Both analyses are based on the same dataset and have overlapping selections, but have completely different background estimation methods.
The best sensitivity is obtained with the 40~\gev\ jet-\pt\ threshold.
Compatible results, albeit with larger uncertainties, are obtained with the 20~\gev\ and 60~\gev\ jet-\pt thresholds. A fit with two independent signal strengths in each lepton category yields consistent values in both categories.


\FloatBarrier
\section{Conclusion}
A search for RPV supersymmetry events with at least one isolated lepton (electron or muon) and high jet multiplicity is presented.
In order to improve the sensitivity of the search, events with two leptons with the same electric charge, and events with at least one lepton are analysed separately.
The selection also relies on the number of $b$-jets in the event.
In order to ensure the highest sensitivity to the electroweak production models, a neural-network-based analysis was introduced.
Data-based techniques are used to estimate the dominant backgrounds from $t\bar{t}$+jets, $W/Z$+jets, diboson, and $\ttW$ production.
The analysis is performed with proton--proton collision data at $\sqrt{s}=13~\TeV$ collected from 2015 to 2018 with the ATLAS detector at the LHC,
corresponding to an integrated luminosity of 139~fb$^{-1}$.
 
With no significant excess over the Standard Model expectation observed, results are interpreted in the framework of simplified models featuring gluino, stop, or electroweakino pair production in RPV supersymmetry scenarios.
In a benchmark model with $\gluino\to t\bar t\ninoone \to  t\bar t t b s$, gluino masses up to 2.38~\TeV\ are excluded at 95\% confidence level.
Top squarks with masses up to 1.36~\TeV\ are excluded in a model with direct stop production and RPV decays of the LSP. In both models, three hypotheses for the LSP are tested: pure bino, pure wino, and pure higgsino.
In a model with $\gluino\to \bar t \stop$ and $\stop\to \bar b \bar s$, gluino masses up to 1.83~\TeV\ are excluded, whereas in a model with $\gluino \to q \bar q \ninoone \to q \bar q q \bar q \ell / \nu$, gluino masses up to 2.25~\TeV\ are excluded.
Direct electroweak production of electroweakinos is also tested, and higgsino (wino) masses between 200 (197)~\gev\ and 320 (365)~\gev\ are excluded.
 
These results improve upon the previously existing LHC limits for the gluino and stop production models considered, owing to the larger luminosity, the dedicated categorization and analysis of events with two leptons with the same electric charge,
and the introduction of multivariate discriminants.
The results for the electroweak production model also improve upon the limits on hadronic RPV decays of electroweakinos from LEP~\cite{Abdallah:2003xc,Heister:2002jc,Achard:2001ek}.
Model-independent limits are also set on the contribution of new phenomena to the signal-region yields.

\clearpage

\section*{Acknowledgements}

We thank CERN for the very successful operation of the LHC, as well as the
support staff from our institutions without whom ATLAS could not be
operated efficiently.
 
We acknowledge the support of
ANPCyT, Argentina;
YerPhI, Armenia;
ARC, Australia;
BMWFW and FWF, Austria;
ANAS, Azerbaijan;
SSTC, Belarus;
CNPq and FAPESP, Brazil;
NSERC, NRC and CFI, Canada;
CERN;
ANID, Chile;
CAS, MOST and NSFC, China;
Minciencias, Colombia;
MSMT CR, MPO CR and VSC CR, Czech Republic;
DNRF and DNSRC, Denmark;
IN2P3-CNRS and CEA-DRF/IRFU, France;
SRNSFG, Georgia;
BMBF, HGF and MPG, Germany;
GSRI, Greece;
RGC and Hong Kong SAR, China;
ISF and Benoziyo Center, Israel;
INFN, Italy;
MEXT and JSPS, Japan;
CNRST, Morocco;
NWO, Netherlands;
RCN, Norway;
MEiN, Poland;
FCT, Portugal;
MNE/IFA, Romania;
JINR;
MES of Russia and NRC KI, Russian Federation;
MESTD, Serbia;
MSSR, Slovakia;
ARRS and MIZ\v{S}, Slovenia;
DSI/NRF, South Africa;
MICINN, Spain;
SRC and Wallenberg Foundation, Sweden;
SERI, SNSF and Cantons of Bern and Geneva, Switzerland;
MOST, Taiwan;
TAEK, Turkey;
STFC, United Kingdom;
DOE and NSF, United States of America.
In addition, individual groups and members have received support from
BCKDF, CANARIE, Compute Canada and CRC, Canada;
COST, ERC, ERDF, Horizon 2020 and Marie Sk{\l}odowska-Curie Actions, European Union;
Investissements d'Avenir Labex, Investissements d'Avenir Idex and ANR, France;
DFG and AvH Foundation, Germany;
Herakleitos, Thales and Aristeia programmes co-financed by EU-ESF and the Greek NSRF, Greece;
BSF-NSF and GIF, Israel;
Norwegian Financial Mechanism 2014-2021, Norway;
NCN and NAWA, Poland;
La Caixa Banking Foundation, CERCA Programme Generalitat de Catalunya and PROMETEO and GenT Programmes Generalitat Valenciana, Spain;
G\"{o}ran Gustafssons Stiftelse, Sweden;
The Royal Society and Leverhulme Trust, United Kingdom.
 
The crucial computing support from all WLCG partners is acknowledged gratefully, in particular from CERN, the ATLAS Tier-1 facilities at TRIUMF (Canada), NDGF (Denmark, Norway, Sweden), CC-IN2P3 (France), KIT/GridKA (Germany), INFN-CNAF (Italy), NL-T1 (Netherlands), PIC (Spain), ASGC (Taiwan), RAL (UK) and BNL (USA), the Tier-2 facilities worldwide and large non-WLCG resource providers. Major contributors of computing resources are listed in Ref.~\cite{ATL-SOFT-PUB-2021-003}.
 

\clearpage
 
\printbibliography

\clearpage 
 
\begin{flushleft}
\hypersetup{urlcolor=black}
{\Large The ATLAS Collaboration}

\bigskip

\AtlasOrcid[0000-0002-6665-4934]{G.~Aad}$^\textrm{\scriptsize 99}$,    
\AtlasOrcid[0000-0002-5888-2734]{B.~Abbott}$^\textrm{\scriptsize 125}$,    
\AtlasOrcid[0000-0002-7248-3203]{D.C.~Abbott}$^\textrm{\scriptsize 100}$,    
\AtlasOrcid[0000-0002-2788-3822]{A.~Abed~Abud}$^\textrm{\scriptsize 34}$,    
\AtlasOrcid[0000-0002-1002-1652]{K.~Abeling}$^\textrm{\scriptsize 51}$,    
\AtlasOrcid[0000-0002-2987-4006]{D.K.~Abhayasinghe}$^\textrm{\scriptsize 91}$,    
\AtlasOrcid[0000-0002-8496-9294]{S.H.~Abidi}$^\textrm{\scriptsize 27}$,    
\AtlasOrcid[0000-0001-5329-6640]{H.~Abramowicz}$^\textrm{\scriptsize 158}$,    
\AtlasOrcid[0000-0002-1599-2896]{H.~Abreu}$^\textrm{\scriptsize 157}$,    
\AtlasOrcid[0000-0003-0403-3697]{Y.~Abulaiti}$^\textrm{\scriptsize 5}$
% ...
\bigskip
\\

% ...
$^{5}$High Energy Physics Division, Argonne National Laboratory, Argonne IL; United States of America.\\
% ...
$^{27}$Physics Department, Brookhaven National Laboratory, Upton NY; United States of America.\\
% ...
$^{34}$CERN, Geneva; Switzerland.\\
% ...
$^{51}$II. Physikalisches Institut, Georg-August-Universit\"{a}t G\"ottingen, G\"ottingen; Germany.\\
% ...
$^{91}$Department of Physics, Royal Holloway University of London, Egham; United Kingdom.\\
% ...
$^{99}$CPPM, Aix-Marseille Universit\'e, CNRS/IN2P3, Marseille; France.\\
$^{100}$Department of Physics, University of Massachusetts, Amherst MA; United States of America.\\
% ...
$^{125}$Homer L. Dodge Department of Physics and Astronomy, University of Oklahoma, Norman OK; United States of America.\\
% ...
$^{157}$Department of Physics, Technion, Israel Institute of Technology, Haifa; Israel.\\
$^{158}$Raymond and Beverly Sackler School of Physics and Astronomy, Tel Aviv University, Tel Aviv; Israel.\\
% ...

\end{flushleft}


\end{document}